\documentclass{article}

\usepackage{arxiv}

\usepackage[utf8]{inputenc} 
\usepackage[T1]{fontenc}    
\usepackage{hyperref}       
\usepackage{url}            
\usepackage{booktabs}       
\usepackage{amsfonts}       
\usepackage{nicefrac}       
\usepackage{microtype}      

\usepackage{lipsum}
\usepackage{graphicx}
\usepackage{multirow}
\usepackage{color,array,dcolumn}
\usepackage{amsmath}
\usepackage{amssymb}
\usepackage[export]{adjustbox}
\usepackage{bbm}
\usepackage{caption}

\newcolumntype{L}[1]{>{\raggedright\let\newline\\\arraybackslash\hspace{0pt}}m{#1}}
\newcolumntype{C}[1]{>{\centering\let\newline\\\arraybackslash\hspace{0pt}}m{#1}}
\newcolumntype{R}[1]{>{\raggedleft\let\newline\\\arraybackslash\hspace{0pt}}m{#1}}

  \usepackage[caption=false,font=normalsize,labelfont=sf,textfont=sf]{subfig}
\begin{document}
\title{Multimodal 4DVarNets for the reconstruction of sea surface dynamics from SST-SSH synergies}
\author{
  R. Fablet\thanks{Corresponding author} \\
  IMT Atlantique, Lab-STICC,
  Brest, FR \\
  \texttt{ronan.fablet@imt-atlantique.fr} \\
   \And
  Q. Febvre \\
  IMT Atlantique, Lab-STICC,
  Brest, FR \\
  \texttt{quentin.febvre@imt-atlantique.fr} \\
   \And
  B. Chapron \\
  Ifremer, LOPS,
  Brest, FR \\
  \texttt{bertrand.chapron@ifremer.fr} \\}

\maketitle

\begin{abstract}
The space-time reconstruction of ocean dynamics from satellite observations is a challenging inverse problem due to the associated irregular sampling of the sea surface. Satellite altimetry provides a direct observation of the sea surface height (SSH), which relates to the divergence-free component of sea surface currents. The associated sampling pattern prevents from retrieving fine-scale dynamics, typically below 10 days. By contrast, other satellite sensors provide higher-resolution observations of sea surface tracers such as sea surface temperature (SST). Multimodal inversion schemes then arise as appealing approaches. Though theoretical evidence supports the existence of an explicit relationship between sea surface temperature and sea surface dynamics under specific dynamical regimes, the generalization to the variety of upper ocean dynamical regimes is complex. Here, we investigate this issue from a physics-informed learning perspective. We introduce a trainable multimodal inversion scheme for the reconstruction of sea surface dynamics from multi-source satellite-derived observations, namely satellite-derived SSH and SST data. The proposed 4DVarNet schemes combine a variational formulation involving trainable observation and {\em a priori} terms with a trainable gradient-based solver.  An observing system simulation experiment for a Gulf Stream region supports the relevance of our approach compared with state-of-the-art schemes. We report a relative improvement greater than 60\% compared with the operational altimetry product in terms of root mean square error and resolved space-time scales. We discuss further the application and extension of the proposed approach for the reconstruction and forecasting of geophysical dynamics from irregularly-sampled satellite observations.
\end{abstract}



%

\section{Introduction}
\label{sec:intro}

Satellite altimeters provide the main source of observation data to inform sea surface dynamics on a regional and global scale \cite{chelton_satellite_2001,abdalla_altimetry_2021}. As illustrated in Fig.\ref{fig:obs data}, the associated daily sampling at sea surface on a global scale remains very scarce for current nadir altimeter constellations. This results in relatively low-resolution reconstructions of sea surface dynamics delivered by operational altimetry-derived products using model-driven and observation-driven schemes \cite{taburet_duacs_2019,benkiran_assessing_2021, le_guillou_mapping_2020}. We can emphasize that ret
rieving mesoscale sea surface dynamics, typically from a few tens of kilometers in terms of horizontal scales, is key for a wide range of scientific topics and applications \cite{chelton_satellite_2001,abdalla_altimetry_2021} such as among others weather and ocean forecasting, climate modeling, ecological studies, maritime traffic routing, offshore activities....

While future altimetry missions such as SWOT will improve the spatial sampling with wide-swath sensors, the associated time sampling will likely remain too scarce for a while to retrieve fine-scale sea surface dynamics \cite{le_guillou_mapping_2020,gaultier_challenge_2015}. Numerous other satellite sensors deliver sea surface observations which may inform sea surface dynamics. Among others, we may cite SST (Sea Surface Temperature) products \cite{donlon_operational_2012} derived from micro-wave and infrared sensors as well as ocean colour (OC) products from multispectral sensors \cite{ciani_ocean_2021}. These wide-swath sensors result in a much denser space-time sampling of the sea surface and the associated satellite-derived fields can reveal sea surface dynamics for finer scales compared with altimetry-derived products. This has motivated a rich literature to exploit SST and OC fields solely or combined with altimeter-derived products to better inform sea surface dynamics \cite{isern-fontanet_potential_2006,turiel_tracking_2009,rio_improving_2016}. The illustration reported in Fig.\ref{fig:obs data} supports such a synergistic approach to jointly exploit satellite-derived SST and altimetry observations to reconstruct sea surface currents, as they clearly share common geometric patterns as stressed by the main meander and the large eddy South to the later. Through the surface quasi-geostrophic (SQG) theory \cite{klein_diagnosis_2009}, analytical derivations further support the relevance of SST-SSH synergies as explored in \cite{isern-fontanet_potential_2006,isern-fontanet_transfer_2014}. The SQG theory only applies to specific dynamical regimes, which limit the operational exploitation of the resulting SST-derived inversion of sea surface dynamics. As such, the design of multimodal inversion schemes which could exploit jointly altimetry and SST observation remains a challenge \cite{abdalla_altimetry_2021}.  

From a methodological point of view, the computation of satellite-derived geophysical products generally relies on optimal interpolation and data assimilation schemes \cite{taburet_duacs_2019,evensen_data_2009,le_guillou_mapping_2020,ubelmann_reconstructing_2021}. Deep learning methods have emerged as appealing approaches to address inverse problems \cite{chen_learning_2015,mccann_convolutional_2017}. Especially, end-to-end neural schemes  provide a plug-and-play solution to train multimodal inversion schemes as they can directly learn from data the underlying multimodal representations \cite{ngiam_multimodal_2011}. The scarce sampling to be dealt with as illustrated in Fig.\ref{fig:obs data} however limits the applicability of off-the-shelf image-to-image translation schemes \cite{zhu_toward_2017}. Recent advances in physics-informed learning \cite{fablet_learning_2021} for inverse problems appear particularly suited to jointly benefit from some physical knowledge and the efficiency of learning approaches. 

Here, we explore these recent advances for the reconstruction of sea surface dynamics, and more particularly sea surface height (SSH) fields, from altimetry and SST observation data and extend 4DVarNet schemes introduced in \cite{fablet_learning_2021} to a multimodal learning-based inversion framework applied to the synergistic exploitation of SSH-SST data. Our key contributions are three-fold:
\begin{itemize}
    \item Formally, we benefit from the underlying variational formulation which combines observation and prior terms, and account for multimodal synergies through a multimodal state and/or multimodal observation terms;  
    \item The resulting multimodal 4DVarNet schemes deliver end-to-end neural architectures with three main trainable components, namely multimodal observation operators, a prior and a gradient-based solver. We show that we can learn these trainable components jointly from data in a supervised manner;
    \item Through an application to the reconstruction of the divergence-free component of sea surface currents from satellite-derived SSH-SST data, we benchmark the proposed schemes w.r.t. state-of-the approaches. We demonstrate that the proposed multimodal inversion framework can significantly improve the reconstruction of finer-scale patterns of SSH fields
    with relative gain greater than 60\%, in terms of mean square error and resolve space-time scales, compared with the operational altimetry-only product.  
\end{itemize}
The remainder is organized as follows. Section \ref{sec:problem statement} briefly reviews related work. We describe the proposed  approach in Section \ref{sec: approach}. Section \ref{sec: results} presents our numerical experiments and We further discuss our contributions and future work in Section \ref{sec: conclusion}

\section{Problem statement and Related work}
\label{sec:problem statement}

The space-time reconstruction of geophysical fields at sea surface from irregularly-sampled satellite observations can be stated as a data assimilation problem \cite{evensen_data_2009}, that is to say the reconstruction of the dynamics of a geophsyical state given some observation data. Classically, given some physical prior on the dynamics, data assimilation relies on a state-space formulation
\begin{equation}
\label{eq: state space}
\left \{\begin{array}{ccl}
    \displaystyle \frac{\partial x(t)}{\partial t} &=& {\cal{M}}\left (x(t) \right )+ \eta(t)\\~\\
    y_m(t) &=& {\cal{H}}_m\left ( x(t) \right ) + \epsilon_m(t), \forall t ,m\\
\end{array}\right.
\end{equation}
where $x$ is the space-time process of interest defined over a space domain ${\cal{D}}$ and a time interval $[0,T]$. $x$ is governed by dynamical model ${\cal{M}}$. $y_m$ is the space-time observation data for observation modality $m$. Observation operator ${\cal{H}}_m$ relates state $x$ to observation $y_m$. $\eta$ and $\{\epsilon_m\}_m$ refer to noise processes, which account for modeling and observation errors. 

From this state-space formulation, we can state the reconstruction of state dynamics $x$ from some observation data $\{y_m\}_{m}=\{y_m(t_i)\}_{i,m}$ sampled at time steps $\{t_i\}_i$ as the minimization of the following variational criterion
\begin{equation}
\label{eq:4dvar model}
\begin{array}{ccl}
\displaystyle U_\Phi\left ( x , \{y_m\}_m \right ) &=& \displaystyle \sum_m \lambda_{m} \sum_i \left \| y_m(t_i)-{\cal{H}}_m \left ( x(t_i) \right )\right \|^2 \\~\\
&+& \displaystyle\gamma \sum_i \left \|x(t_i) - \Phi(x)(t_i) \right \|^2
\end{array}
\hspace*{-0.5cm}
\end{equation}
where $\Phi(x)(t_i)$ is the time integration of dynamical model  ${\cal{M}}$ over time interval $[t_{i-1},t_i]$ from state $x(t_{i-1})$.  $\{\lambda_{m}\}_m$ and $\gamma$ are the weighing parameters of the observation and prior terms. This variational formulation is referred to as a weak-constrained 4DVar scheme in the data assimilation literature \cite{evensen_data_2009}. Numerically speaking, the minimization of variational cost (\ref{eq:4dvar model}) classically involves gradient-based solvers using adjoint approaches \cite{evensen_data_2009}. Within a discrete-time formulation of (\ref{eq: state space}), ensemble and Kalman methods are also among the state-of-the-art schemes \cite{evensen_data_2009}. Model-driven reconstructions of sea surface dynamics from satellite-derived observations exploit these approaches with dynamical models given by ocean circulation models \cite{abdalla_altimetry_2021,benkiran_assessing_2021}. We may point out that such approaches address the reconstruction of 3{\sc d}+t ocean state series, which include sea surface variables. It results in a much more complex problem that the reconstruction of the sole sea surface dynamics.

Optimal interpolation (OI) \cite{cressie_statistics_2015} is a particular case of the above formulation with  the following additional assumptions:  operators $\{{\cal{H}}_m\}$ are masking operators, dynamical prior ${\cal{M}}$ is a linear operator and noise processes $\eta$ and $\{\epsilon_m\}_m$ are Gaussian processes. Under these hypotheses, one can derive the analytical solution of the minimization of the discrete-time formulation of (\ref{eq:4dvar model}). The state-of-the-art altimeter-derived SSH product \cite{taburet_duacs_2019} relies on such an optimal interpolation scheme. The resolved space-time scales are in the same range as those resulting from the assimilation of ocean circulation models \cite{benkiran_assessing_2021}, meaning that horizontal scales below 100km cannot be retrieved in general.

When considering multimodal observation data as considered here, we may explore two different options from state-space formulation (\ref{eq: state space}). On the one hand, we may include in state $x$ all the observed geophysical parameters (here, both SSH and SST fields) such that  the synergies between the geophysical parameters shall be accounted for through dynamical model ${\cal{M}}$ or associated flow operator $\Phi$. On the other hand, we may also explore observation models which could inform the relationship between the geophysical parameter to be reconstructed (here, the SSH) and the one which is observed (here, the SST). As an example, for SSH-SST synergies, the SQG framework \cite{isern-fontanet_potential_2006,klein_diagnosis_2009} provides a theoretical motivation to this second option. As detailed in the next section, we benefit from the proposed trainable framework to explore and compare both options.
 
As mentioned above, we can consider various optimization algorithms for the minimization of variational cost (\ref{eq:4dvar model}). Interestingly, when considering numerical implementations in deep learning and differentiable frameworks (e.g., pytorch) of all the operators in play in (\ref{eq:4dvar model}), we can benefit from the embedded automatic differentiation tools to implement gradient descent algorithms. This also opens avenues for considering pre-trained operators without analytical derivations of adjoint operators. This 
has recently been explored for computational imaging and signal processing problems both with plug-and-play priors and pre-trained observation operators \cite{wei_tuning-free_2020,zhang_plug-and-play_2021}. We may point out however that in such schemes, there is no guarantee for the pre-trained operators to be fully-relevant for the considered inversion task. End-to-end learning approaches can address these shortcomings as one may learn an inverse model using some reconstruction performance metrics in the training loss \cite{aggarwal_modl_2019,gastineau_generative_2022,lucas_using_2018}. Deep learning methods for space-time inpainting issues \cite{kim_deep_2019} do not apply directly given the very high missing data rates to be accounted for with ocean remote sensing data. Physics-driven learning schemes naturally arise as appealing approaches to benefit from prior physical knowledge on sea surface dynamics and associated satellite-derived observations. While one may complement classic end-to-end neural architecture with physics-informed training losses as illustrated in \cite{gastineau_generative_2022} for pan-sharpening applications, we here explore neural approaches which explicitly rely on a variational formulation similar to (\ref{eq:4dvar model}) \cite{fablet_learning_2021,kobler_total_2020}. Such approaches make explicit the exploitation of an underlying state-space formulation. More specifically, as detailed in the next section, we extend our prior work \cite{fablet_learning_2021} to multimodal inversion problems with a view to exploiting SSH-SST synergies for the reconstruction of sea surface dynamics.

\section{Proposed approach}
\label{sec: approach}

This section introduces the proposed multimodal learning-based inversion scheme. We first present the proposed multimodal data assimilation formulation before introducing the resulting end-to-end learning scheme.

\subsection{Multimodal data assimilation formulation}
\label{ss:4dvar model}

As detailed below, we benefit from the versatility of the end-to-end learning framework introduced in \cite{fablet_learning_2021} and explore SST-SSH synergies to enhance the reconstruction of sea surface dynamics multimodal SST-SSH observation terms. 

We first introduce formally the SSH-only and SSH-SST state-space formulations as follows. 
\begin{itemize}
    \item {\bf SSH-only state-space:} Here, state $x$ in (\ref{eq: state space}) only involves the SSH. Following \cite{beauchamp_end--end_2022,fablet_end--end_2019}, SSH component $x_{SSH}$ decomposes as a coarse-scale component $\bar{x}_{SSH}$ and two fine-scale components $\delta^o x_{SSH}$ and $\delta^r x_{SSH}$.
    We assume to be provided with two altimeter-derived data sources: the irregularly-sampled altimeter-derived SSH data denoted as $y_{SSH}$ and an optimally-interpolated product from altimeter-derived data denoted as $\bar{y}_{SSH}$ \footnote{Here, we will consider the operational processing referred to as DUACS which applies an optimal interpolation to altimeter-derived data to deliver gap-free SSH fields \cite{taburet_duacs_2019}.}. From (\ref{eq:4dvar model}), we derive the following matrix-based variational data assimilation formulation:  
    \begin{equation}
    \label{eq:4dvar model ssh-only}
    \begin{array}{c}
    \displaystyle U_\Phi\left ( x , y_{SSH},\bar{y}_{SSH} \right ) = \displaystyle \lambda_{1} \left \| \bar{y}_{SSH}-\bar{x}_{SSH}\right \|^2\\~\\
    + \displaystyle \lambda_{2} \left \| H\left(y_{SSH}\right) \cdot \left(y_{SSH}-\bar{x}_{SSH}-\delta^o x_{SSH}\right )\right \|^2\\~\\
    + \displaystyle\gamma \left \|x - \Phi(x) \right \|^2
    \end{array}
    \end{equation}
    where $x=(\bar{x}_{SSH},\delta^o x_{SSH},\delta^f x_{SSH})$. $H\left(y_{SSH}\right)$ is the masking operator to account for the sampling pattern of altimeter-derived SSH data. In this state-space formulation, the reconstructed SSH field is given by $\widehat{x}_{SSH}=\bar{x}_{SSH}+\delta^r x_{SSH}$. Component $\delta^o x_{SSH}$ appears only in the observation term, whereas component $\delta^r x_{SSH}$ only appears in the reconstruction equation. This parameterization is proven efficient to remove geometric artifacts associated with the sampling patterns of nadir altimeter data. We let the reader refer to \cite{beauchamp_end--end_2022} for an additional discussion on this point.
    \item {\bf SST-SSH state-space:} Here, we  benefit from the versatility of trainable schemes so that state variable $x$ includes both a SSH component $x_{SSH}$ and a SST component $x_{SST}$. We consider the same parameterization as above for SSH component $x_{SSH}$ such that $x_{SSH} = (\bar{x}_{SSH},\delta^o x_{SSH},\delta^r x_{SSH})$. In this  SST-SSH state-space formulation, we naturally complement variational formulation (\ref{eq:4dvar model ssh-only}) with an additional SST-specific observation term such that   
    \begin{equation}
    \label{eq:4dvar model ssh-sst}
    \begin{array}{c}
    \displaystyle U_\Phi\left ( x , y_{SSH},\bar{y}_{SSH} \right ) = \displaystyle \lambda_{1} \left \| \bar{y}_{SSH}-\bar{x}_{SSH}\right \|^2\\~\\
    + \displaystyle \lambda_{2} \left \| H\left(y_{SSH}\right) \cdot \left(y_{SSH}-\bar{x}_{SSH}-\delta^o x_{SSH}\right )\right \|^2\\~\\
    + \displaystyle \lambda_{3} \left \| y_{SST}-x_{SST} \right \|^2 
    + \displaystyle\gamma \left \|x - \Phi(x) \right \|^2
    \end{array}
    \end{equation}
    In this formulation, prior term $\|x - \Phi(x) \|^2$ will account for SSH-SST synergies.   
\end{itemize}
In both formulations, operator $\Phi$ states the prior onto the state to be reconstructed. While in a model-driven configuration it derives from known governing equations for state $x$, our previous work suggests that the parameterization of $\Phi$ using state-of-the-art neural network architectures such as U-Nets might lead to better inversion performance as the form of the prior can adapt to the considered inversion problem and observation patterns \cite{beauchamp_end--end_2022,fablet_learning_2021}.  Overall, in these two state-space formulations, the trainable parameters refer to the trainable parameters of operator $\Phi$ as well as weighing factors $\lambda_{1,2,3}$. Regarding the parameterization of operator $\Phi$, we follow the same approach as in \cite{beauchamp_end--end_2022,fablet_learning_2021}. We consider a two-scale residual U-Net architecture \cite{cicek_3d_2016} with bilinear blocks to account for the non-linearities expected in upper ocean dynamics. Given $T$-day time windows and a $W\times W$ spatial grid, state $x$ for the SSH-only state-space (resp. the SST-SSH state-space), is given as a $(3*T)\times W \times W$ (resp. $(4*T)\times W \times W$) tensor to apply 2-dimensional convolution layers. 
    
{\bf Multimodal observation term:} as advocated by the SQG theory \cite{isern-fontanet_potential_2006,lapeyre_dynamics_2006}, we also investigate a multimodal observation term to explicitly state that SST observations may inform SSH fields. 
We consider the following synergistic term added to (\ref{eq:4dvar model ssh-only}) or (\ref{eq:4dvar model ssh-sst}) depending on the considered state-space
\begin{equation}
\label{eq: multimodal obs}
     U_{MM}\left ( x , y_{SST} \right ) = \left  \|{\cal{G}}^1_{MM} \left ( y_{SST} \right) - {\cal{G}}^2_{MM} \left ( x \right) \right \|^2
\end{equation}
with ${\cal{G}}^1_{MM}$ and ${\cal{G}}^2_{MM}$ convolutional operators acting respectively on SST observation $y_{SST}$ and state $x$. We may remind that $y_{SST}$ and $x$ refer to space-time tensors. 
The SQG theory would lead to parameterize as ${\cal{G}}^1_{MM}$ as a pass-band filter and ${\cal{G}}^2_{MM}$ as a combination of a pass-band filter and of a fractional Laplacian operator such that (\ref{eq: multimodal obs}) would lead to
\begin{equation}
\label{eq: multimodal obs SQG}
     \left \| {\cal{F}}^1 * y_{SST} - {\cal{F}}^2 * \left ( \Delta ^{1/2} (\bar{x}_{SSH}+\delta^f x_{SSH}) \right) \right \|^2
\end{equation}
with ${\cal{F}}^{1,2}$ linear pass-band filters to select the scale range which the SQG theory applies to and  $\Delta ^{1/2}$ the fractional Laplacian operator which is a linear filter defined in the spectral domain. Overall, this parameterization results in the extraction of linear features to match SST and SSH patterns. Here, we investigate a generalization of the SQG-based parameterization, where operators ${\cal{G}}^{1,2}_{MM}$ are trainable linear or non-linear operators. As detailed hereafter, we consider simple ConvNets with a single layer in the linear case and 4 layers with $\tanh$ activations in the non-linear case. Importantly, whereas the SQG theory involves space-only filters, our trainable operators may exploit space-time filters. One may exploit this multimodal observation term both in the SSH-only state-space formulation and in the SSH-SST one. 

\subsection{End-to-end inversion model and associated learning scheme}

From the proposed multimodal variational formulation with trainable components, we design an end-to-end inversion scheme which implements a gradient-based iterative solver as proposed in \cite{fablet_learning_2021}. For a given state-space formulation and associated variational cost $U_\Phi$, the proposed end-to-end neural architecture performs a predefined number of iterations of an  iterative gradient-based update. More precisely, at iteration $k$, we apply: 
\begin{equation}
\label{eq: lstm update}
\left \{\begin{array}{ccl}
     h^{(k+1)} , c^{(k+1)} &=&   LSTM \left[ \cdot \nabla_x U \left ( x^{(k)},\{y_m\}_m \right),  h^{(k)} , c^{(k)} \right ]  \\~\\
     x^{(k+1)} &=& x^{(k)} - {\cal{L}}  \left( h^{(k+1)} \right )  \\
\end{array} \right.\hspace*{-0.2cm}
\end{equation} where $h^{(k)}$ and $ c^{(k)}$ are the internal states of the LSTM cell at iteration $k$ and $x^{(k+1)}$ the updated state. ${\cal{L}}$ is a linear layer to map the LSTM state to the space spanned by state $x$   This gradient-based iterative update is similar to neural parameterization for the learning of optimizers \cite{hospedales_meta-learning_2020}. As state $x$ is implemented as a multivariate 2d tensor, we consider 2{\sc d} convolutional LSTM cells. Experimentally, we cross-validated the use of 150-dimensional convolutional LSTM cells.

Overall, the resulting end-to-end architecture uses as inputs observation data $y_1=y_{SSH}$, $y_2=\bar{y}_{SSH}$ and $y_3=y_{SST}$ as well as some state initialisation $x^{(0)}$ to output the reconstructed state. This architecture implements a predefined number $K$ of the above gradient steps, typically from 5 to 15. We denote by $\widehat{x}=\Psi _{\Theta} \left ( x^{(0)} , \{y_m\}_m \right )$ the output of the end-to-end inversion scheme after $K$ iterations of (\ref{eq: lstm update}) ({\em i.e.}, $\widehat{x}= x^{(K)}$) where $\Theta$ stands for the set of all trainable parameters. Depending on the considered multimodal configuration, the trainable components of the architecture comprise those of prior $\Phi$ and of the LSTM-based solver, possibly complemented by those of multimodal observation operators ${\cal{G}}^{1,2}_{MM}$.

We exploit a supervised learning strategy to train our model. Given a training dataset comprising triplets of true states, observation data and initial conditions $\{x^{true}_{SSH,n},\{y_{m,n}\}_{m},x^{(0)}_n\}_n$, the training loss typically involves a weighted sum of the reconstruction error for reconstructed state $x_{SSH}$ and its gradient
\begin{equation}
{\cal{L}}_{x} =  \sum_n \left \| x^{true}_{SSH,n} - \widehat{x}_{SSH,n} \right \|^2
\end{equation}
\begin{equation}
{\cal{L}}_{\nabla x} = \sum_n \left \| \nabla x^{true}_{SSH,n} - \nabla\widehat{x}_{SSH,n} \right \|^2
\end{equation}
As proposed in \cite{fablet_learning_2021}, we also include additional regularisation terms
\begin{equation}
\begin{array}{ccl}
{\cal{L}}_{\Phi} &=&  \sum_n \left \| x^{true}_{n} - \Phi \left ( x^{true}_n  \right ) \right \|^2 \\~\\ &+&  \sum_i \left \| \widehat{x}_{n}- \Phi \left ( \widehat{x}_{n} \right ) \right \|^2
\end{array} 
\end{equation}
such that the overall training loss is computed as a weighted sum of these different terms: $\nu_{ x}{\cal{L}}_{ x}+\nu_{\nabla x}{\cal{L}}_{\nabla x}+\nu_{\Phi}{\cal{L}}_{\Phi}$ with $\nu_{ x}$, $\nu_{ \nabla x}$ and $\nu_{ \Phi}$ weighing parameters. The training procedure exploits Adam optimizer with a 1e-3 learning rate over 400 epochs. We select the best model according to reconstruction metrics evaluated on the validation dataset at each training epoch. The Pytorch code of our implementation, include all parameter values,  is available along with trained models \cite{fablet_release_2023}.  

\section{Results}
\label{sec: results}

This section details the numerical experiments we run to evaluate the reconstruction performance of the proposed approach. We first detail the considered experimental setting. We then present our results, including a comparison to state-of-start schemes.  

\subsection{Dataset and experimental setting}

With a view to assessing the relevance of the proposed multimodal inversion framework, we consider the benchmarking setting introduced in \cite{le_guillou_mapping_2020}. It relies on an Observing System Simulation Experiment (OSSE) for nadir and wide-swath satellite altimetry data. We exploit one-year NATL60 numerical simulation dataset \cite{ajayi_spatial_2020} for a 10$^\circ$x10$^\circ$ area along the Gulf Stream from October 2012 to September 2013 with a daily time resolution and a $1/20^\circ$ spatial resolution. Nadir altimetry data involves the space-time sampling of a real 4-altimeter configuration, whereas wide-swath SWOT data rely on SWOT simulator \cite{gaultier_challenge_2015}. In both cases, we consider noise-free observations. For SST observations, we assume gap-free observations of daily sea surface temperature (SST) fields. During the training stage, we consider SST observations with a $1/20^\circ$ spatial resolution. For evaluation purposes, we also investigate subsampled versions with resolutions of $1/10^\circ$, $1/5^\circ$, $1/4^\circ$ and $1/2^\circ$.

Regarding the training procedure and the evaluation framework, we split the OSSE dataset into training, validation and test datasets as follows. We use as training dataset the data from February 2013 to September 2013 and as validation dataset data from January 2012. We use the validation dataset to monitor performance metrics and select the best model during the training procedure. The overall evaluation procedure  relies on performance metrics computed for the test dataset which refers to the 40-day period from October, 22 2012 to December, 2 2012. As evaluation metrics, we first consider the metrics introduced in \cite{le_guillou_mapping_2020} to benchmark the proposed schemes w.r.t.  the state-of-the-art approaches \footnote{We refer the reader to the following link for the detailed presentation of the evaluation experiment and benchmarked approaches \url{https://github.com/ocean-data-challenges/2020a_SSH_mapping_NATL60}}:
\begin{itemize}
    \item $\mu$, the normalized root-mean-square-error-based metrics equals 1 for a perfect reconstruction;
    \item $\lambda_t$, the minimum time scale resolved in days; 
    \item $\lambda_x$, the minimum spatial scale resolved in degrees.
\end{itemize}
As described in \cite{le_guillou_mapping_2020}, the last two metrics are computed in the spectral domain. With a view to enhancing the differences between the different configurations of the proposed multimodal inversion scheme, we also asses the relative improvement w.r.t. the operational baseline (DUACS) \cite{taburet_duacs_2019} using the following three metrics:
\begin{itemize}
    \item $\tau_{SSH}$, the relative gain with respect to  DUACS baseline for the MSE of the reconstruction of the SSH;
    \item $\tau_{\nabla SSH}$, the relative gain with respect to  DUACS baseline for the MSE of the reconstruction of the gradient of the SSH;
    \item $\tau_{\Delta SSH}$, the explained variance for the Laplacian of the reconstructed SSH fields.
\end{itemize}
We compute all these metrics from the reconstruction error of the SSH fields w.r.t. the daily-averaged SSH fields of the numerical simulations. 

\begin{figure*}[htb]
    \centering
    \begin{tabular}{C{4.5cm}C{4.5cm}C{4.5cm}}
    \includegraphics[trim={250 100 300 100},clip,width=4.5cm]{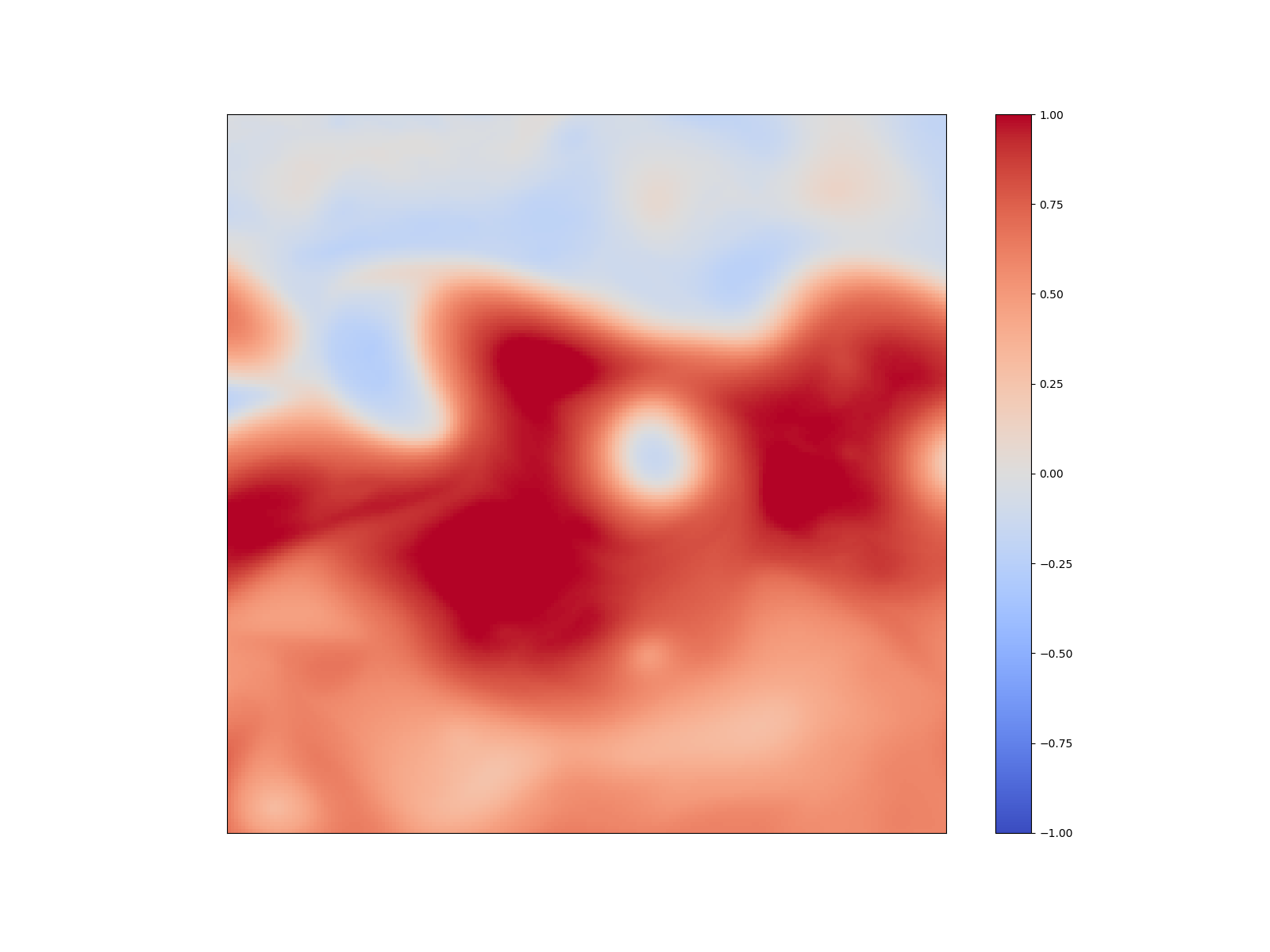}&
    \includegraphics[trim={250 100 300 100},clip,width=4.5cm]{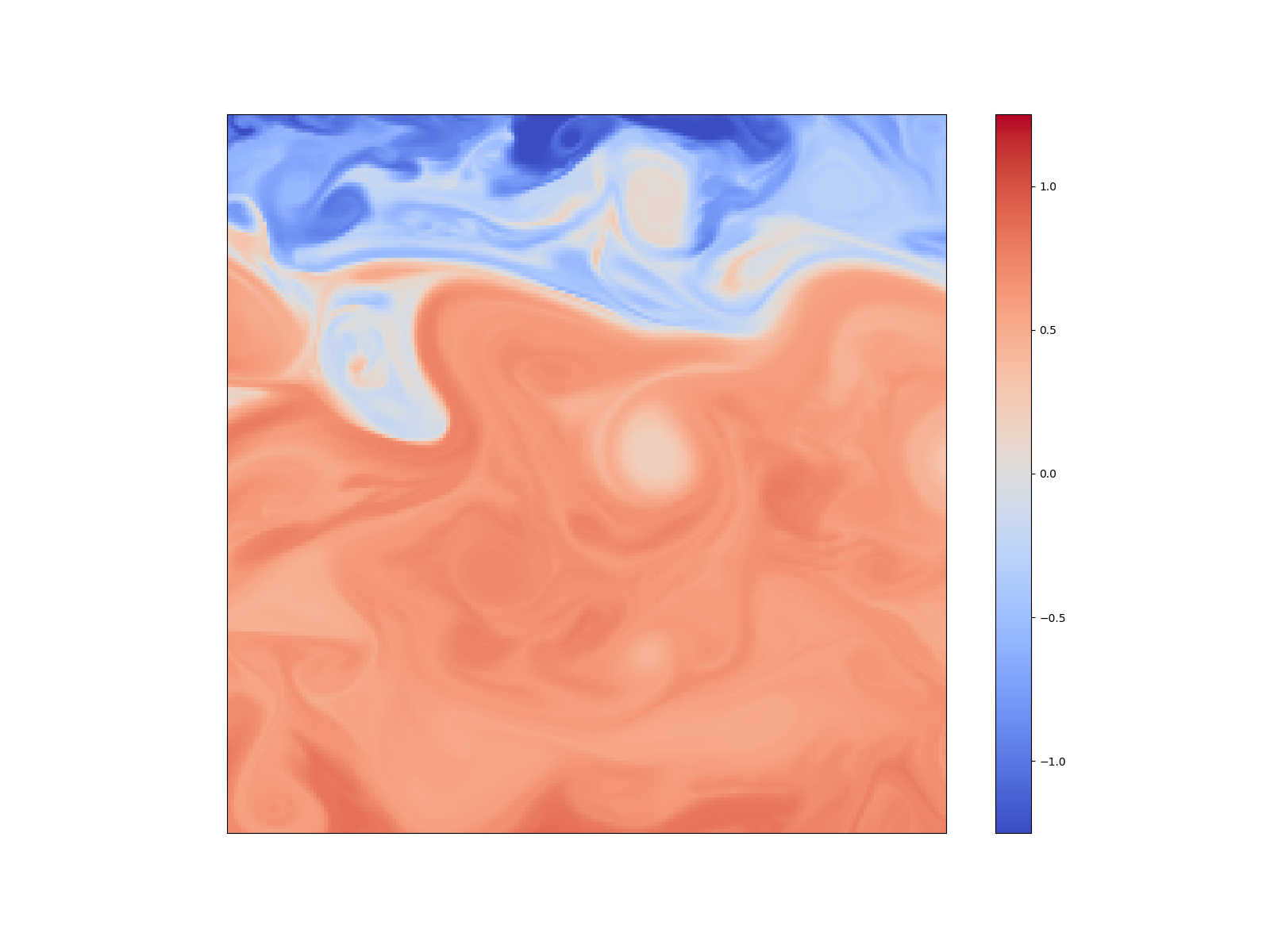}&
    \includegraphics[trim={250 100 300 100},clip,width=4.5cm]{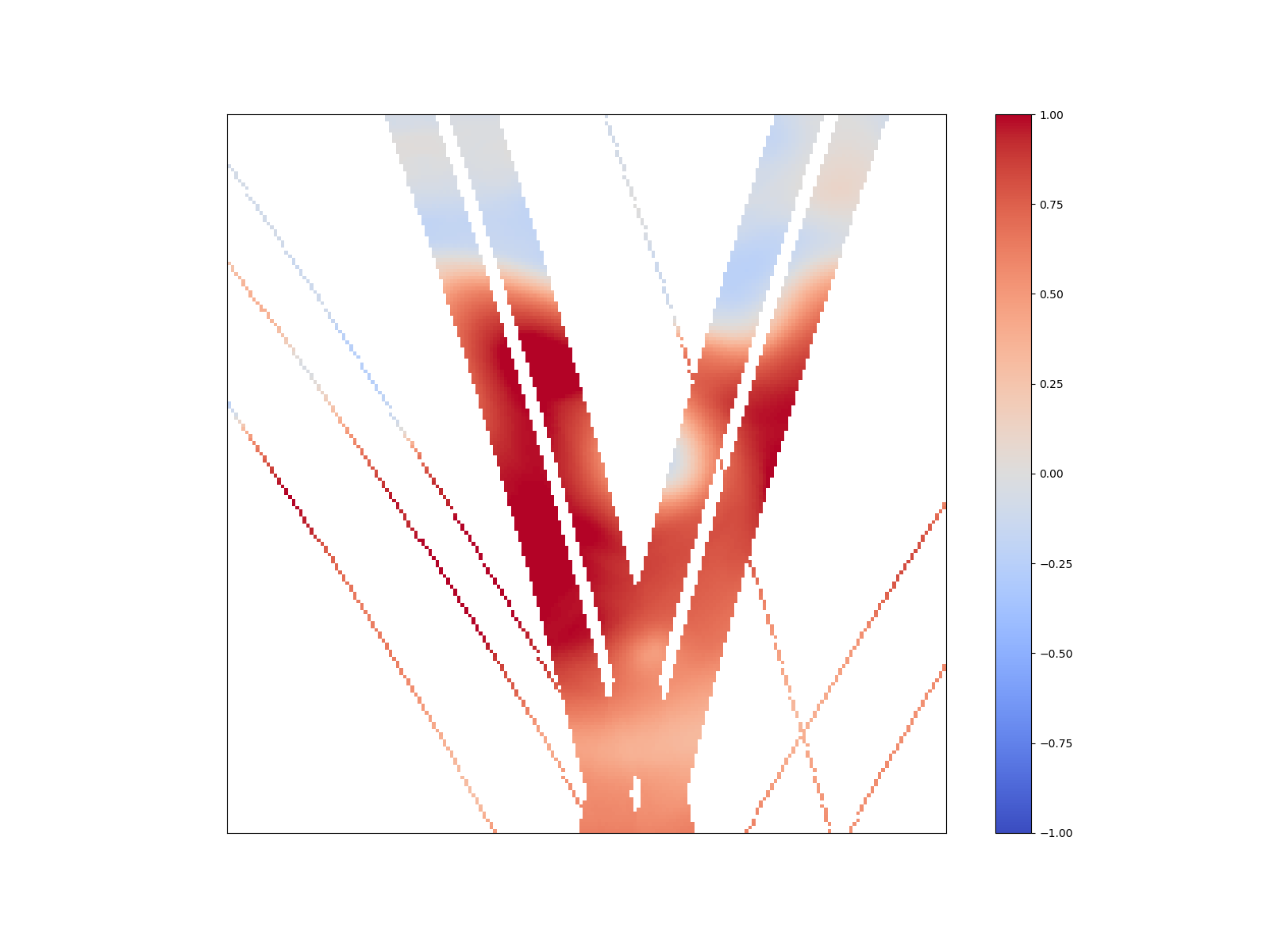}\\
    {\bf true SSH}&{\bf true SST}&{\bf Altimetry data}\\
    \end{tabular}
    \caption{{\bf Illustration of the considered OSSE case-study:} from left to right, NATL60 SSH field on October 25$^{th}$ 2012, associated SST field and altimetry data. The altimetry data combine nadir altimeter tracks and wide-swath SWOT data.}
    \label{fig:obs data}
\end{figure*}

\subsection{Impact of the parameterization of the multimodal framework}

We first investigate how the parameterization of the proposed multimodal framework affects the reconstruction performance. It includes both the parameterization of the multimodal observation term (\ref{eq: multimodal obs}) as well as complementing  or not the reconstructed state with the SST (\ref{eq: state space}). Regarding multimodal observation term (\ref{eq: multimodal obs}), we consider both linear and non-linear parameterizations for operators $\{{\cal{G}}^{1,2}_{2,n}\}_n$:
\begin{itemize}
    \item {\bf Linear parameterization:} In the linear setting, we exploit linear $7\times3\times3$ space-time convolution kernels for operators $\{{\cal{G}}^{1}_{2,n}\}_n$ applied to SST observations. Similarly, operators $\{{\cal{G}}^{2}_{2,n}\}_n$  extract linear features from state $x$ with $3\times3$ kernels. We vary the number of extracted features from 1 to 50.
    \item {\bf Non-linear parameterization:} In the non-linear setting, operators $\{{\cal{G}}^{1,2}_{2,n}\}_n$ are convolutional networks with four layers and $\tanh$ activation functions. We also tested classic $ReLu$ activations which led to slightly worse reconstruction performance. We also vary the number of extracted features from 1 to 50.
\end{itemize}
 
In Tab.\ref{tab:res MM model}, we report the synthesis of the reconstruction performance of the different configurations of the proposed multi-modal framework (\ref{eq: multimodal obs}). For the configurations with multi-modal observation terms (\ref{eq: multimodal obs}), we report the performance of the best configuration, here a 20-dimensional non-linear observation operator. We further analyse below how this parameterization affects the performance (see Tab.\ref{tab:res obs model}). As comparison baseline, we consider the 4DVarNet scheme using only altimetry data and no SST observations. Overall, all multimodal configurations clearly outperform the altimetry-only baseline with a significant gain up to 8\% to 16\% in terms of the reconstruction of the SSH and its derivatives. The greatest improvement occurs for the resolved time scale (up to 2.47 days vs. 5.30 days). These results point out that the best performance comes from a SSH-only state-space where we exploit SST data in a trainable multi-modal observation term (\ref{eq: multimodal obs}). This suggests that we better account for SSH-SST synergies through  multi-modal observation term (\ref{eq: multimodal obs}) than through a U-Net parameterization for prior $\Phi$ in the SST-SSH state-space formulation. The lower performance of the latter combined with a multi-modal observation term (\ref{eq: multimodal obs}) may relate to overfitting issues as this parameterization is more complex. Larger training datasets and data augmentation may provide relevant solutions to overcome these issues. 

\begin{table*}[tb]
    \footnotesize
    \centering
    \begin{tabular}{|C{1.75cm}|C{1.75cm}|C{1.5cm}|C{1.5cm}|C{1.5cm}|C{1.5cm}|C{1.5cm}|C{1.5cm}|C{1.5cm}|C{1.5cm}|}
    \toprule
    \toprule
     \bf State &\bf MM term (\ref{eq: multimodal obs})&
     \bf $\mu$ & $\lambda_x$ ($^\circ$)& $\lambda_t$ (days)&
     $\tau_{SSH}$&$\tau_{\nabla SSH}$&$\tau_{\Delta SSH}$\\
    \toprule
    \toprule
     SSH  & No & 0.96 & 0.67 & 5.3 &71.4\% & 65.1\% & 83.7\% \\    
       & Yes & \bf 0.97 & \bf 0.50& \bf 2.47  & \bf 83.8\% & \bf 81.4\% & \bf 91.8\%\\ 
    \toprule
     SST+SSH  & No & 0.96 & 0.60 & 3.18 & 75.7\% & 73.9\% & 88.4\% \\   
       & Yes & 0.96 & 0.57 & 2.49 & 78.7\%& 75.9\%& 89.2\% \\    
    \bottomrule
    \bottomrule
    \end{tabular}
    \caption{{\bf Reconstruction performance of the proposed 4DvarNet
    framework for different configurations}: when the state in (\ref{eq: state space}) only relates to SSH, the multimodal component comes from MM term (\ref{eq: multimodal obs}); we also consider a multimodal setting where the state in (\ref{eq: state space}) comprises both SSH and SST. In the latter configuration, one may use or not the multimodal term. We refer the reader to the main text for the definition of the performance metrics. We highlight in bold the best score.
    }
    \label{tab:res MM model} 
\end{table*}

We further analyze the sensitivity of the reconstruction performance with respect to the parameterization of multimodal observation term (\ref{eq: multimodal obs}). As reported in Tab.\ref{tab:res obs model}, we vary the dimensionality of the observation term for both linear and non-linear observation terms from 1-dimensional operators to 50-dimensional ones. Interestingly, one-dimensional parameterizations already leads to very good performance. Whereas the performance is very similar from 1-to-5-dimensional linear operators, more complex linear operators lead to a poorer performance which indicate some over-parameterization. By contrast, when considering non-linear operators, we observe an increasing performance trend from a 1-dimensional operator to a 20-dimensional one. This likely reflects the ability of the non-linear operators to learn relevant and specific SST features. As mentioned above, we consider hyperbolic tangent activation functions. When considering other activation functions such as ReLu activatons, we worsen the reconstruction performance. This likely relates to the regularity of the the higher-order derivatives of geophysical fields which are rather low-contrast fields compared to natural images.

\begin{table*}[tb]
    \footnotesize
    \centering
    \begin{tabular}{|C{1.75cm}|C{1.0cm}|C{1.0cm}|C{1.5cm}|C{1.5cm}|C{1.5cm}|C{1.5cm}|C{1.5cm}|}
    \toprule
    \toprule
     \bf MM term (\ref{eq: multimodal obs}) & \bf $N_{Feat}$& 
     \bf $\mu$ & $\lambda_x$ ($^\circ$)& $\lambda_t$ (days)&
     $\tau_{SSH}$&$\tau_{\nabla SSH}$&$\tau_{\Delta SSH}$\\
    \toprule
    \toprule
     Linear  & 1 & 0.96 & \textit{\textbf{0.50}} & \textit{\textbf{2.99}} & 79.6\%&  78.0\%  & 90.9\% \\
       & 2 & 0.97 & 0.60 & 3.12 & \textit{\textbf{82.0}} \%  & 78.7\% & 90.5 \% \\
       & 3 & \textit{\textbf{ 0.97}} & 0.57 & 3.14 & 80.9 \%  & \textit{\textbf{79.1}}\% &  \textit{\textbf{91.3\%}}  \\
       & 5 & 0.97 & 0.60 & 2.50 & 80.3\% & 76.7\% &  89.5\%  \\
       & 10 & 0.96 & 0.75 & 29.4 & 79.0\% & 75.6\% &  89.1\%  \\
       & 20 & 0.97 & 0.61 & 3.20 & 81.3\% & 77.5\% &  89.9\%  \\
       & 50 & 0.96 & 0.76 & 29.8 & 77.3\% &   74.1\% &  88.7\%  \\
     \toprule
     Non-linear & 1 & 0.97 & 0.60 & 3.23 & 81.5\%  & 77.8\% & 90.3\% \\
       & 2 &  0.97 & 0.59 & 3.31 &  81.6\% &  77.5\% &  89.7\% \\
       & 3 &  0.97 & 0.55 & 2.63 &  81.3\% &  78.6\% &  90.6\% \\
       & 5 & 0.97 & 0.58 & 3.12 & 81.1\% &  78.0\% &  90.0\% \\
       & 10 & 0.97 & 0.54 & 2.46 &  81.5\% &  79.5\% &  91.1\% \\
       & 20 & \bf 0.97 & \bf 0.50 & \bf 2.47 & \bf 83.8 \% &  \bf 81.4\%  &  \bf 91.8\% \\
       & 50 & 0.97 & 0.53 & 2.49 &  82.9\% &  79.5\%  &  91.0\%\\
    \bottomrule
    \bottomrule
    \end{tabular}
    \caption{{\bf Reconstruction performance for different parameterizations of multimodal (MM) term (\ref{eq: multimodal obs}):} we report the performance metrics of the proposed 4DVarNet framework for linear and non-linear operators in (\ref{eq: multimodal obs}) with 1, 2, 3, 5 or 10 multimodal features ($N=\{1,2,3,5,10\}$).  We consider the same performance metrics as in Tab.\ref{tab:res all}. }
    \label{tab:res obs model} 
\end{table*}

\begin{table*}[tb]
    \footnotesize
    \centering
    \begin{tabular}{|C{2.cm}|C{1.9cm}|C{1.cm}|C{1.25cm}|C{1.25cm}|C{1.25cm}|C{1.25cm}|C{1.25cm}|}
    \toprule
    \toprule
     \bf Approach& \bf Data used&\bf $\mu$ & $\lambda_x$ ($^\circ$)& $\lambda_t$ (days)&$\tau_{SSH}$&$\tau_{\nabla SSH}$&$\tau_{\Delta SSH}$\\
    \toprule
    \toprule
     DUACS \cite{taburet_duacs_2019} & SSH only& 0.92 & 1.22 & 11.15 & 0\% & 0\%&44.9\%\\    
     DYMOST \cite{ubelmann_dynamic_2014} & SSH only& 0.93 & 1.20 & 10.07  & - & - & -\\    
     MIOST \cite{ubelmann_reconstructing_2021} & SSH only& 0.94 & 1.18 & 10.14  & - & - & - \\    
     BFN \cite{le_guillou_mapping_2020} & SSH only & 0.93 & 0.8 & 10.09   & - & - & - \\    
     SQG \cite{isern-fontanet_potential_2006} & SSH and SST & 0.93 & 1.12 & 11.16  & - & - & - \\    
    \toprule
      U-Net & SSH only  & 0.94 &  1.21 & 10.21  & 38.4\% & 36.2\% & 70.7\% \\
      & SSH and SST & 0.95 & 1.09 & 37.0  & 54.6\% & 56.(\% & 79.6\% \\
    \toprule
     4DVarNet  & SSH only & 0.96 & 0.67 & 5.3  & 71.4\% & 65.1\% & 83.7\% \\
     (ours) & SSH-SST-L  & 0.97 & 0.57& 3.14  & 80.9\% & 79.1\% & 91.3\%\\
      & SSH-SST-NL  & \bf 0.97 & \bf 0.50& \bf 2.47  & \bf 83.8\% & \bf 81.4\% & \bf 91.8\%\\
    \bottomrule
    \bottomrule
    \end{tabular}
    \caption{{\bf Synthesis of the reconstruction performance of the benchmarked approaches:} we report the performance metrics of the benchmarked approaches for the reconstruction of image time series of sea surface currents from satellite data. We refer the reader to the main text for the description of the different metrics. We highlight in bold the best score. }
    \label{tab:res all}
\end{table*}

\subsection{Comparison to state-of-the-art schemes}

Based on the previous experiments, we synthesize the performance of the best configurations of the proposed multimodal framework with respect to that of state-of-the-art schemes according to the benchmarking framework presented in \cite{le_guillou_mapping_2020}. As detailed in the associated data challenge 
\footnote{\url{https://github.com/ocean-data-challenges/2020a_SSH_mapping_NATL60}}, wee first include  approaches which only rely on altimetry data. They include the operational optimal-interpolation-based method (DUACS) \cite{taburet_duacs_2019}, model-driven interpolations using variational data assimilation schemes \cite{le_guillou_mapping_2020,ubelmann_dynamic_2014} and a multiscale interpolation approach \cite{ubelmann_reconstructing_2021}. We also implement a SQG-based inversion schemes to complement the optimally-interpolated SSH fields for horizontal scales below 1.2$^\circ$. The last category of approaches we consider in our benchmarking experiment refers to direct learning-based inversion schemes. Here, we train U-Net architectures \cite{cicek_3d_2016} to reconstruct the SSH fields from gappy data using an initial zero-filling strategy. We consider SSH-only and SSH-SST configurations to asses the relevance of 4VarNet architectures. Regarding the parameterization of the U-Nets, we evaluate both the U-Net architecture considered for prior $\Phi$ in the implemented 4DVarNet schemes as well as a standard 3-scale U-Net architecture with ReLu activations \cite{cicek_3d_2016}. We only report the results for the former parameterization which led to the best reconstruction performance.  
 
We report in Tab.\ref{tab:res all} the synthesis of the performance metrics for all the benchmarked approaches. Among all the methods using only altimetry data, the 4DVarNet scheme clearly leads to the best reconstruction performance with a very significant gain w.r.t. the operational optimal interpolation baseline greater than 50\% for MSE scores and 40\% for the resolved space-time scale. Multimodal 4DVarNet schemes further improve the reconstruction score. We obtain the greatest improvement for the resolved time scale (2.47d vs. 5.30d for the altimeter-only 4DVarNet and 11.15d for the operational baseline) and the SSH gradient (81.4\% vs. 65.1\% for the altimeter-only 4DVarNet regarding the relative gain w.r.t. the operational baseline). We also observed some improvement though smaller for the resolved spatial scale (0.50$^\circ$ vs 0.68$^\circ$ and 1.22$^\circ$). These results also highlight the benefits of 4DVarNet schemes compared to the application of state-of-the-art image-to-image neural architectures. These results support the relevance of the proposed variational formulation to make explicit underlying observation and prior operators. Especially, the 4DVarNet scheme using SSH-only data delivers better reconstruction metrics than U-net-based inversion using jointly SSH and SST data. The latter leads to a very high value of the resolved time scale which indicates the presence of artifacts in the associated interpolation.  

\begin{figure*}[htb]
    \centering
    \begin{tabular}{C{4.5cm}C{4.5cm}C{4.5cm}}
    \includegraphics[trim={250 100 300 100},clip,width=4.5cm]{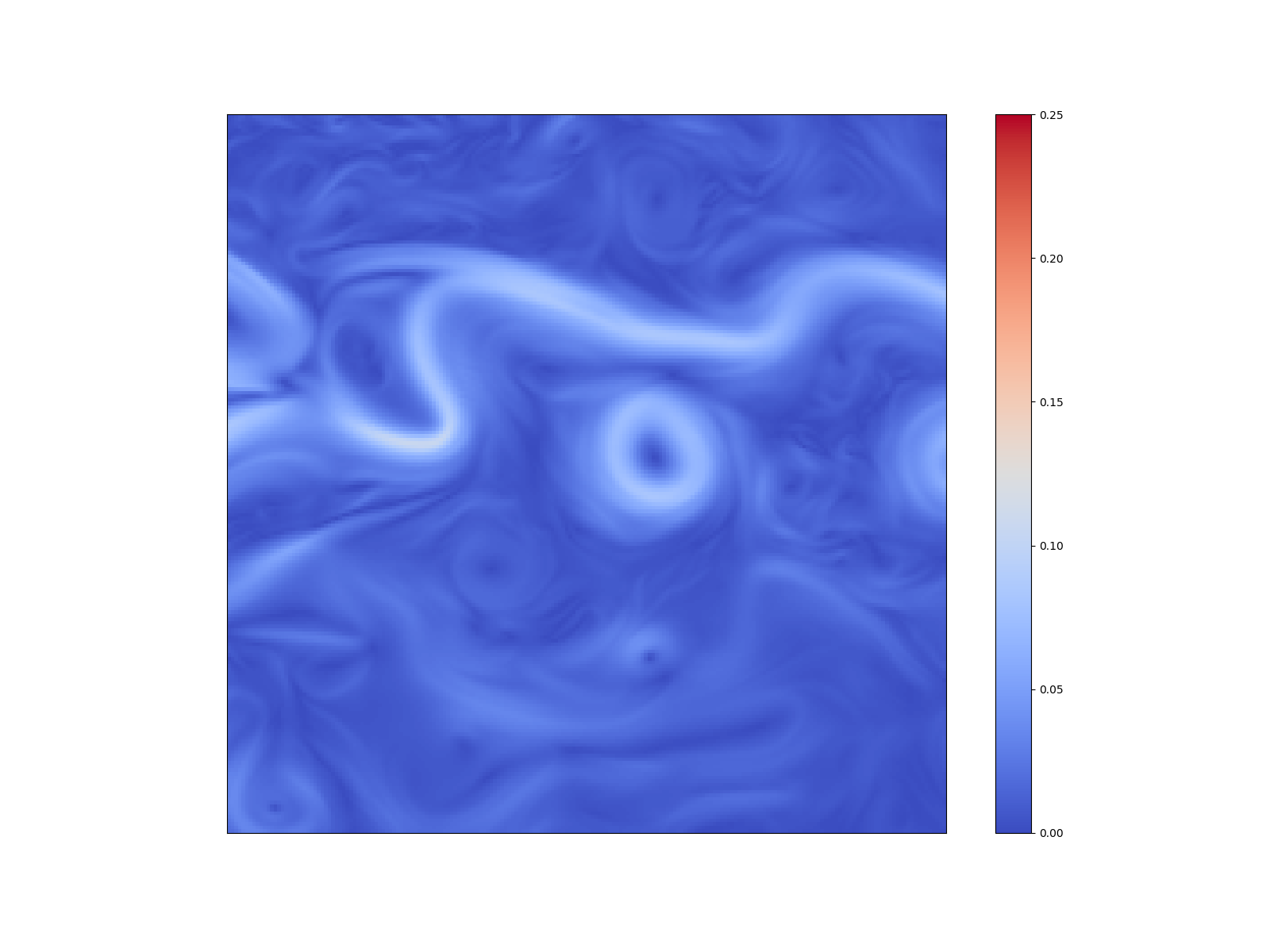}&
    \includegraphics[trim={250 100 300 100},clip,width=4.5cm]{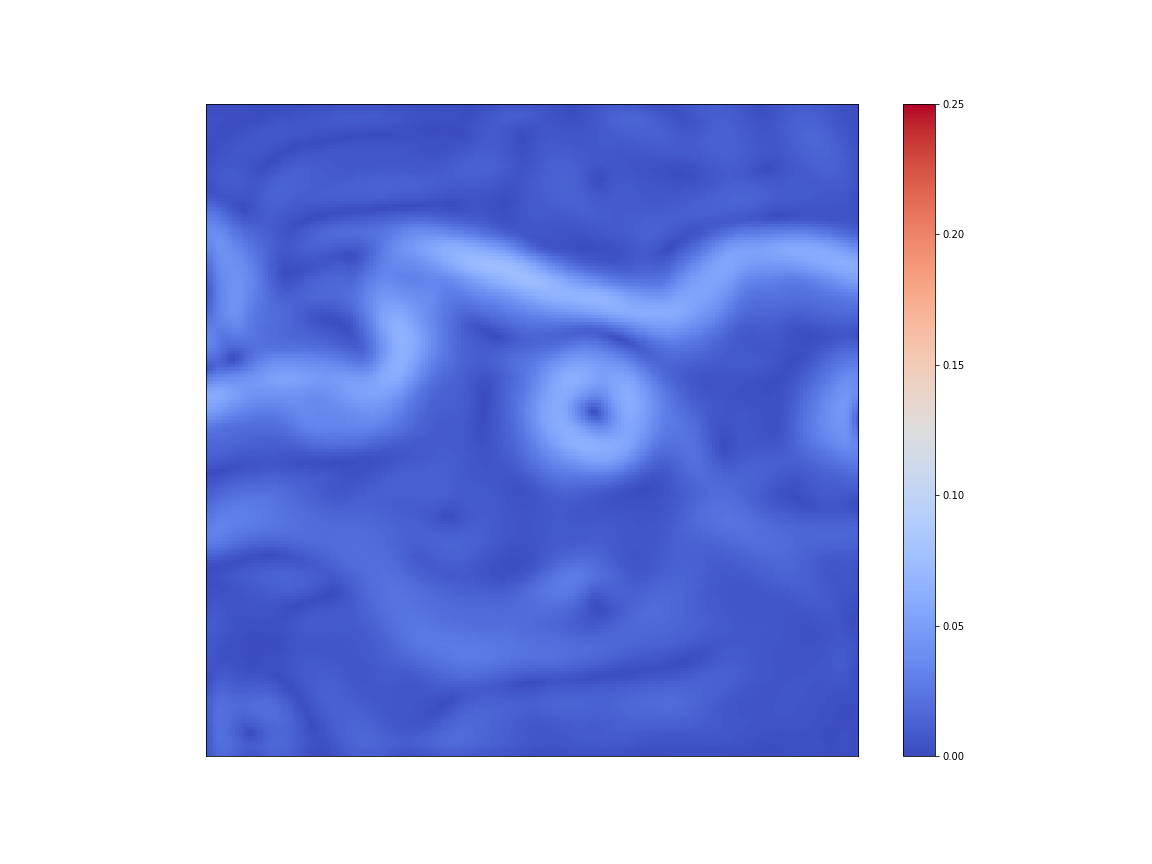}&
    \includegraphics[trim={250 100 300 100},clip,width=4.5cm]{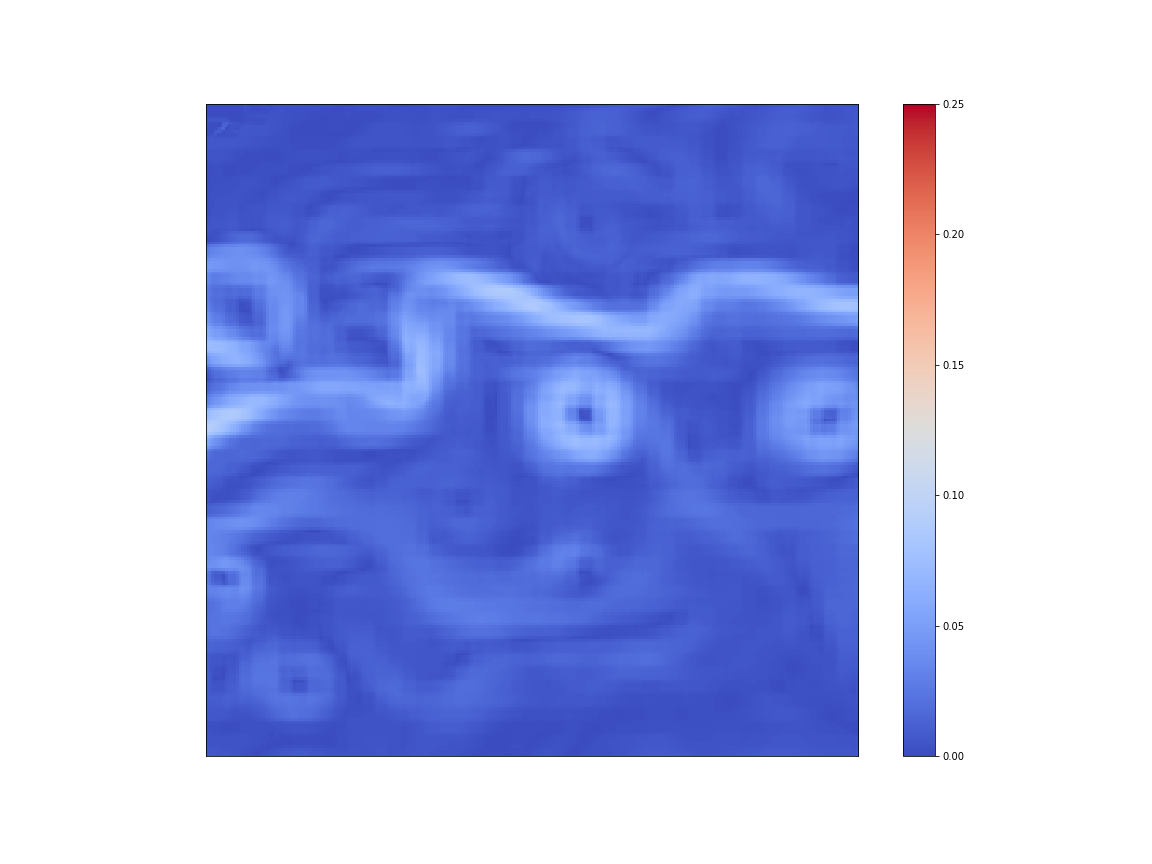}\\
    {\bf Target}&{\bf DUACS}&{\bf SQG}\\
    \includegraphics[trim={250 100 300 100},clip,width=4.5cm]{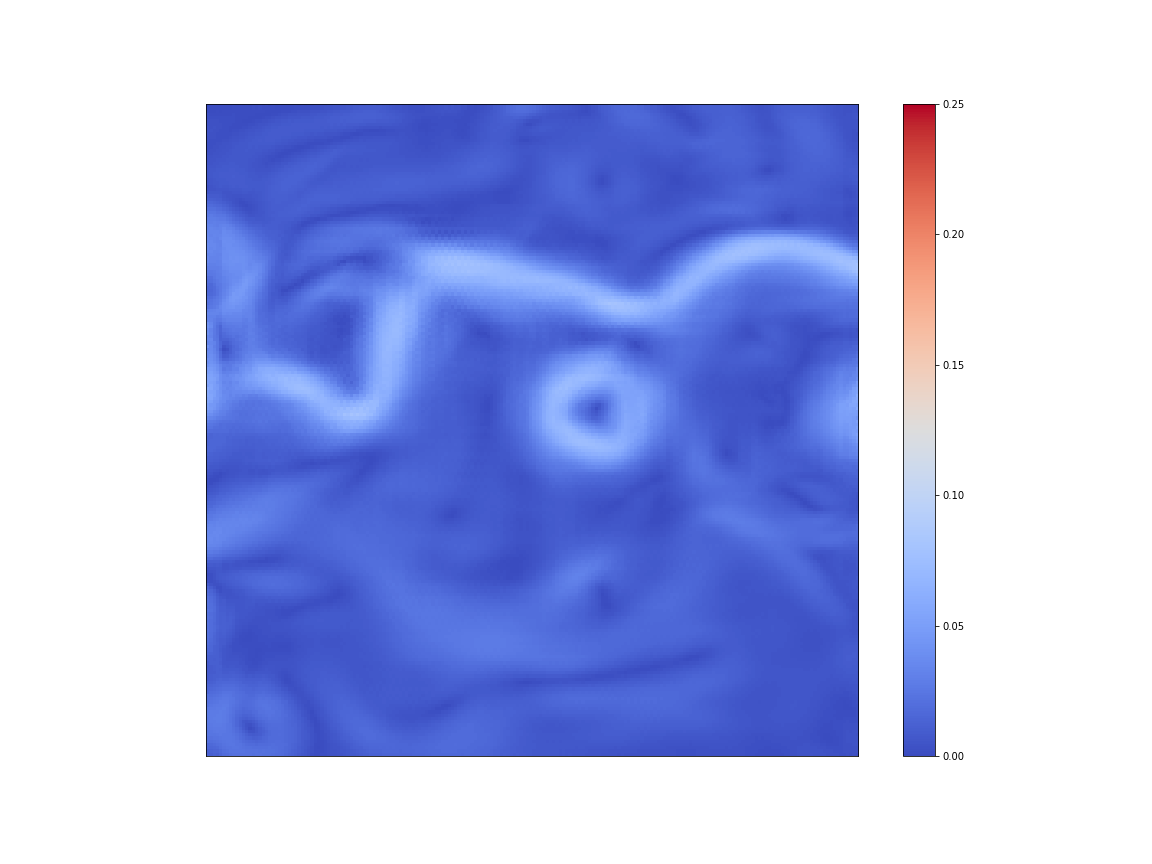}&
    \includegraphics[trim={250 100 300 100},clip,width=4.5cm]{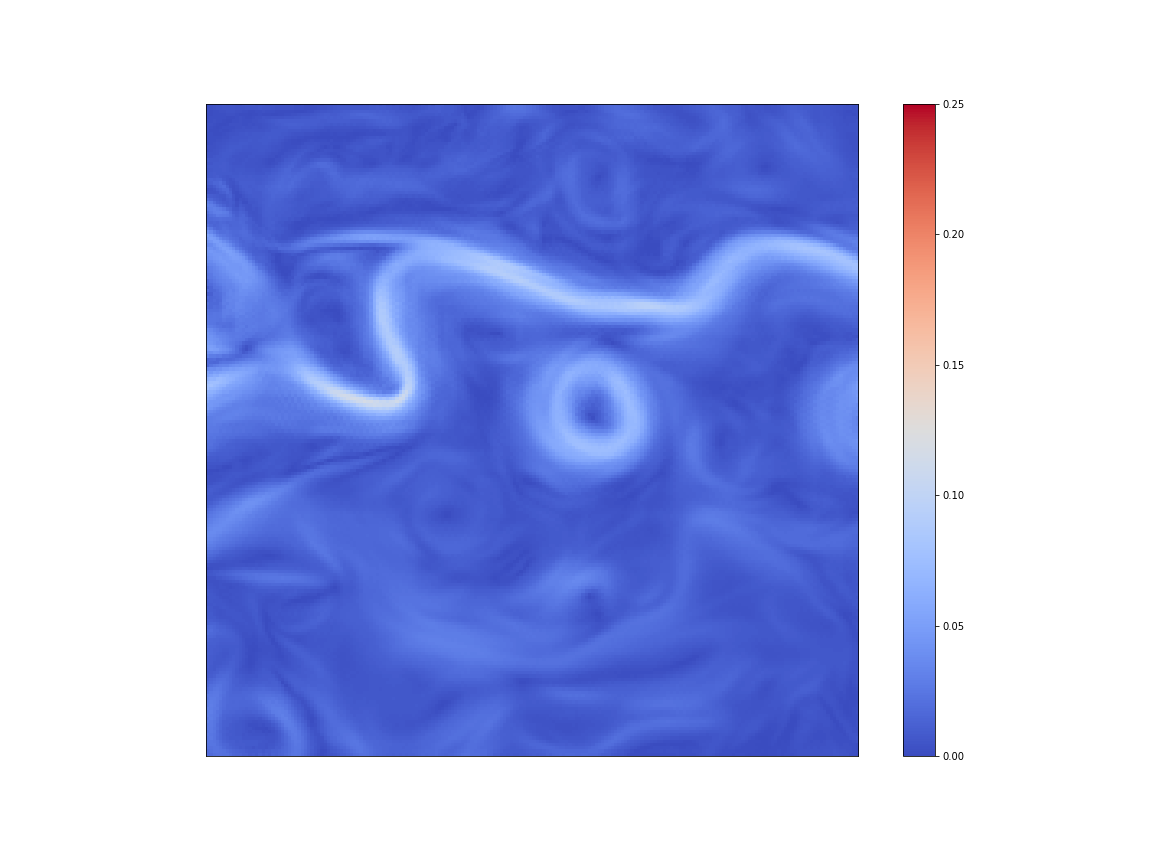}&
    \includegraphics[trim={250 100 300 100},clip,width=4.5cm]{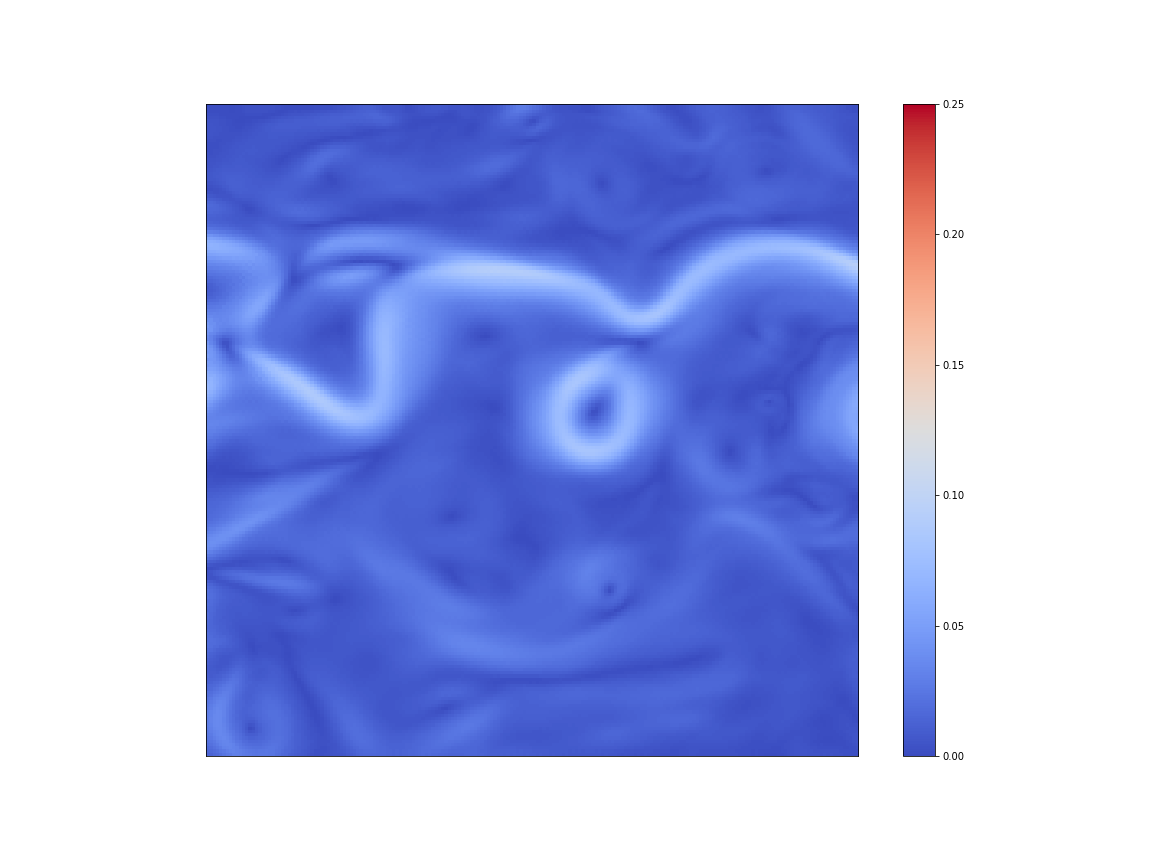}\\
    {\bf U-Net-SSH-only}&{\bf U-Net-SSH-SST}&{\bf  4DVarNet-SSH-only}\\
    \includegraphics[trim={250 100 300 100},clip,width=4.5cm]{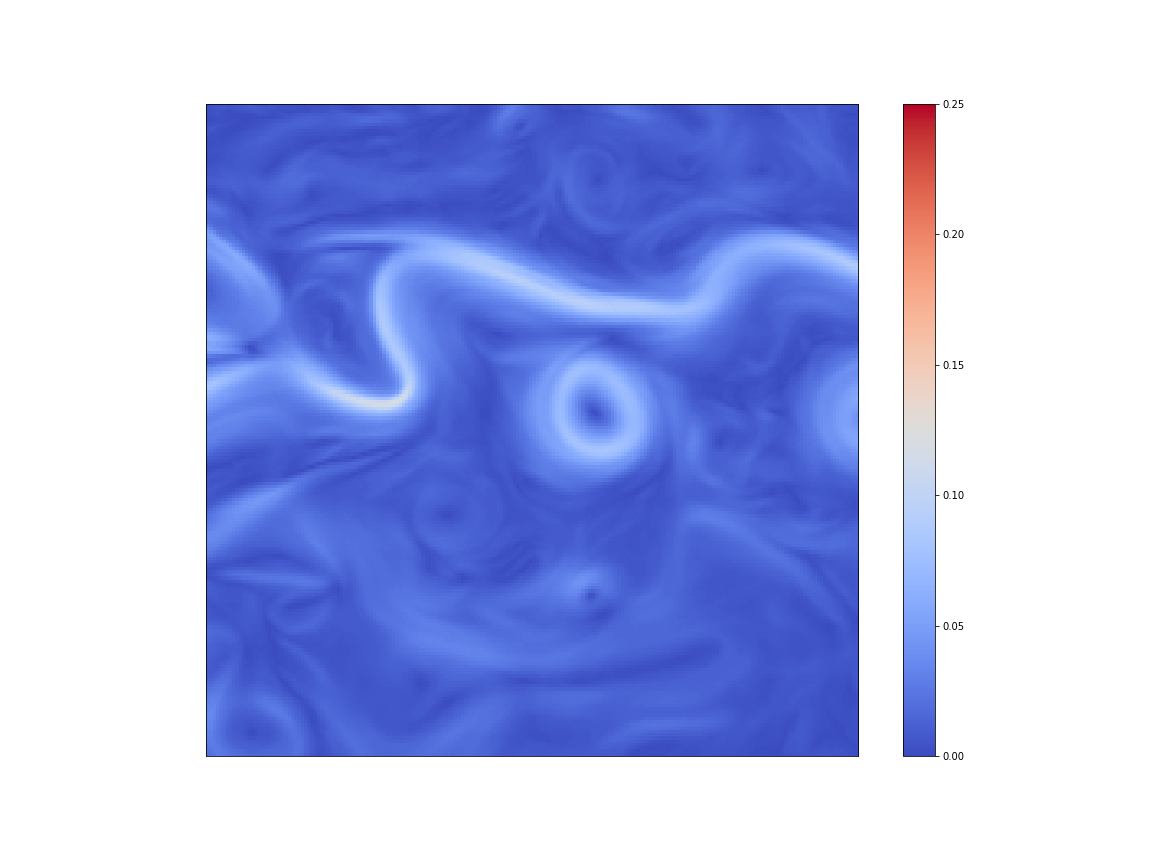}&
    \includegraphics[trim={250 100 300 100},clip,width=4.5cm]{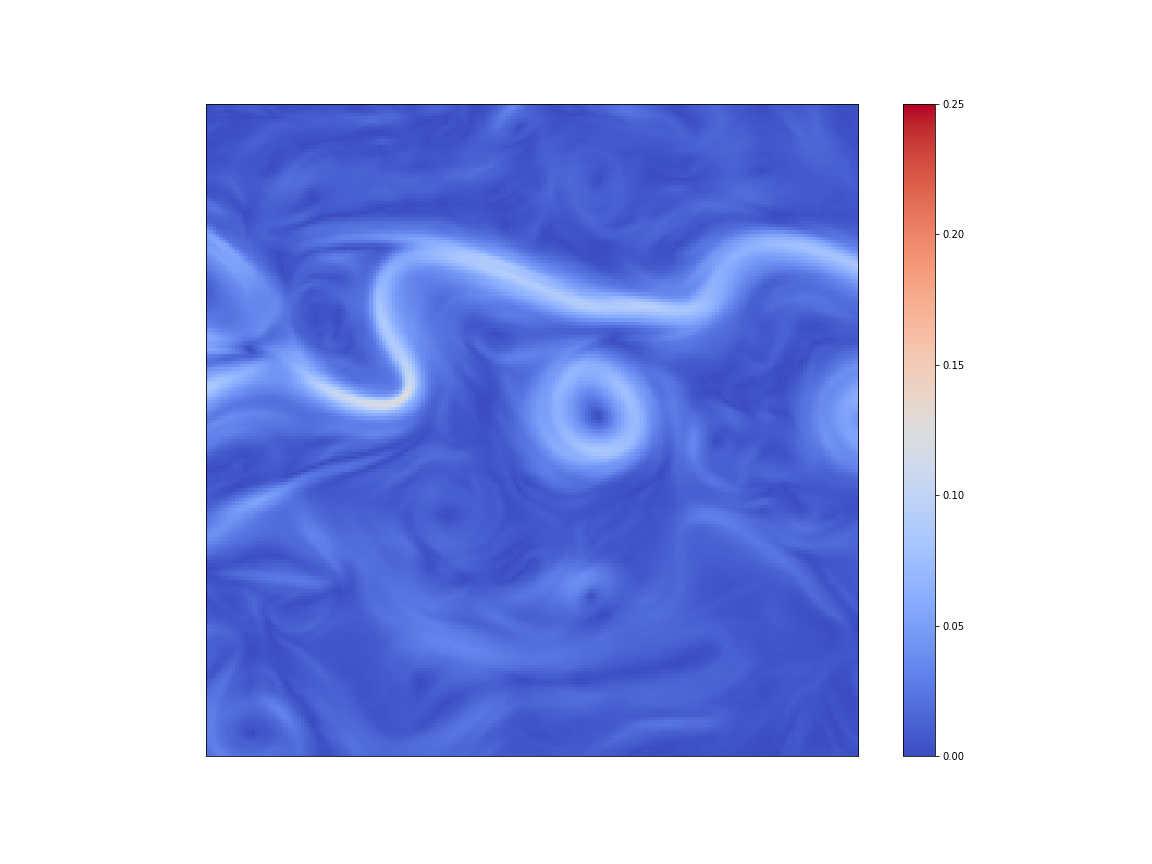}&
    \includegraphics[trim={250 100 300 100},clip,width=4.5cm]{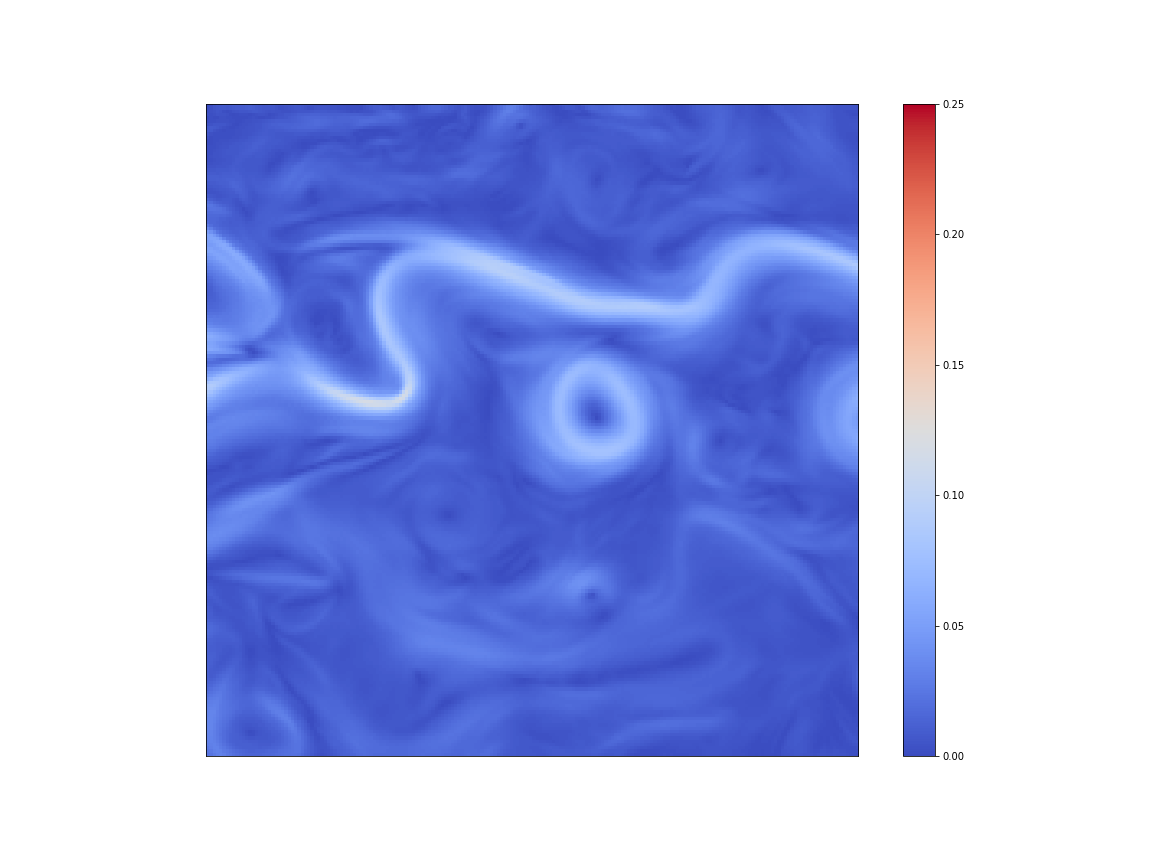}\\
    {\bf 4DVarNet-MMState}&{\bf  4DVarNet-MMObs-Lin}&{\bf 4DVarNet-MMObs-NoLin}\\
    \end{tabular}
    \caption{{\bf Comparison of the gradient of the reconstructed SSH fields on October 25$^{th}$ 2012: } from left to right and top to bottom, we compare the map of the norm of the gradient of the true SSH field (top left) to those of interpolated SSH fields for benchmarked schemes, namely DUACS product  \cite{taburet_duacs_2019}, a SQG-based inversion \cite{isern-fontanet_potential_2006}, direct end-to-end neural inversion using U-Nets \cite{cicek_3d_2016} with altimetry-only data (UNet-SSH-only) and multi-modal input data (UNet-SSTH-SST),  4DVarNet scheme using altimetry-only data (4DVarNet-SSH-only) \cite{} and proposed multimodal  4DVarNet schemes using a multi-modal state (4DVarNet-MMState) and muti-modal observation terms with linear operators (4DVarNet-MMObs-Lin) and non-linear ones (4DVarNet-MMObs-NonLin). We use the same colorbar for all fields. We let the reader to the main text for the description of the different schemes.}
    \label{fig:grad}
\end{figure*}

\begin{figure*}[htb]
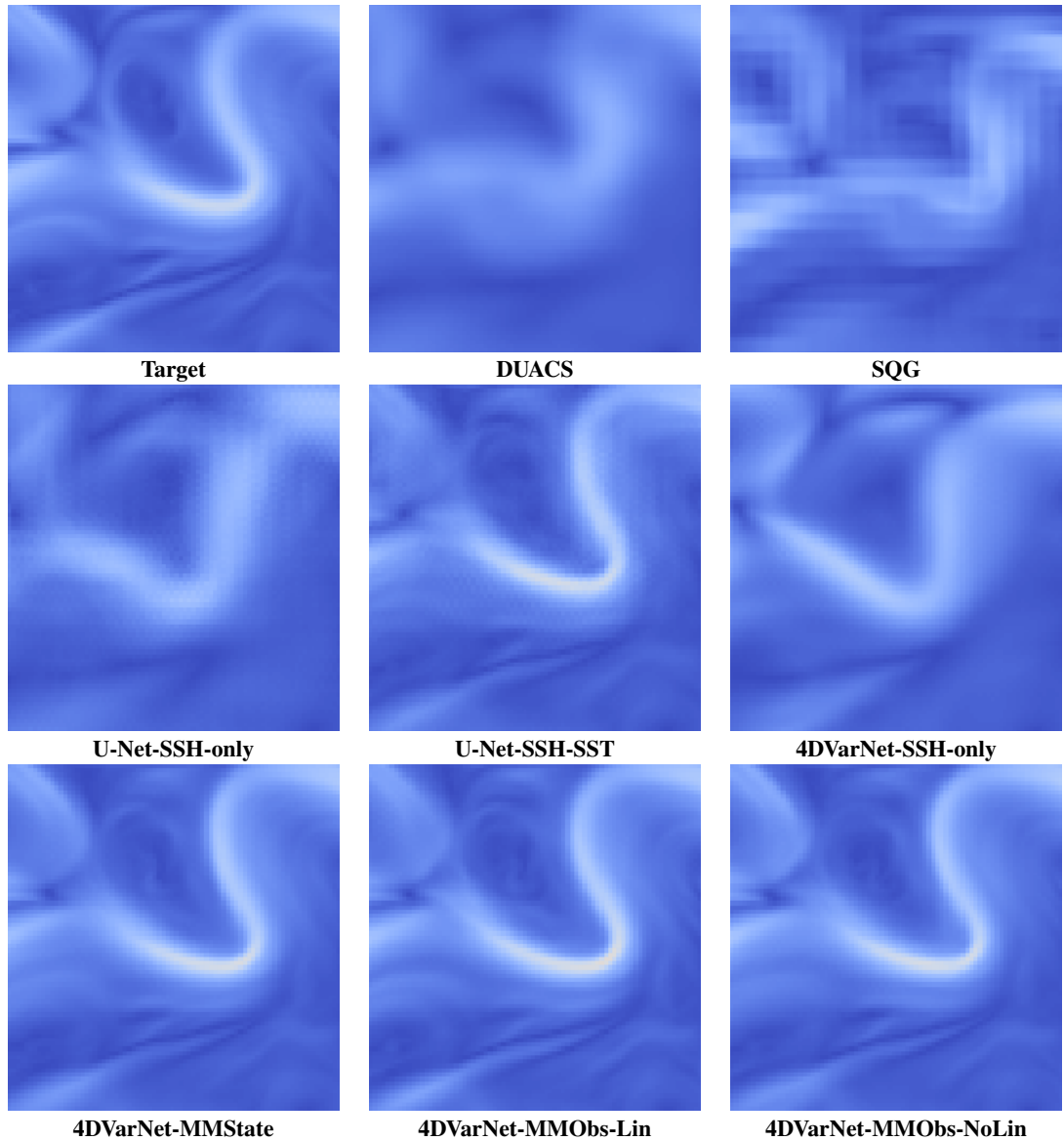

    \centering
    {\footnotesize
    \begin{tabular}{C{4.5cm}C{4.5cm}C{4.5cm}}
    \includegraphics[trim={220 350 680 250},clip,width=4.5cm]{FigBestNoLin4DVarNet/ssh_grad_true_d25.png}&
    \includegraphics[trim={220 350 680 250},clip,width=4.5cm]{FigBestNoLin4DVarNet/ssh_grad_oi_d25.png}&
    \includegraphics[trim={220 350 680 250},clip,width=4.5cm]{FigBenchmarkedModels/ssh_grad_recSQG_d25.png}\\
    {\bf Target}&{\bf DUACS}&{\bf SQG}\\
    \includegraphics[trim={220 350 680 250},clip,width=4.5cm]{FigDirectInvUnetnoSST/ssh_grad_rec_SLA_d25.png}&
    \includegraphics[trim={220 350 680 250},clip,width=4.5cm]{FigDirectInvUnetwSST/ssh_grad_rec_SLA_d25.png}&
    \includegraphics[trim={220 350 680 250},clip,width=4.5cm]{Fig4DVarNetSLAOnly/ssh_grad_4dvarnet_SLA_d25.png}\\
    {\bf U-Net-SSH-only}&{\bf U-Net-SSH-SST}&{\bf  4DVarNet-SSH-only}\\
    \includegraphics[trim={220 350 680 250},clip,width=4.5cm]{Fig4DVarNetSLAMMState/ssh_grad_rec_d25.png}&
    \includegraphics[trim={220 350 680 250},clip,width=4.5cm]{FigBestLin4DVarNet/ssh_grad_4dvarnet_SLASST_d25.png}&
    \includegraphics[trim={220 350 680 250},clip,width=4.5cm]{FigBestNoLin4DVarNet/ssh_grad_4dvarnet_SLASST_d25.png}\\
    {\bf 4DVarNet-MMState}&{\bf  4DVarNet-MMObs-Lin}&{\bf 4DVarNet-MMObs-NoLin}\\
    \end{tabular}}
    \caption{{\bf Zoom on reconstructed SSH gradient fields on October 25$^{th}$ 2012:} we display zooms for a upper-left subregion of the fields displayed in  Fig.\ref{fig:grad}.}
    \label{fig:grad zoom}
\end{figure*}

\begin{figure*}[htb]
    \centering
    {\footnotesize
    \begin{tabular}{C{4.5cm}C{4.5cm}C{4.5cm}}
    \includegraphics[trim={220 350 680 250},clip,width=4.5cm]{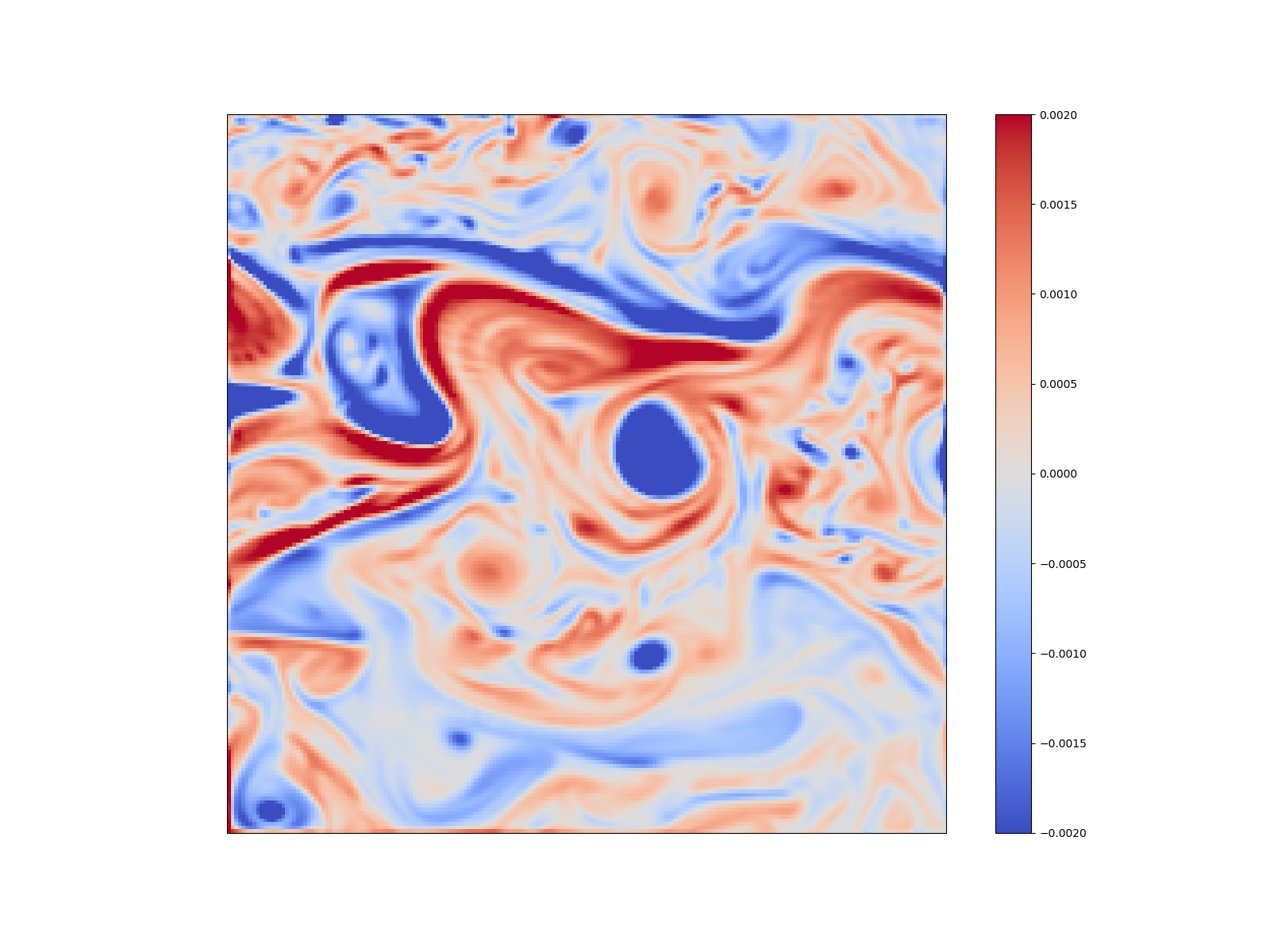}&
    \includegraphics[trim={220 350 680 250},clip,width=4.5cm]{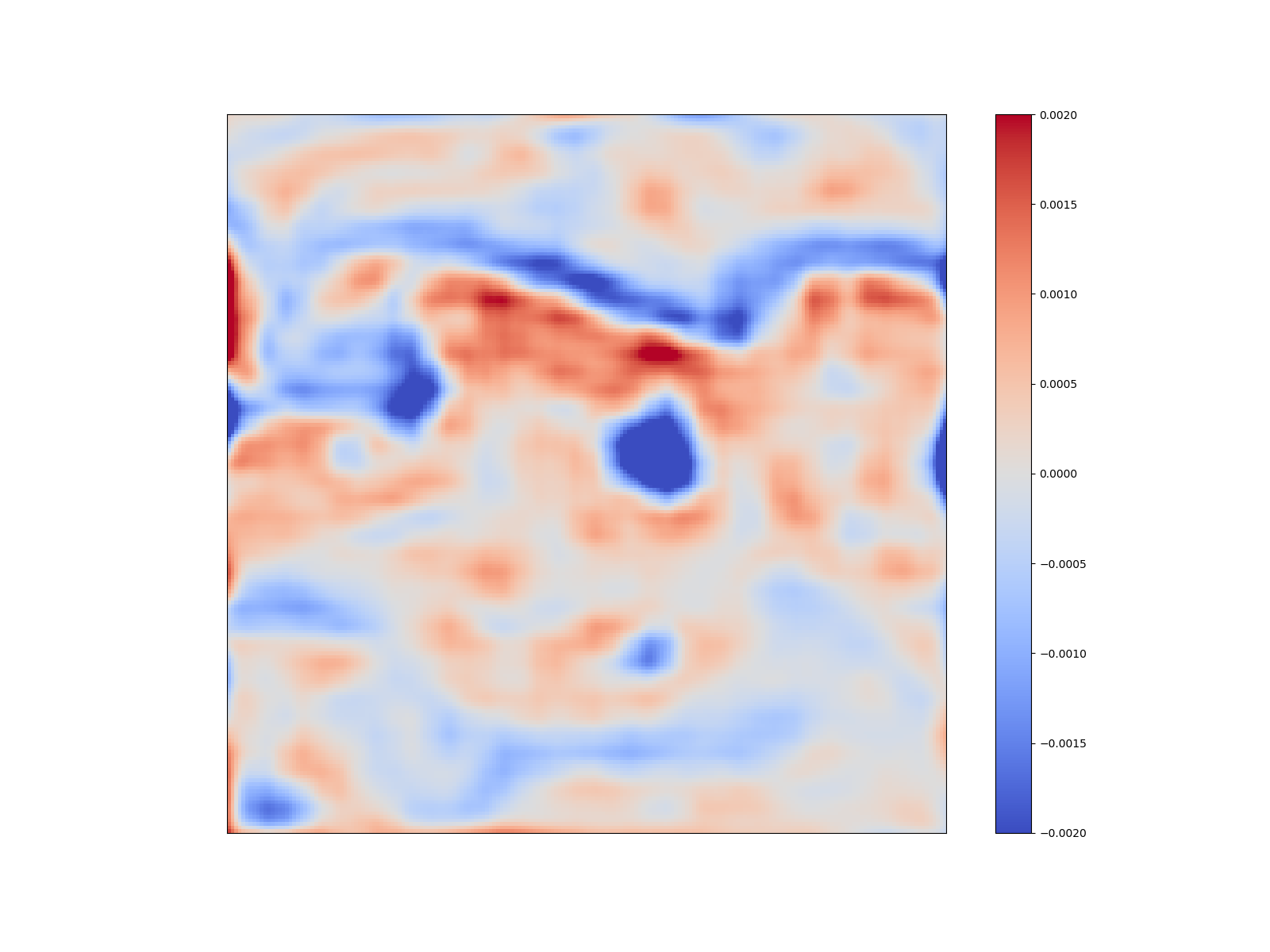}&
    \includegraphics[trim={220 350 680 250},clip,width=4.5cm]{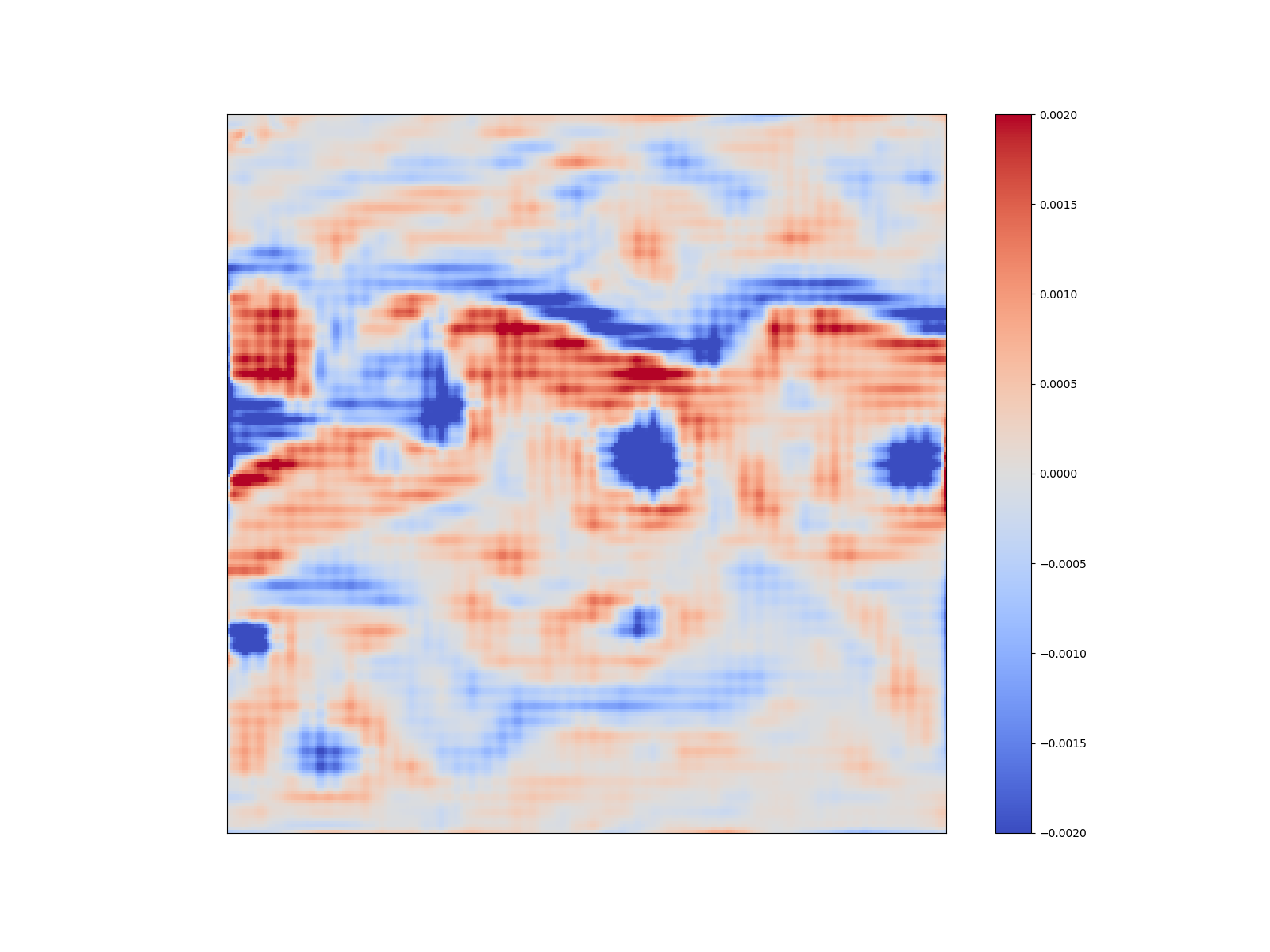}\\
    {\bf Target}&{\bf DUACS}&{\bf SQG}\\
    \includegraphics[trim={220 350 680 250},clip,width=4.5cm]{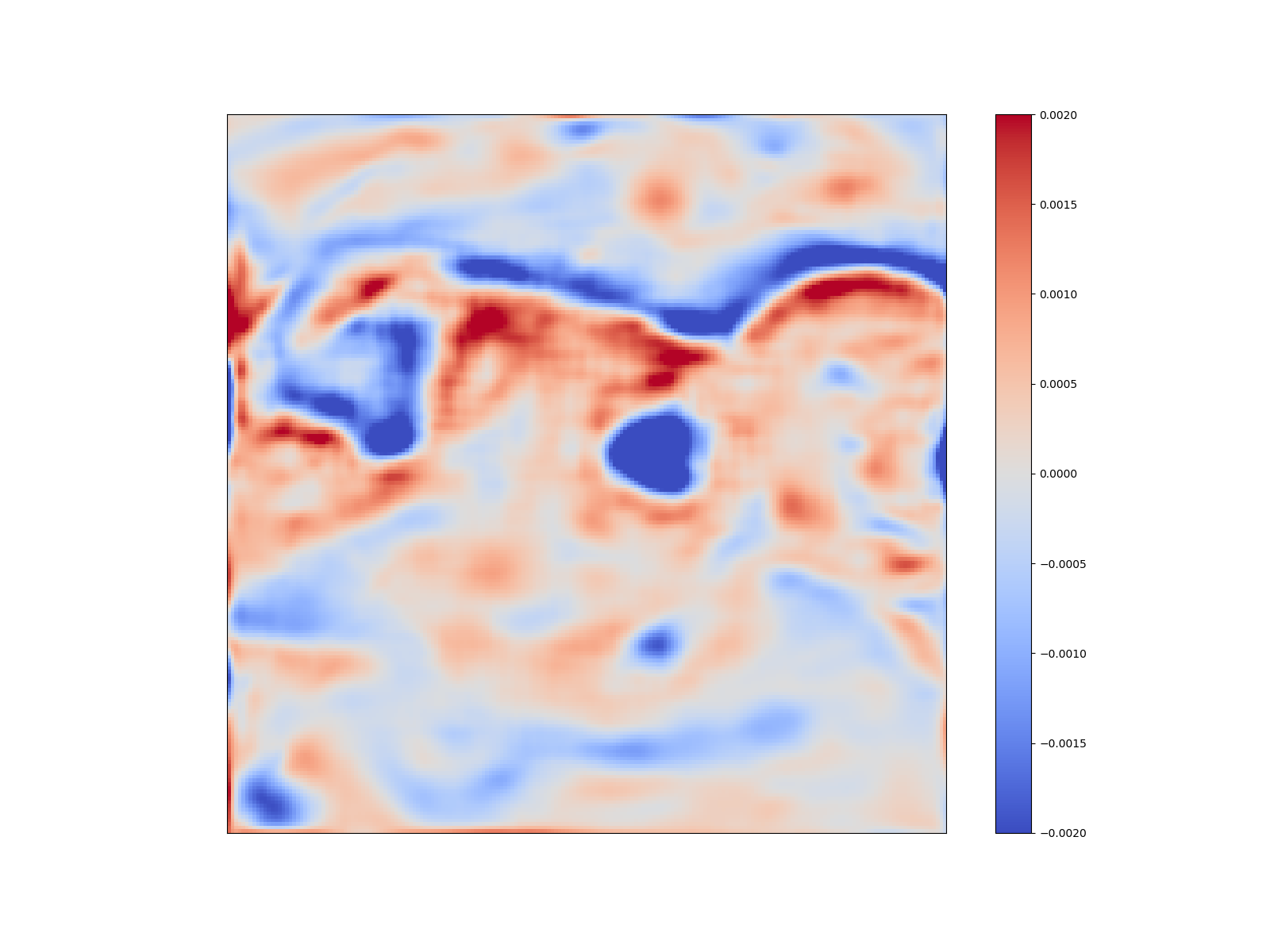}&
    \includegraphics[trim={220 350 680 250},clip,width=4.5cm]{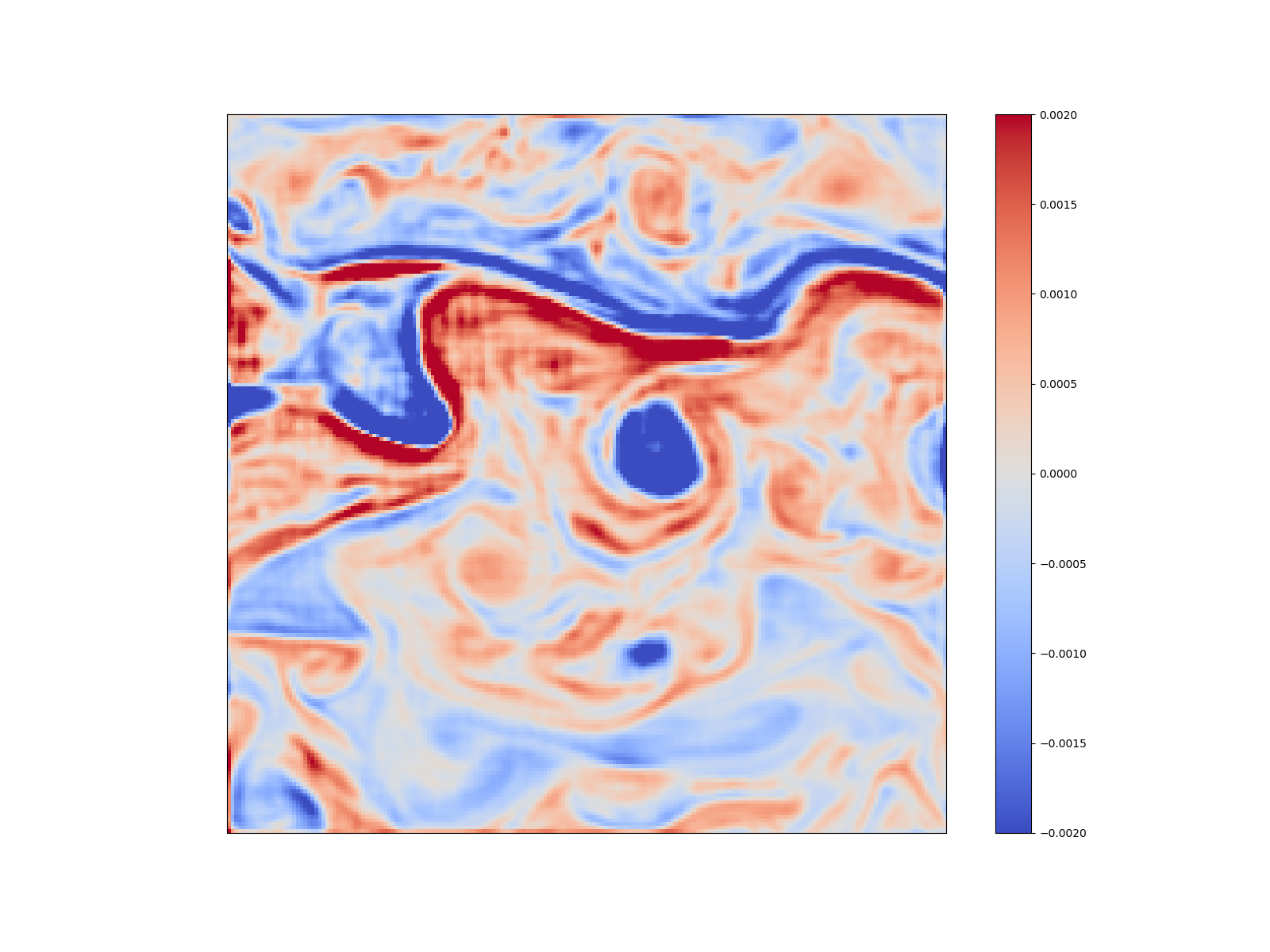}&
    \includegraphics[trim={220 350 680 250},clip,width=4.5cm]{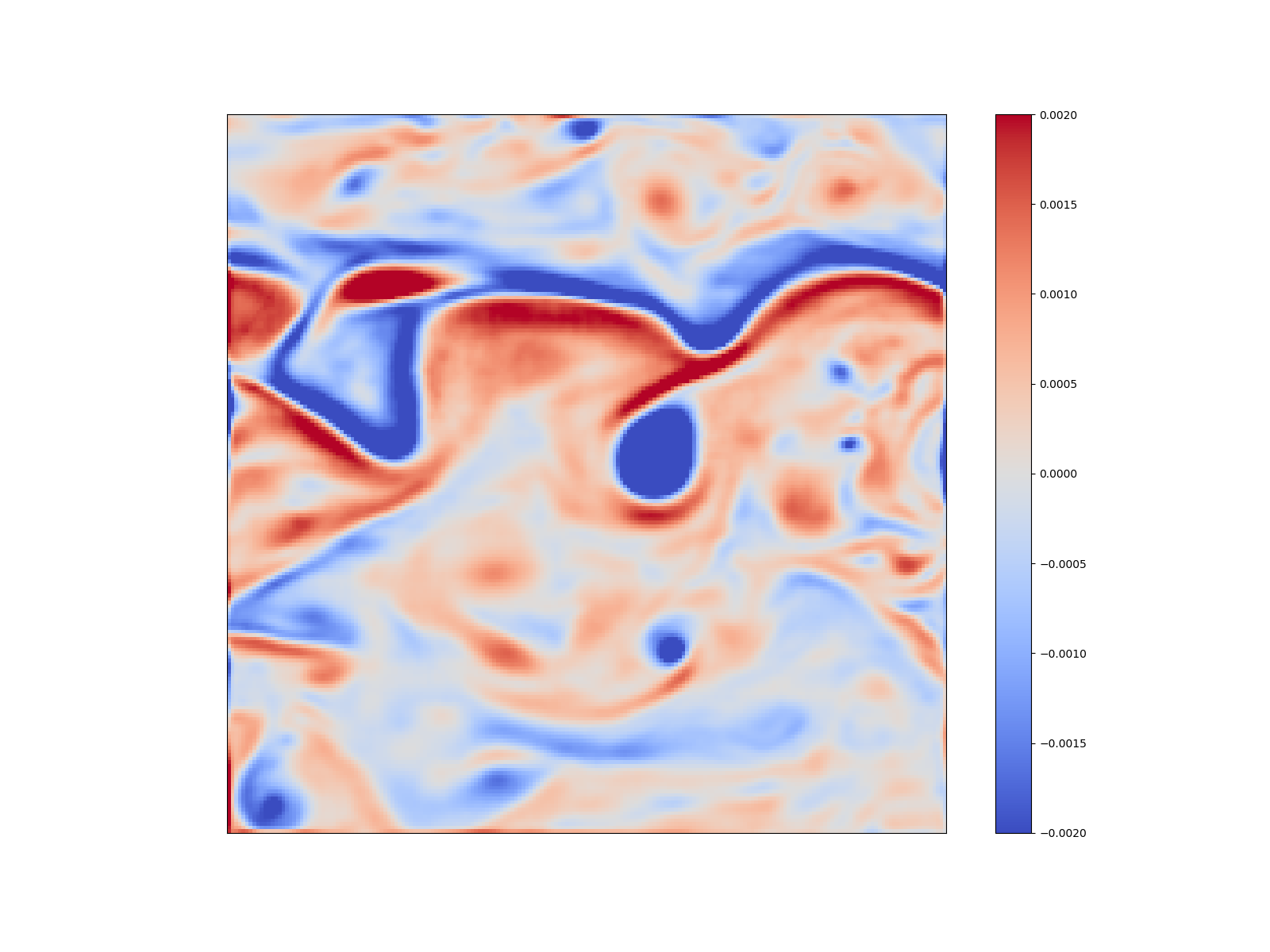}\\
    {\bf U-Net-SSH-only}&{\bf U-Net-SSH-SST}&{\bf  4DVarNet-SSH-only}\\
    \includegraphics[trim={220 350 680 250},clip,width=4.5cm]{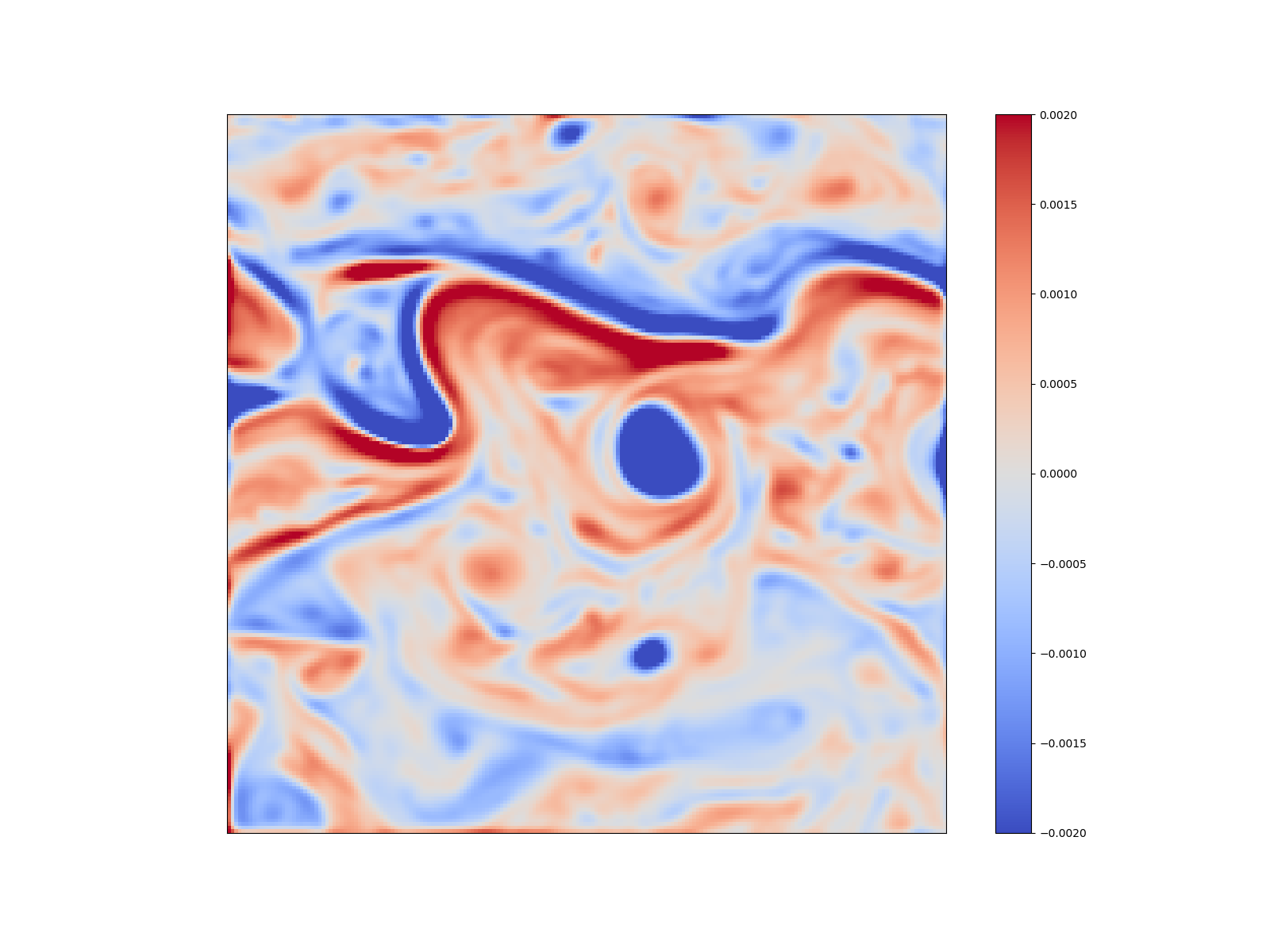}&
    \includegraphics[trim={220 350 680 250},clip,width=4.5cm]{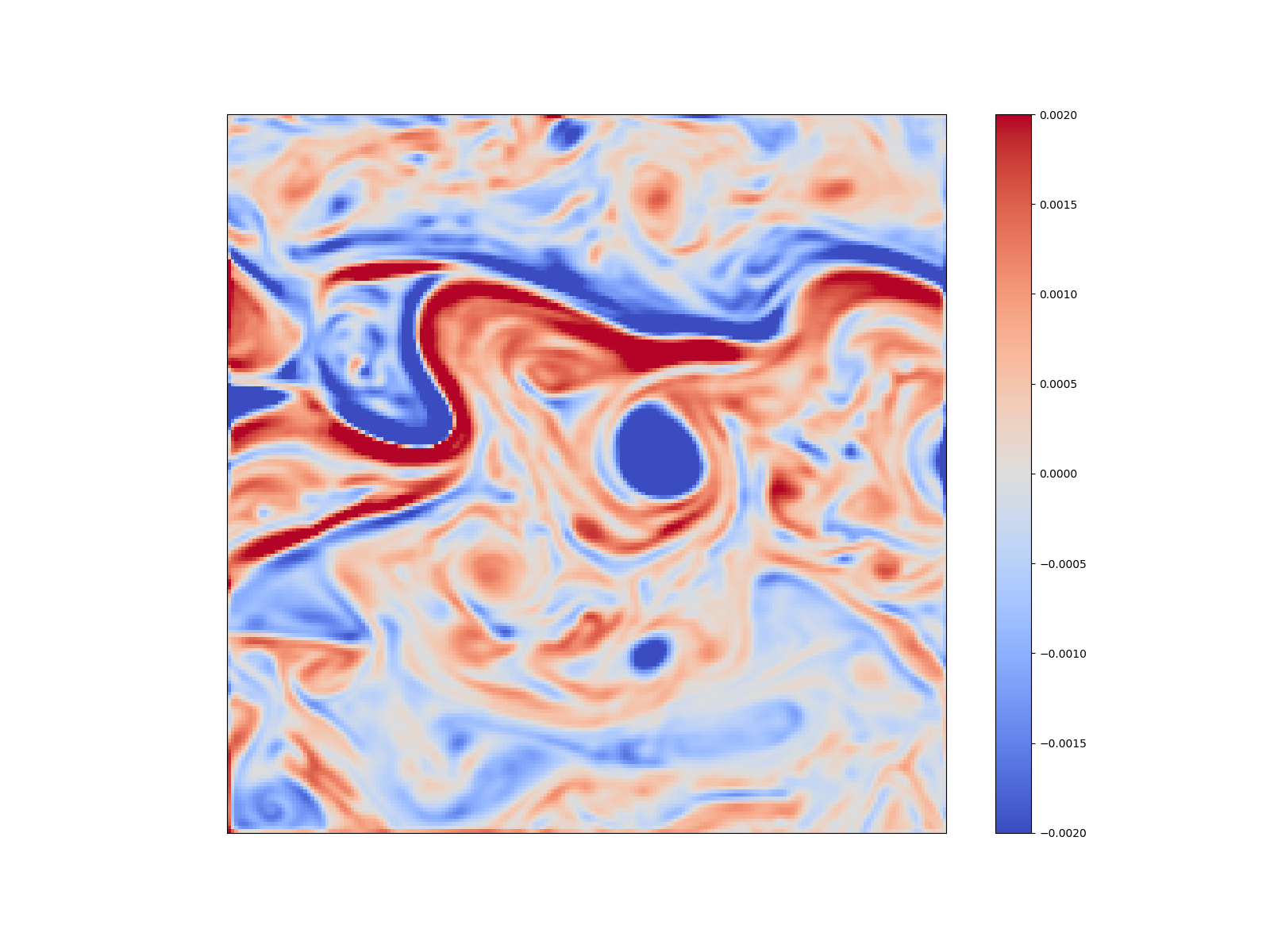}&
    \includegraphics[trim={220 350 680 250},clip,width=4.5cm]{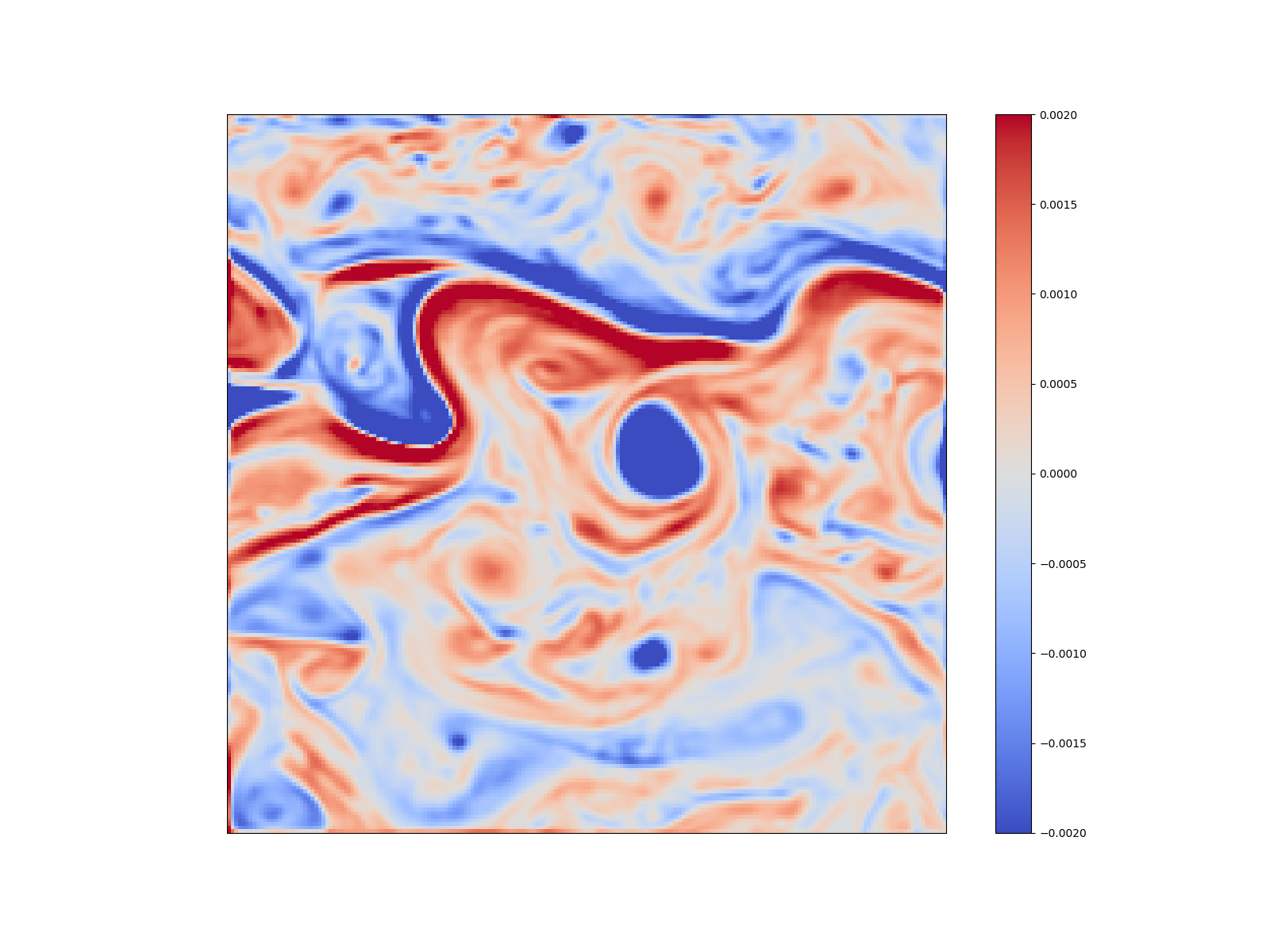}\\
    {\bf 4DVarNet-MMState}&{\bf  4DVarNet-MMObs-Lin}&{\bf 4DVarNet-MMObs-NoLin}\\
    \end{tabular}}
    \caption{{\bf Zoom on reconstructed SSH Laplacian fields on October 25$^{th}$ 2012:} we depict zooms for a top-left subregion of the gradient norm fields displayed in Fig.\ref{fig:grad}.}
    \label{fig:lap zoom}
\end{figure*}

As an illustration, we display in Fig.\ref{fig:grad} the norm of the gradient of the reconstructed SSH fields. Visually, the operational processing \cite{taburet_duacs_2019} leads to a blurry reconstruction as stressed by the resolved spatial scale above 1$^\circ$. By contrast, all 4DVarNet schemes lead to much sharper gradients which are more similar to those of the true SSH fields. We draw similar observation from the reconstructed Laplacian fields. This example also provides a clear illustration of the added value of SSH-SST synergies. When considering the SSH-only 4DVarNet scheme, we cannot perfectly recover the geometry of the main meander (see zoom on upper-left region in Fig.\ref{fig:grad zoom} and  
\ref{fig:lap zoom}) as well as the orientation of the large eddy South of the main meander due to the scarce sampling of the altimeter data. We may notice that the SST field in Fig.\ref{fig:obs data} clearly reveals these geometrical patterns. Interestingly, multimodal 4DVarNet schemes successfully extract these features to improve the reconstruction of the SSH field. Zooms in Fig.\ref{fig:grad zoom} and  
\ref{fig:lap zoom} further emphasize the better performance of the 4DVarNet scheme with a non-linear multimodal observation term as well as local artifacts generated by a direct inversion using U-Nets. 

We also visualize the SST features learnt by the best linear and non-linear multimodal 4DVarNet schemes (resp. using 3-dimensional and 20-dimensional multimodal observation terms (\ref{eq: multimodal obs})). All features enhance fine-scale patterns. Non-linear features involve even finer-scale patterns and seem to reveal a greater diversity of patterns. For instance, some features exhibit the large eddy South of the main meander while others not. In line with the SQG theory \cite{lapeyre_dynamics_2006,isern-fontanet_potential_2006}, it seems that we can interpret the learnt SST feature extraction step as a combination of pass-band filters and of template-driven detection filters. 

\begin{figure*}[htb]
    \centering
    \begin{tabular}{C{2.75cm}C{2.75cm}C{2.75cm}C{2.75cm}C{2.75cm}}
    &\multicolumn{3}{c}{\bf Learnt SST feature maps}&\\
    &\includegraphics[trim={250 100 300 100},clip,width=2.75cm]{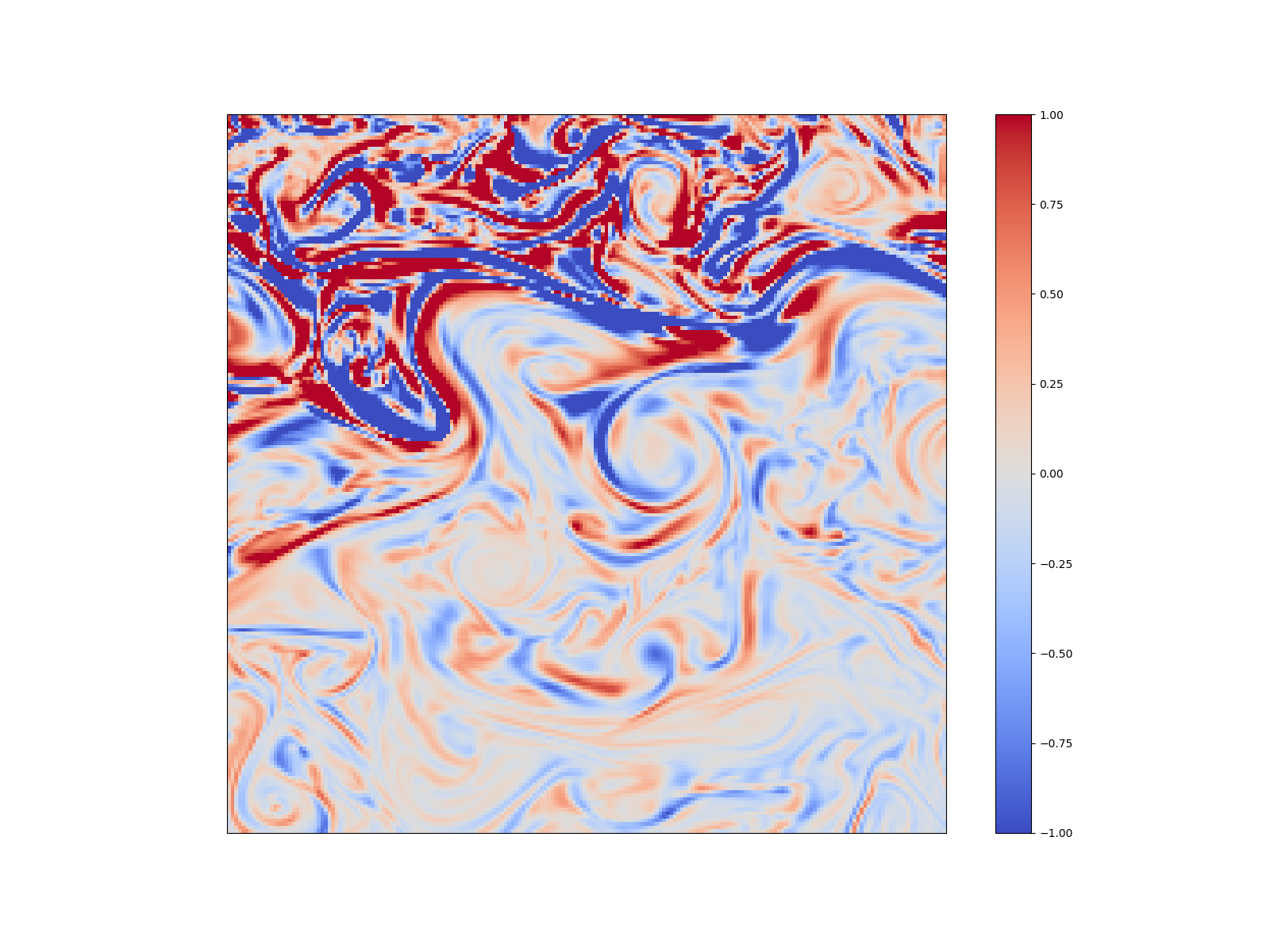}&
    \includegraphics[trim={250 100 300 100},clip,width=2.75cm]{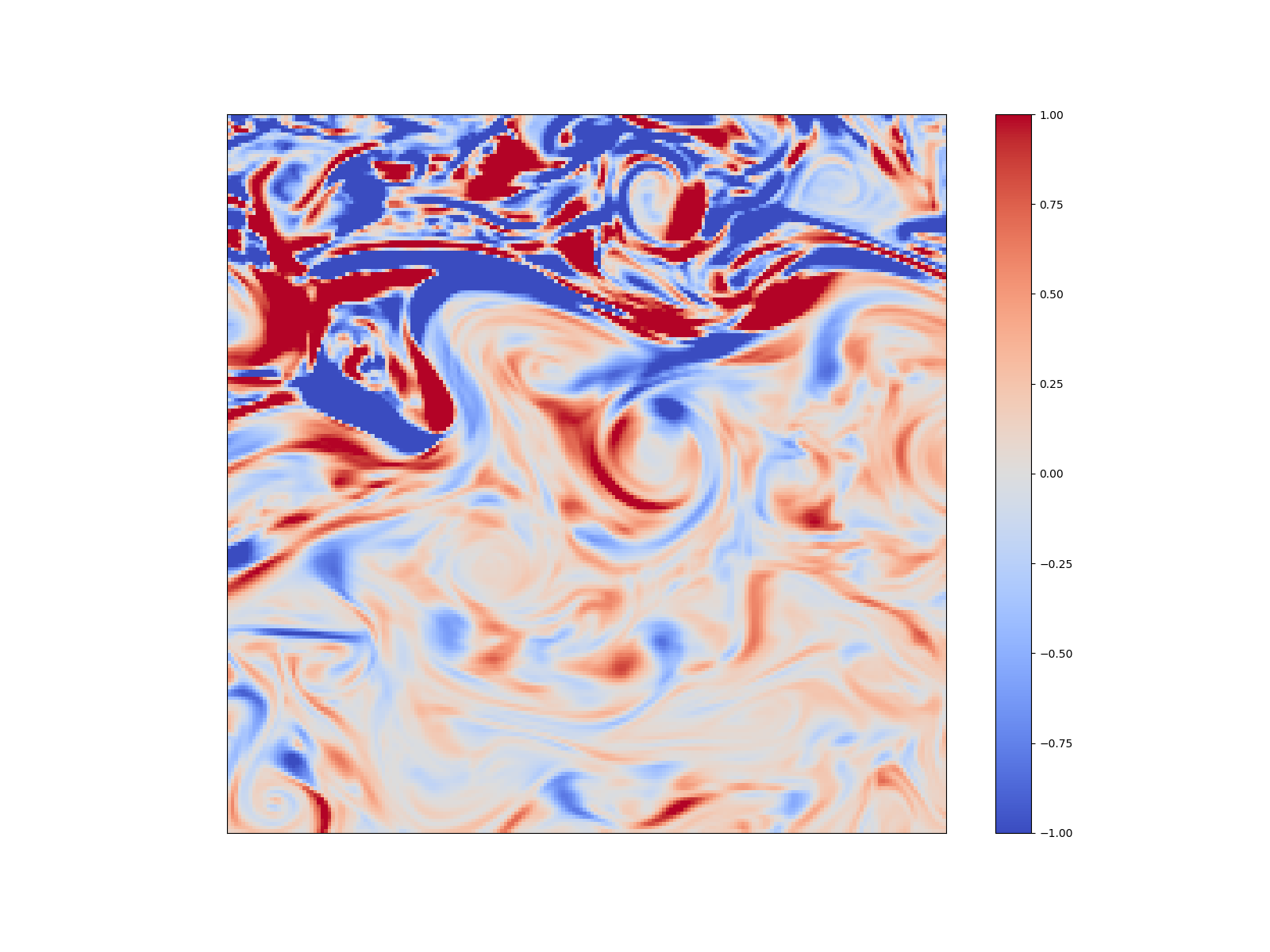}&
    \includegraphics[trim={250 100 300 100},clip,width=2.75cm]{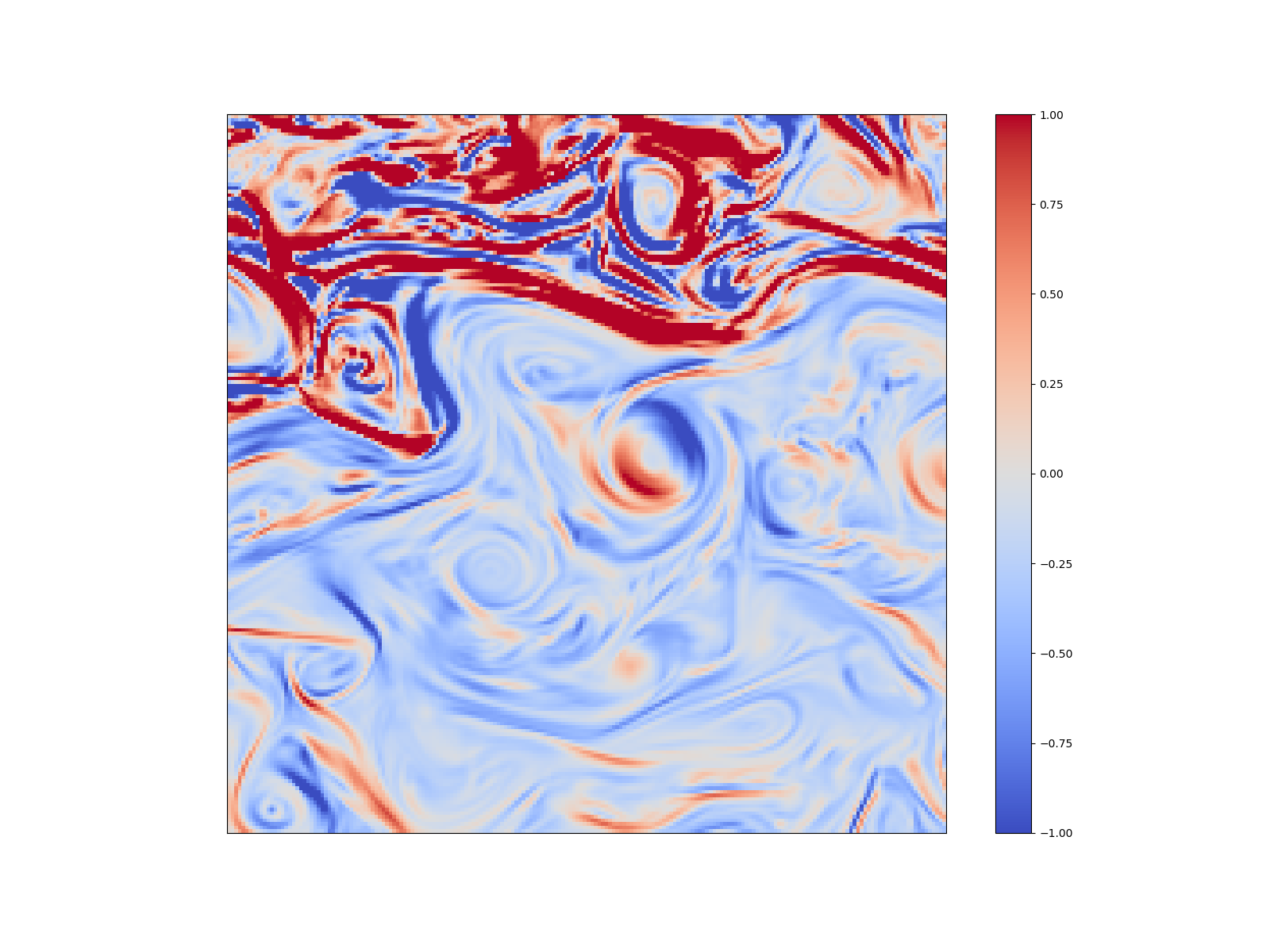}&\\
    ~\\
    \multicolumn{5}{c}{\bf Learnt non-linear SST feature maps}\\
    \includegraphics[trim={250 100 300 100},clip,width=2.75cm]{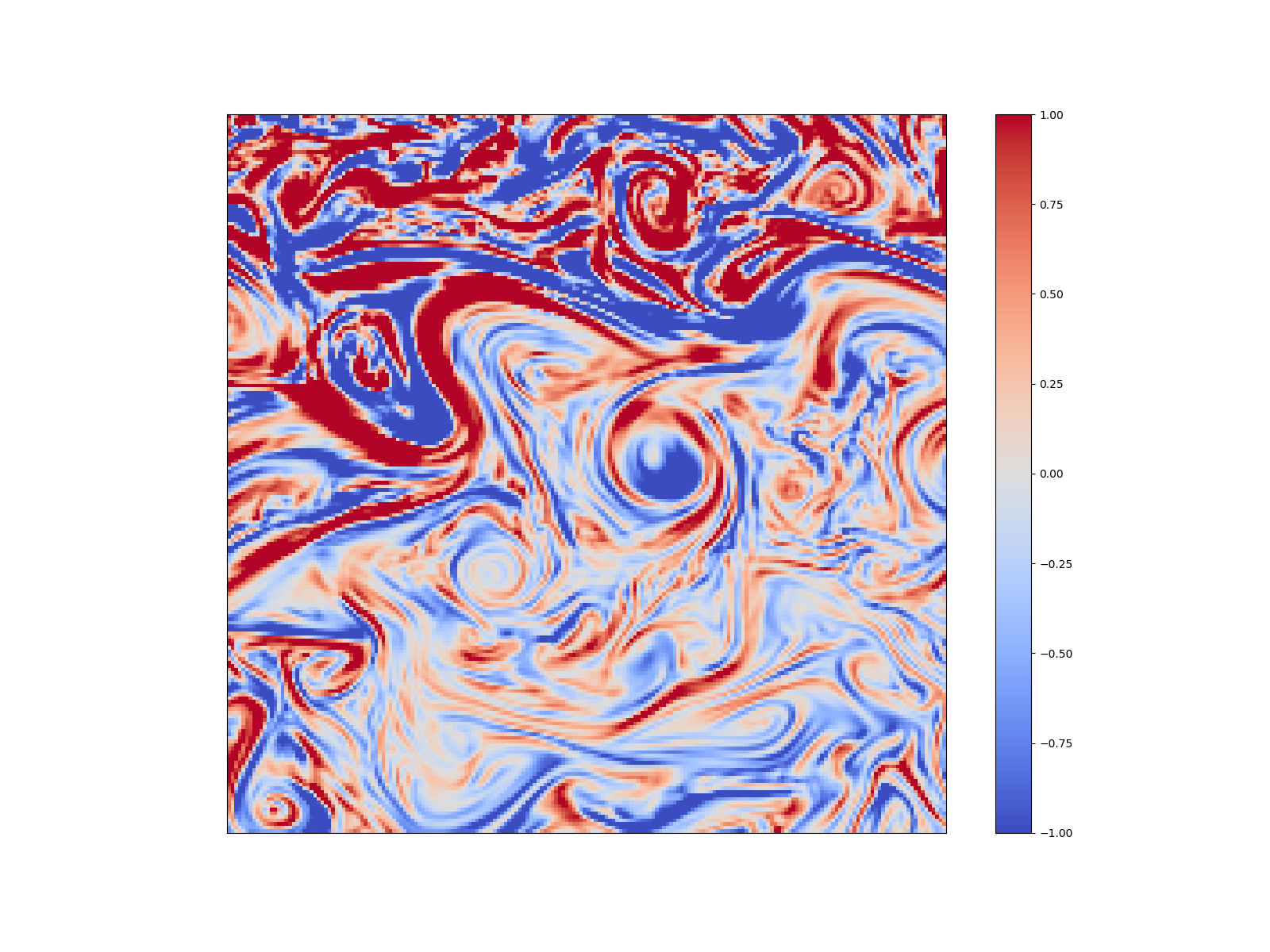}&
    \includegraphics[trim={250 100 300 100},clip,width=2.75cm]{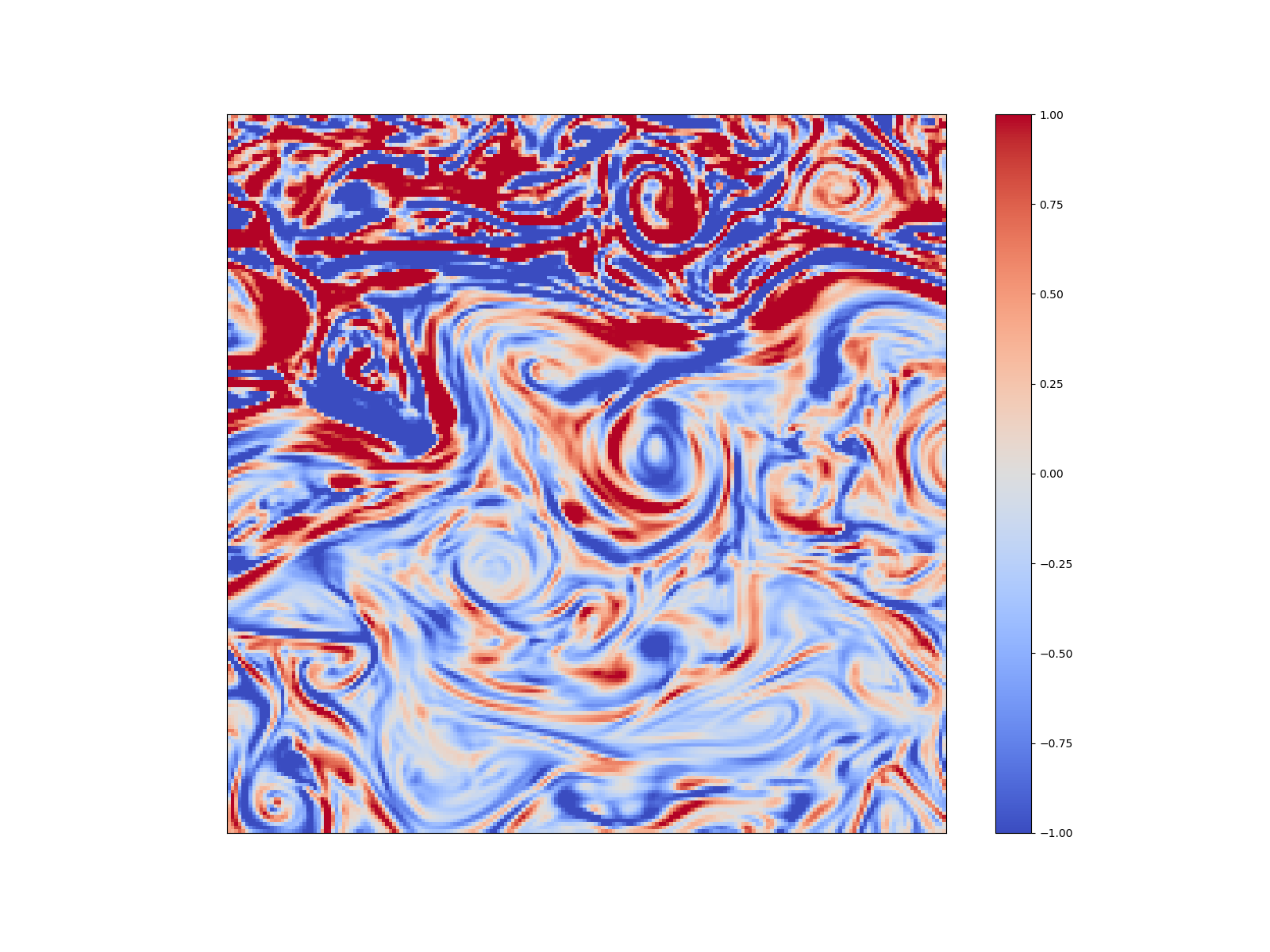}&
    \includegraphics[trim={250 100 300 100},clip,width=2.75cm]{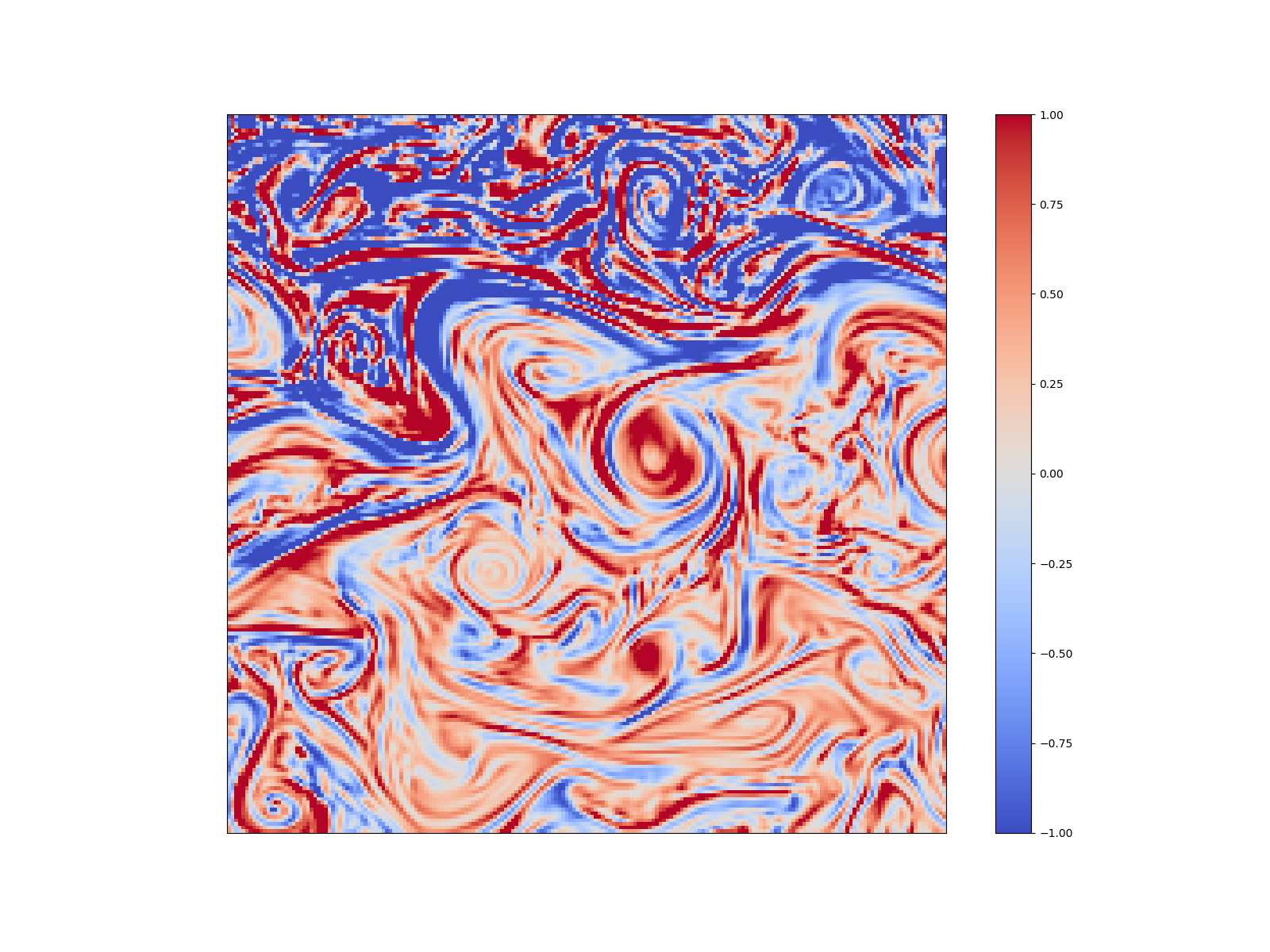}&
    \includegraphics[trim={250 100 300 100},clip,width=2.75cm]{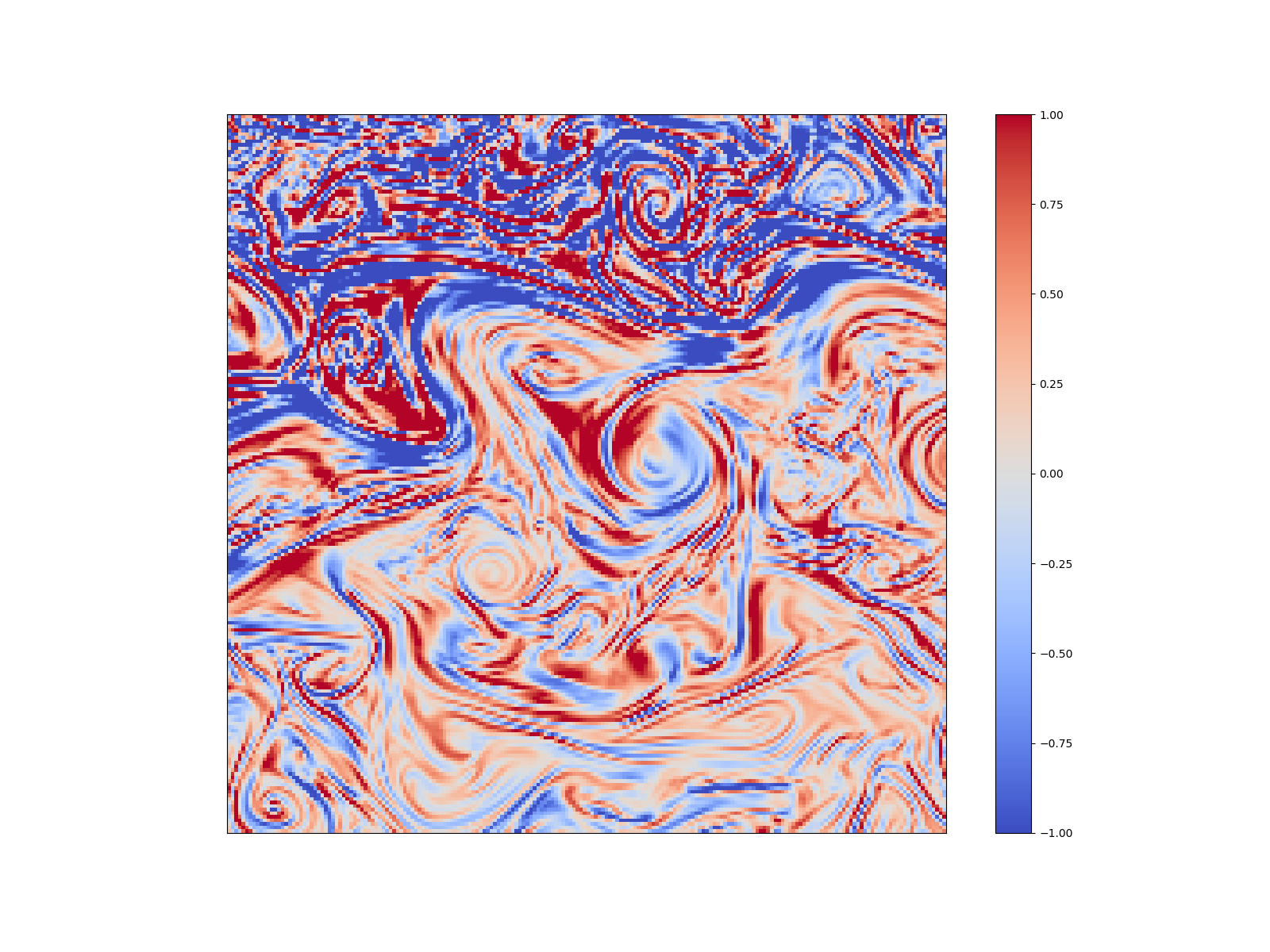}&
    \includegraphics[trim={250 100 300 100},clip,width=2.75cm]{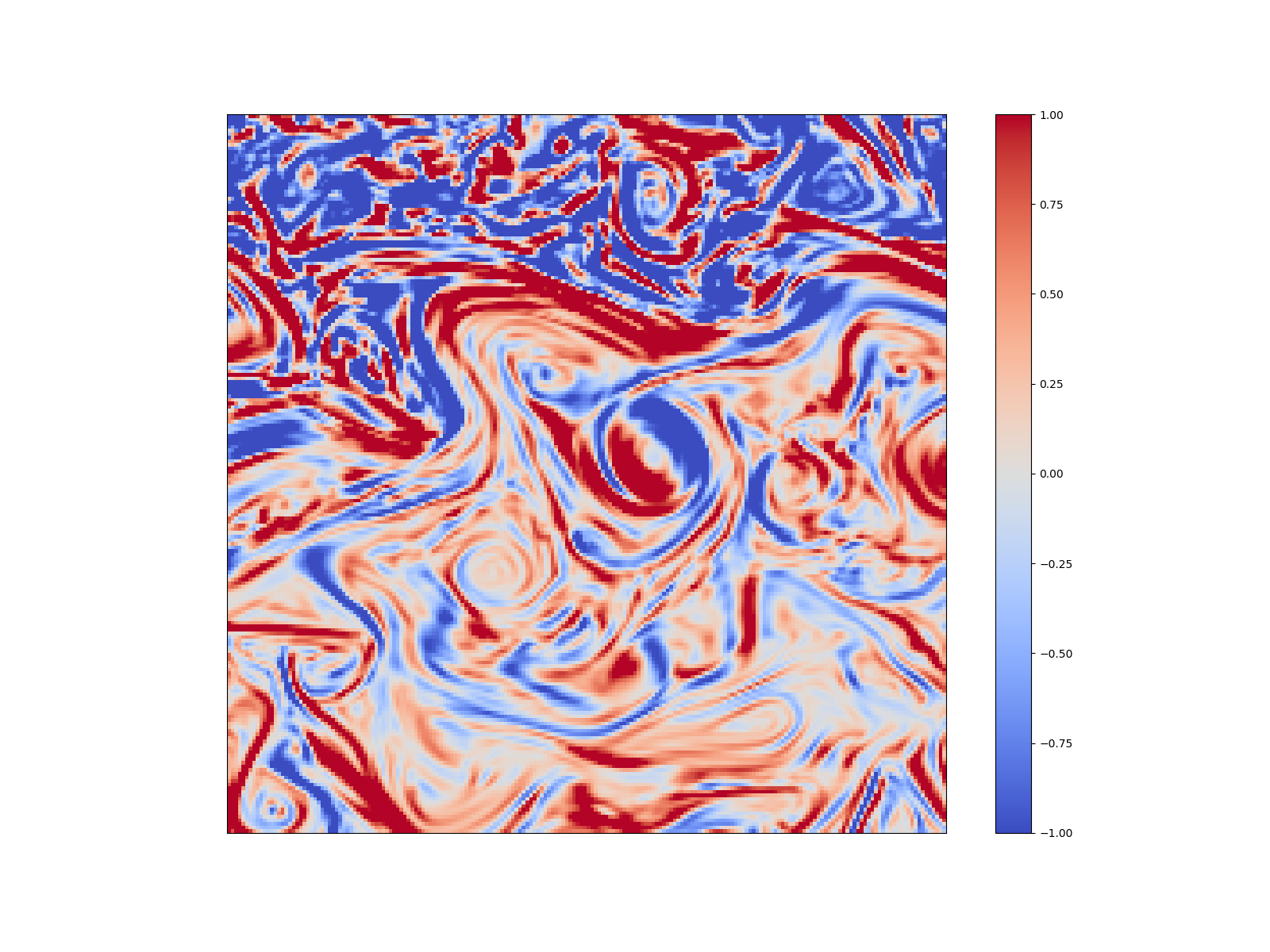}\\
    \includegraphics[trim={250 100 300 100},clip,width=2.75cm]{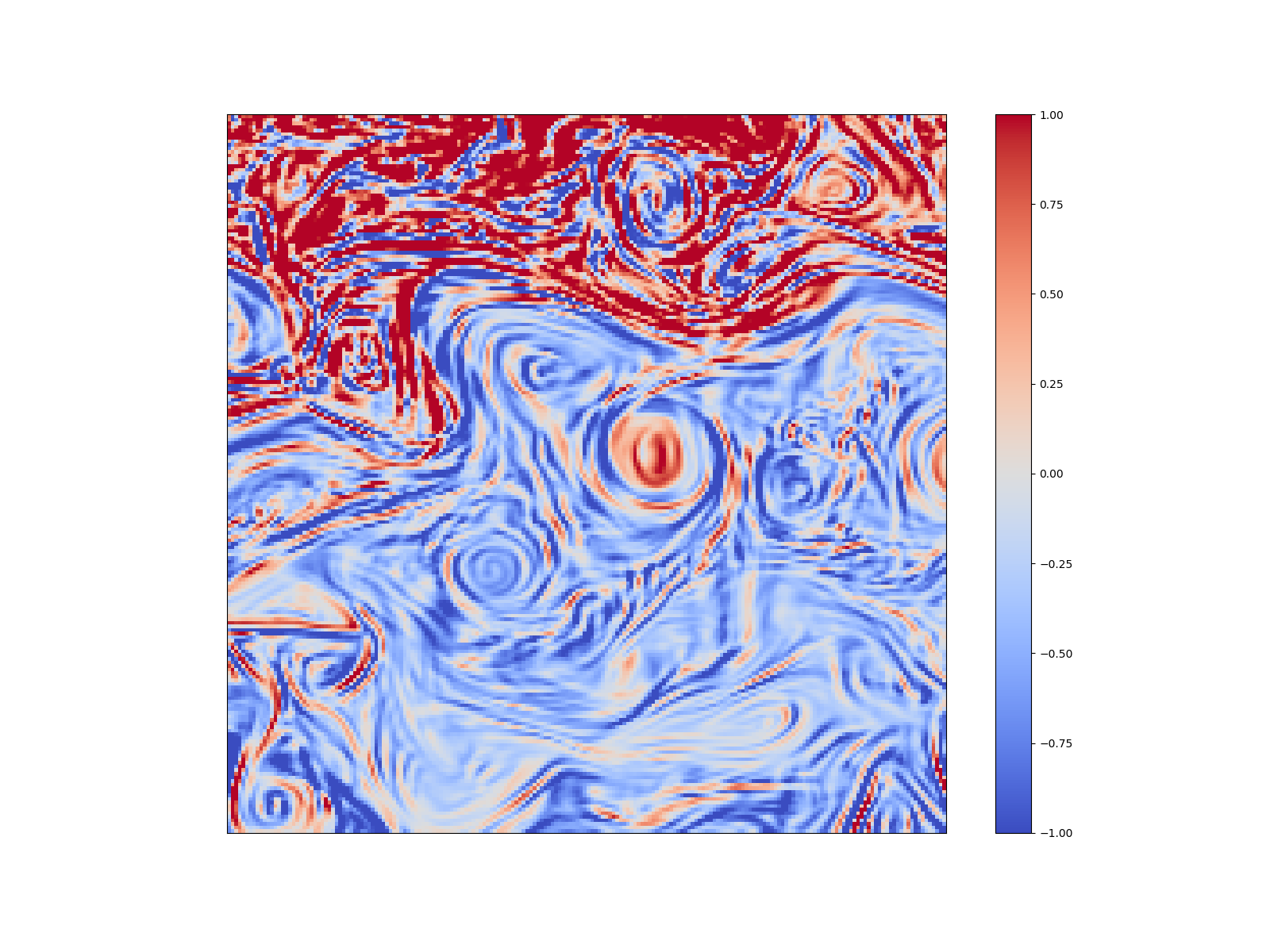}&
    \includegraphics[trim={250 100 300 100},clip,width=2.75cm]{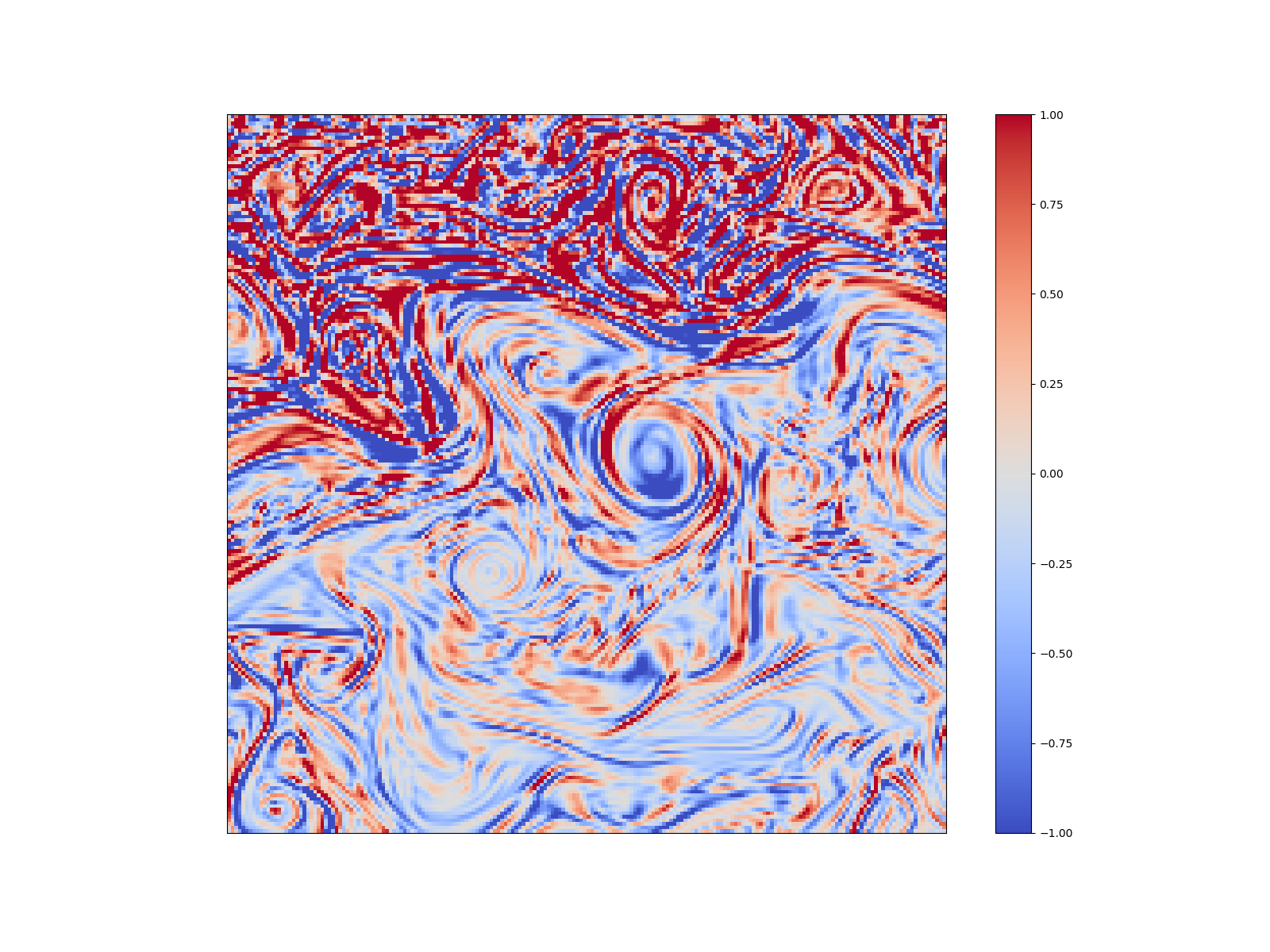}&
    \includegraphics[trim={250 100 300 100},clip,width=2.75cm]{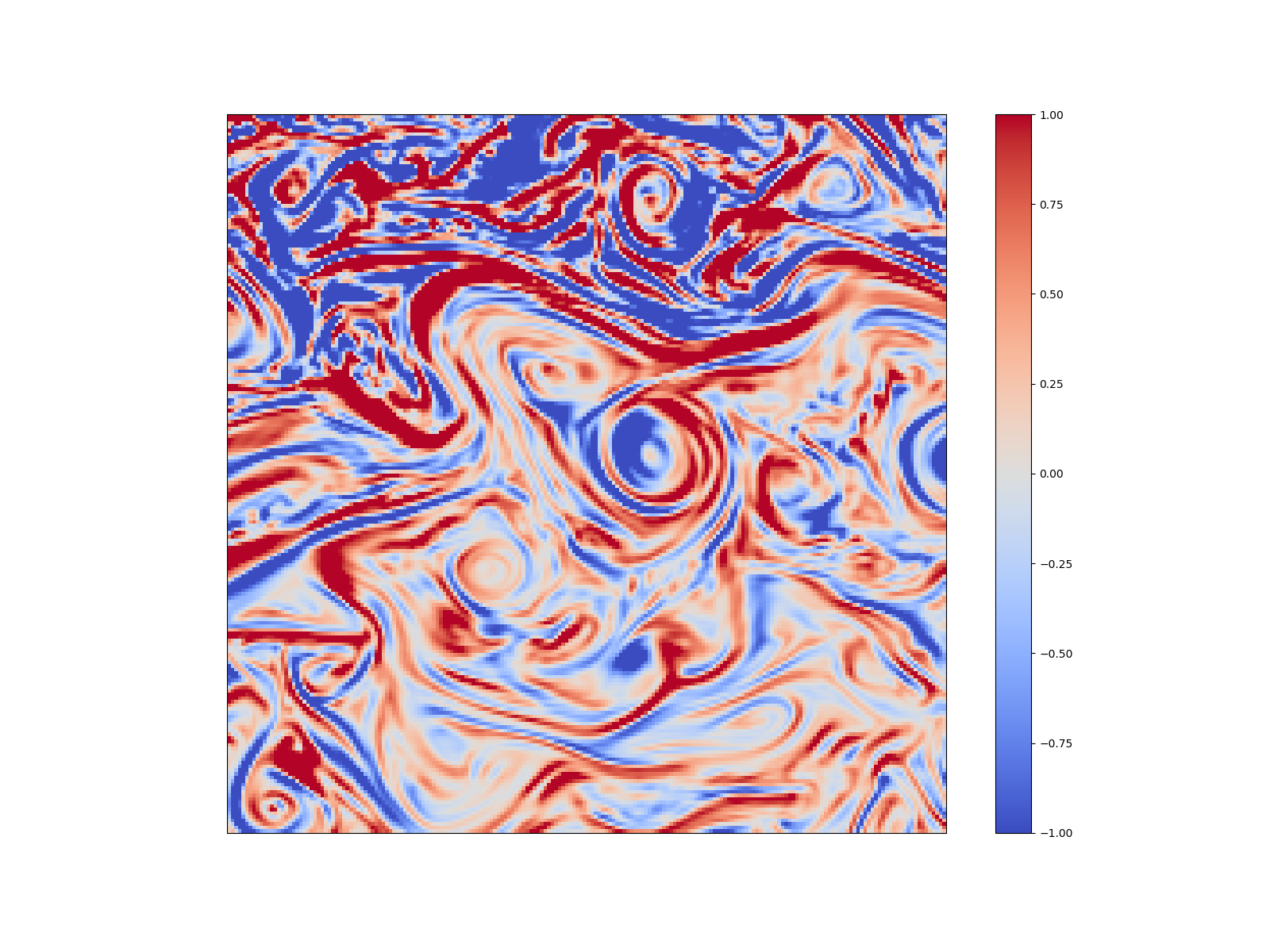}&
    \includegraphics[trim={250 100 300 100},clip,width=2.75cm]{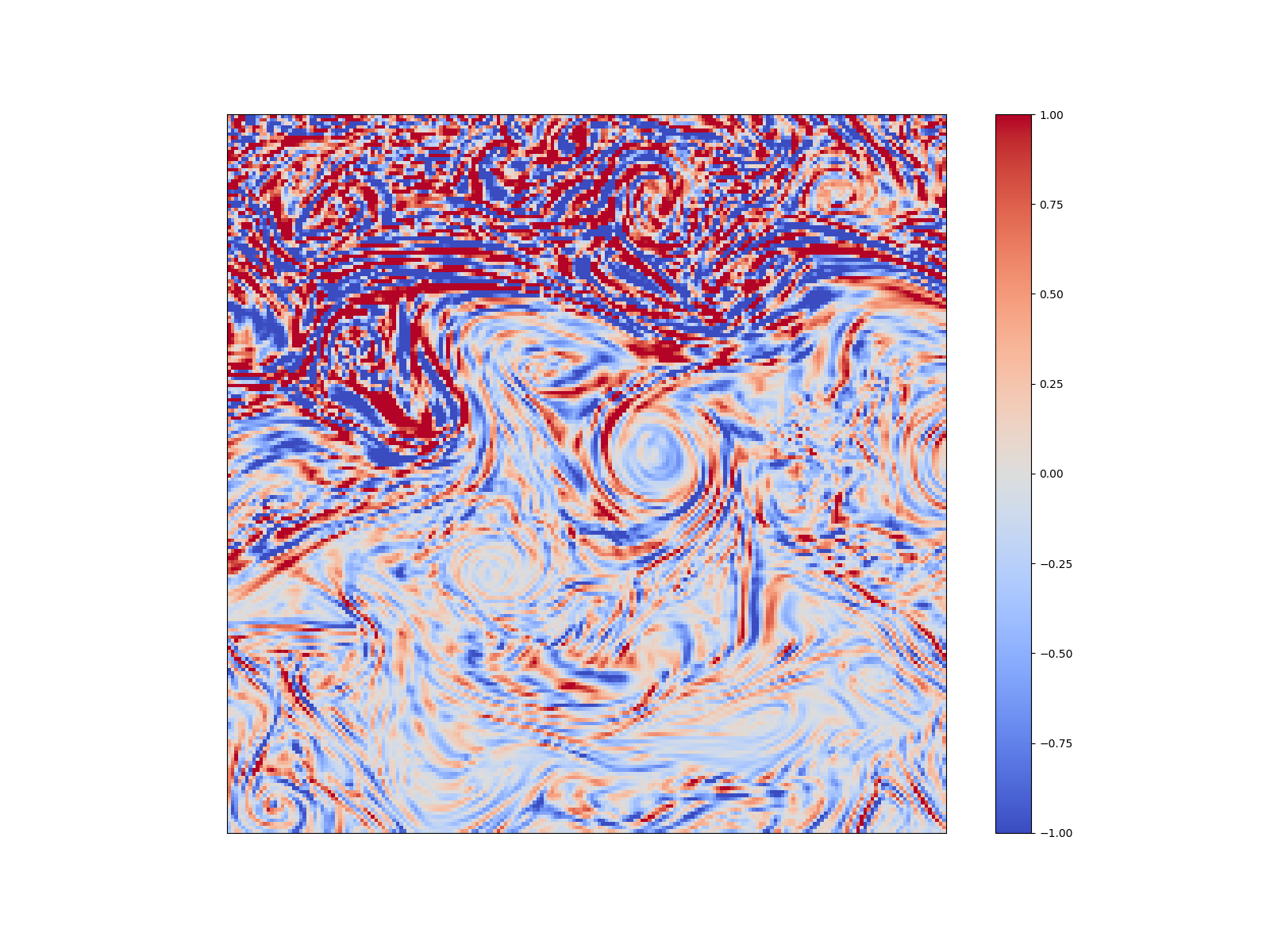}&
    \includegraphics[trim={250 100 300 100},clip,width=2.75cm]{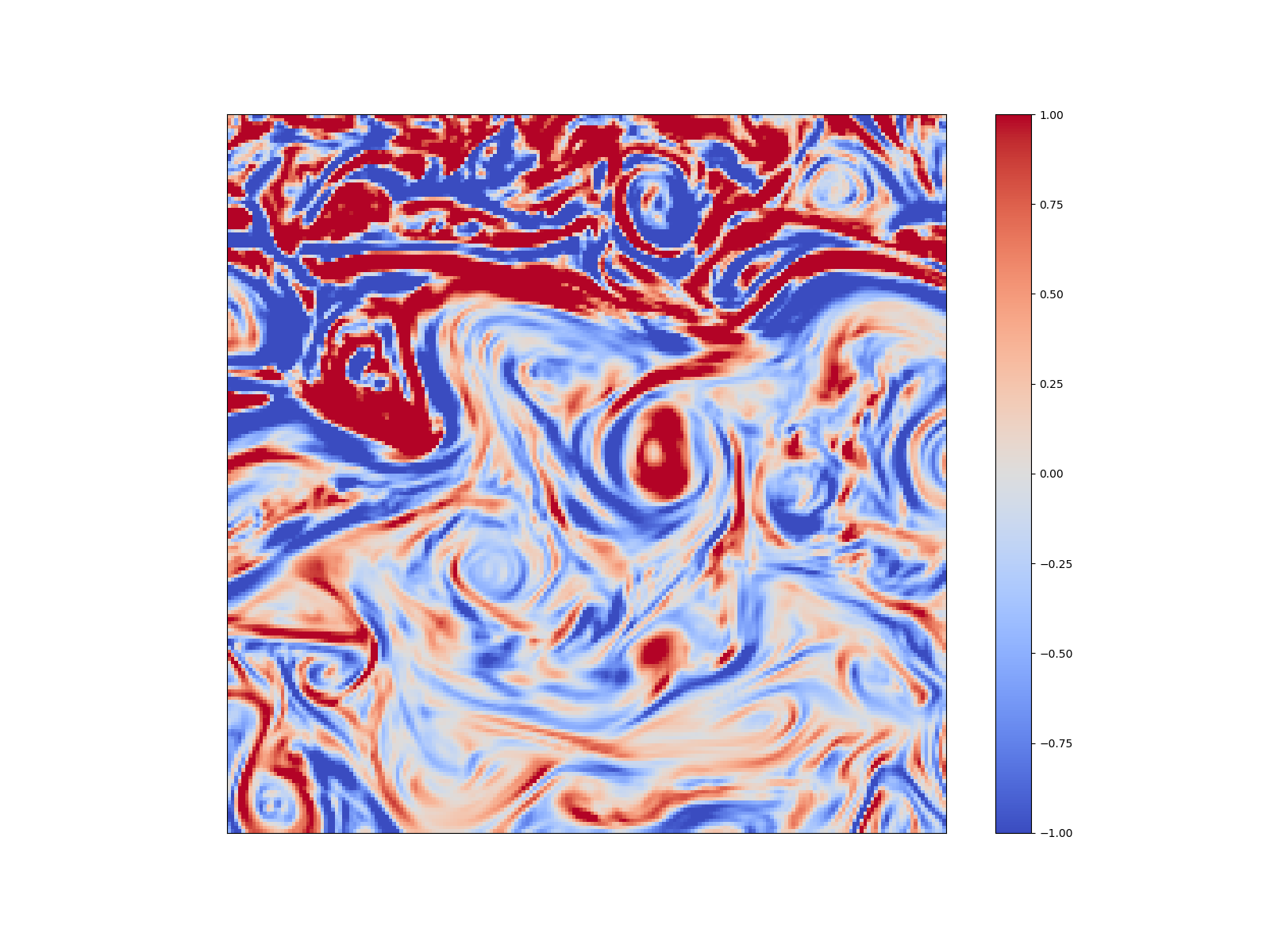}\\
    \includegraphics[trim={250 100 300 100},clip,width=2.75cm]{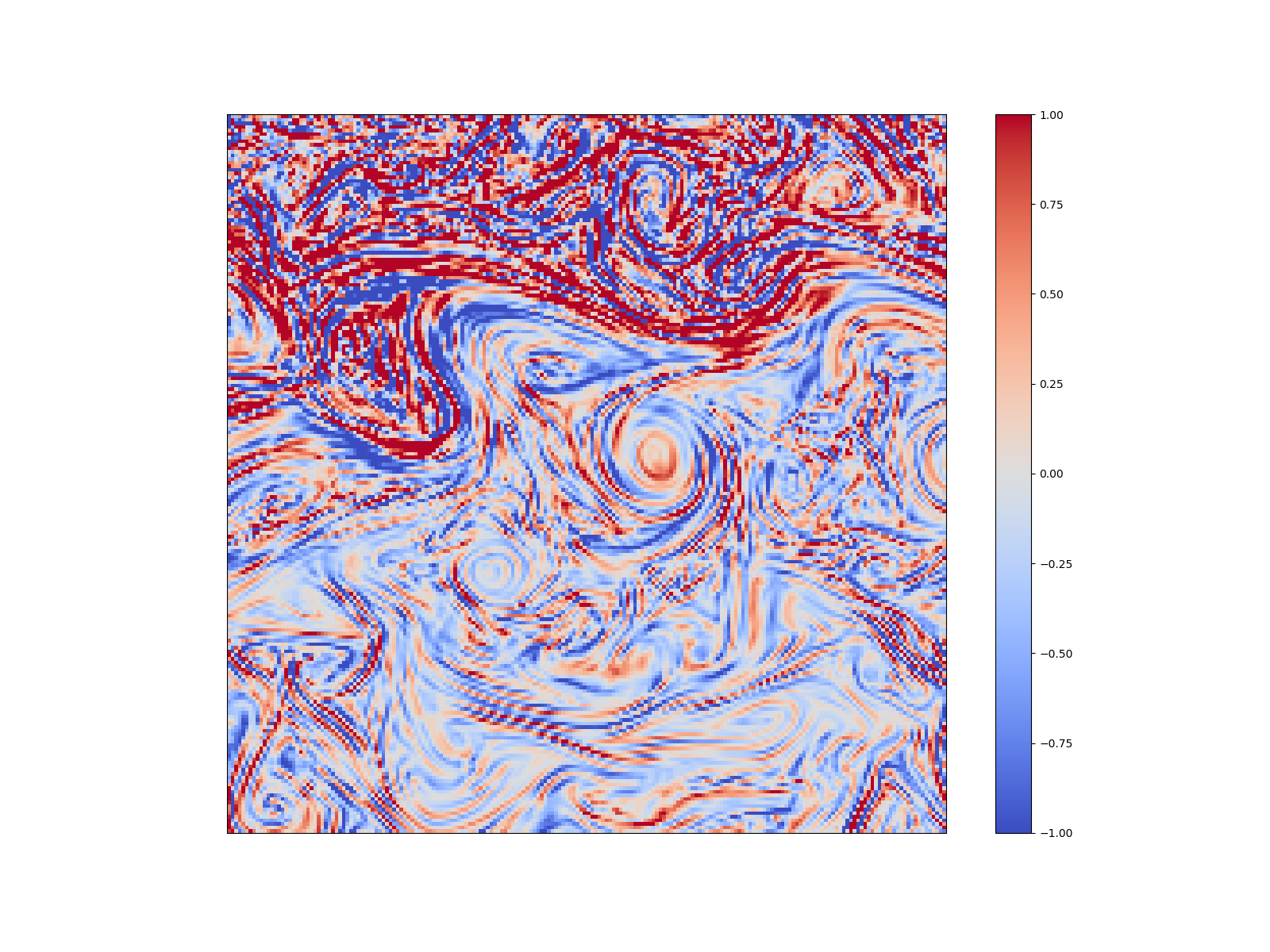}&
    \includegraphics[trim={250 100 300 100},clip,width=2.75cm]{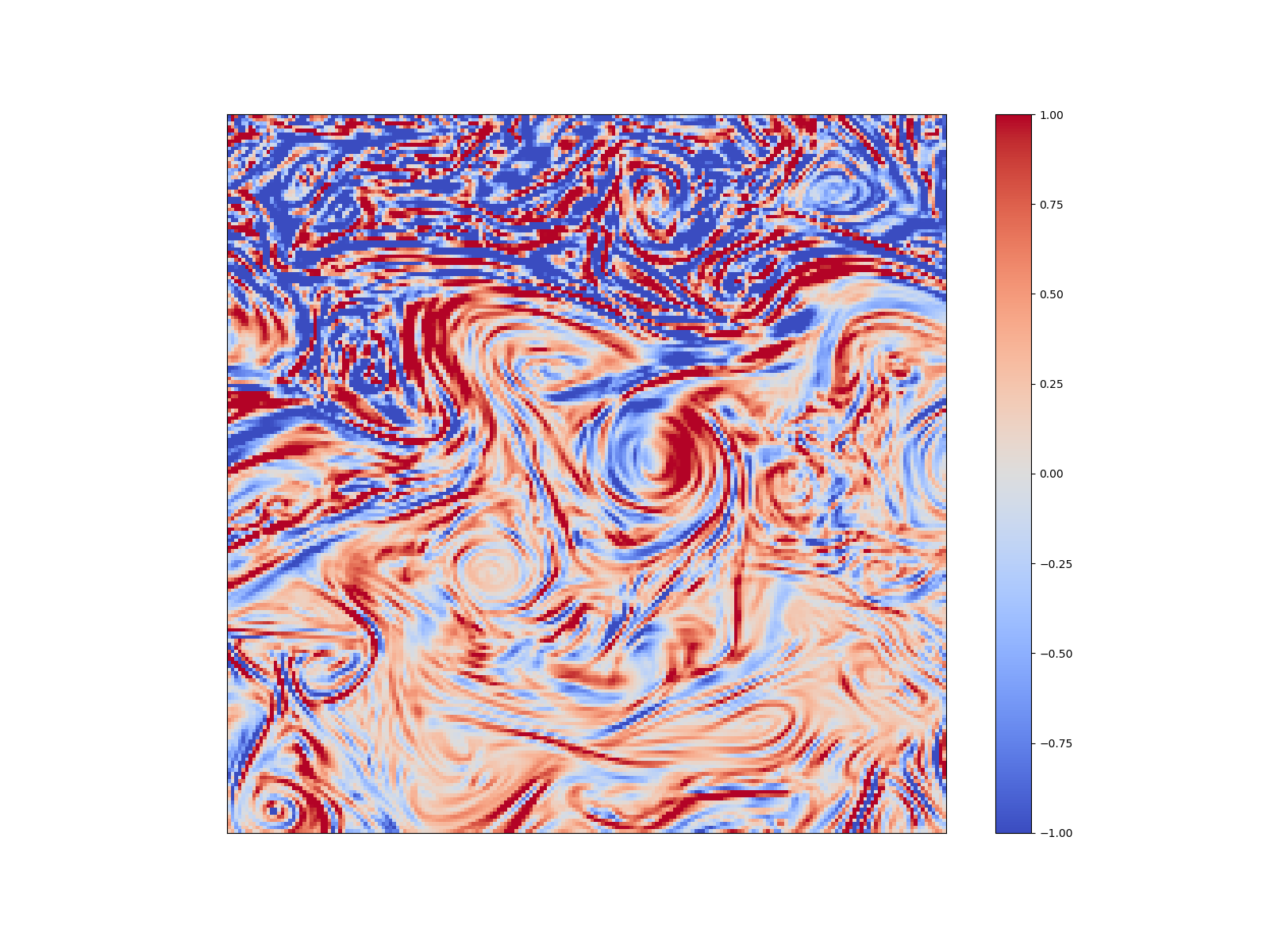}&
    \includegraphics[trim={250 100 300 100},clip,width=2.75cm]{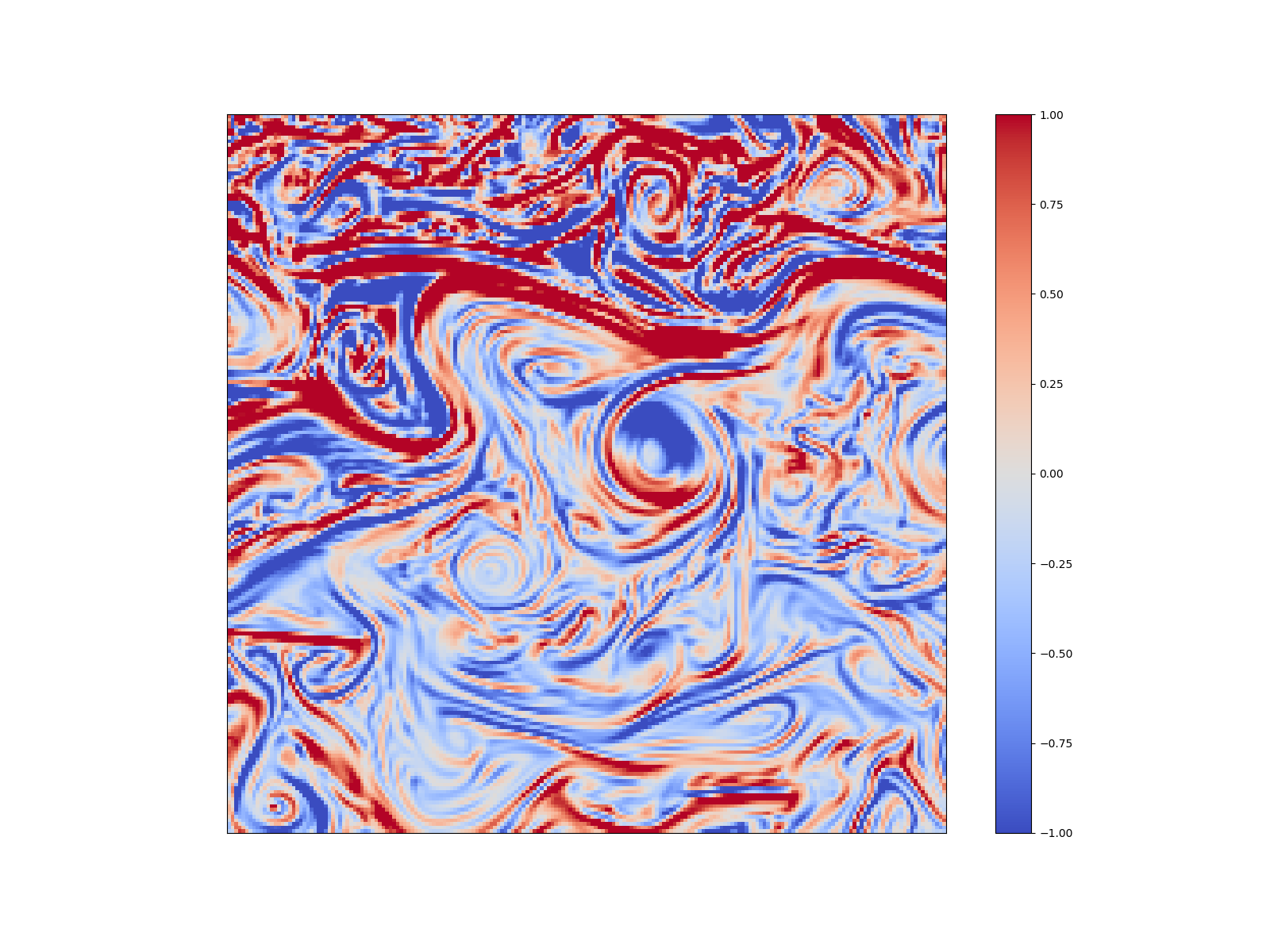}&
    \includegraphics[trim={250 100 300 100},clip,width=2.75cm]{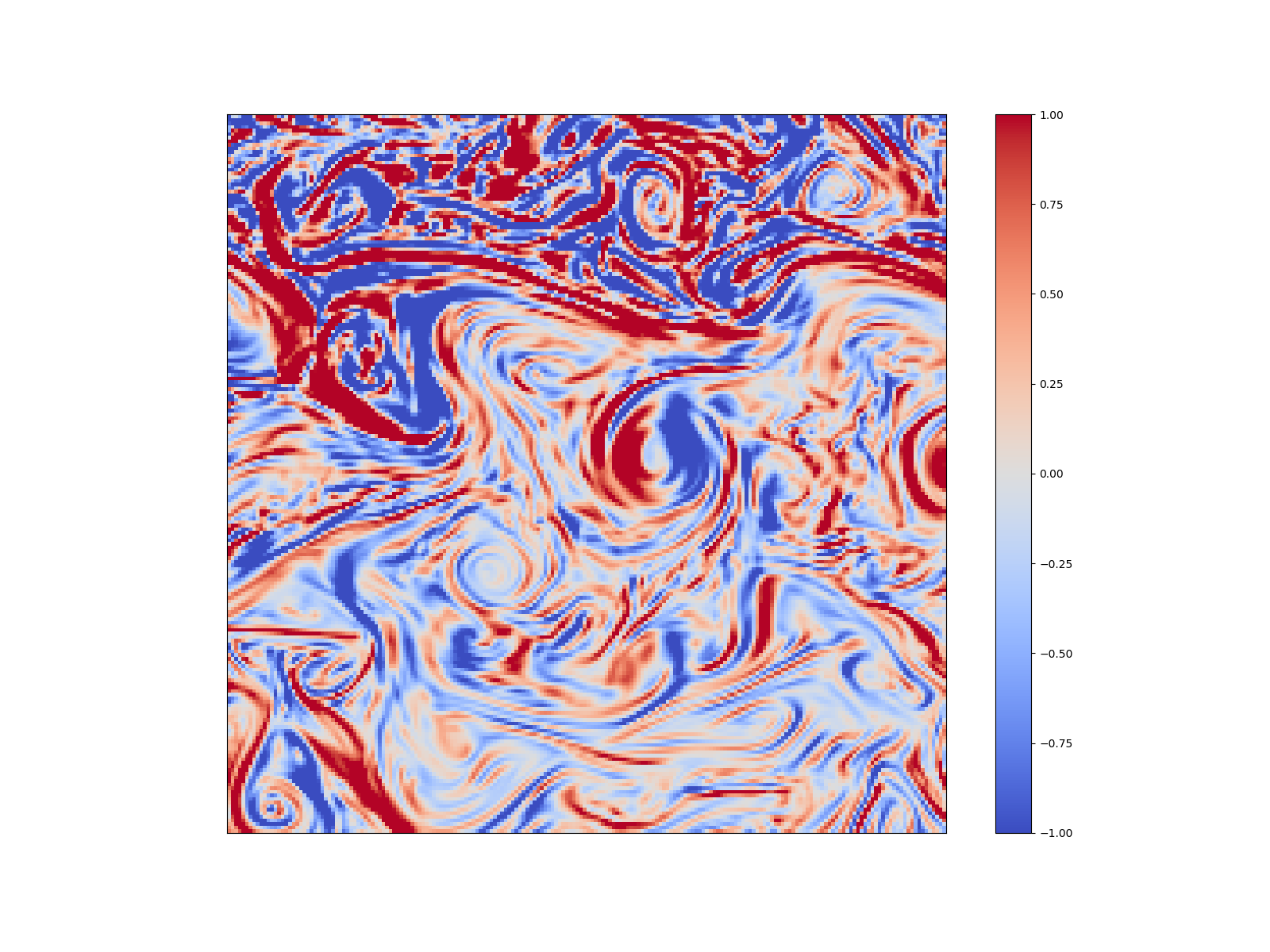}&
    \includegraphics[trim={250 100 300 100},clip,width=2.75cm]{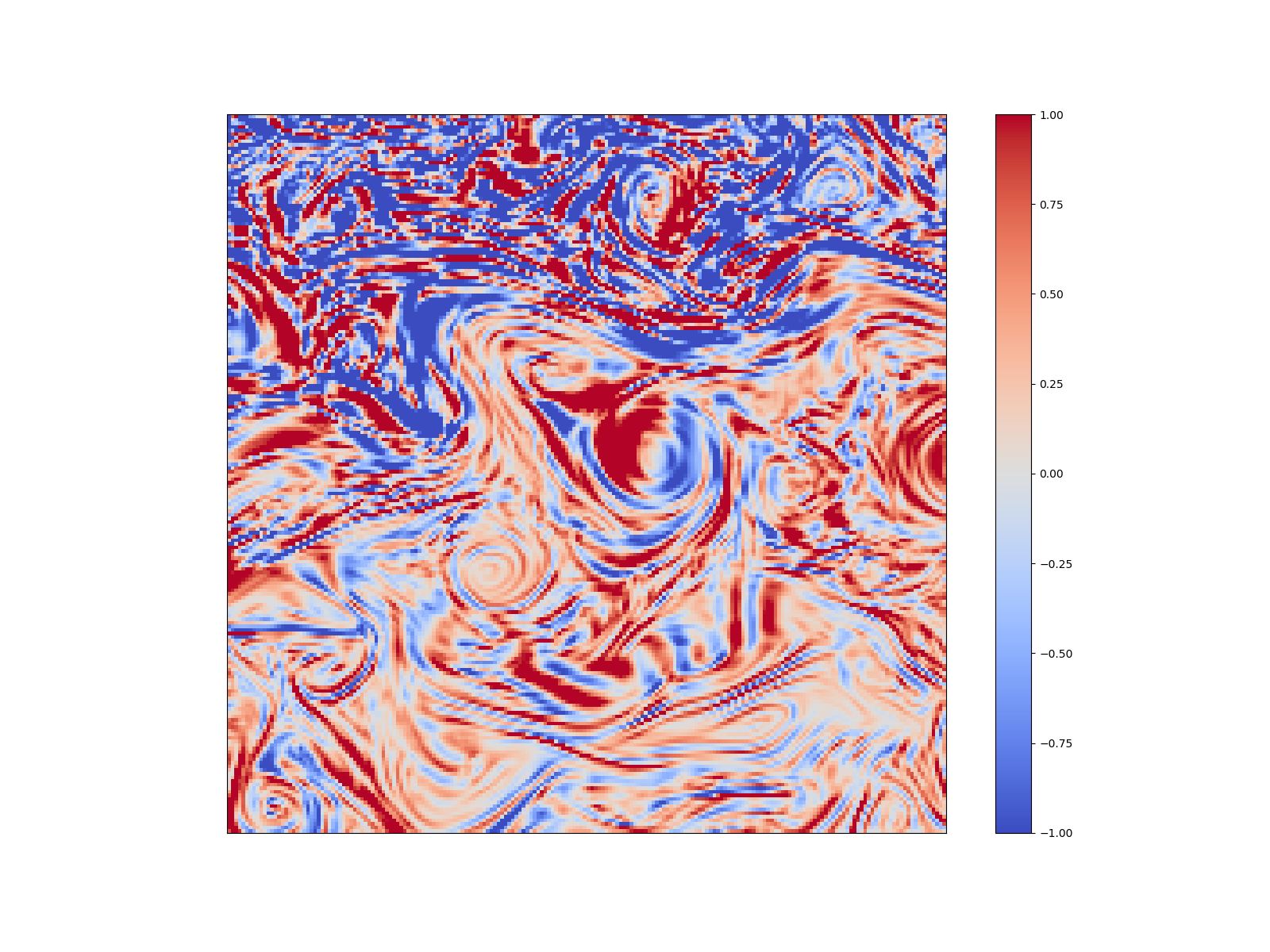}\\
    \includegraphics[trim={250 100 300 100},clip,width=2.75cm]{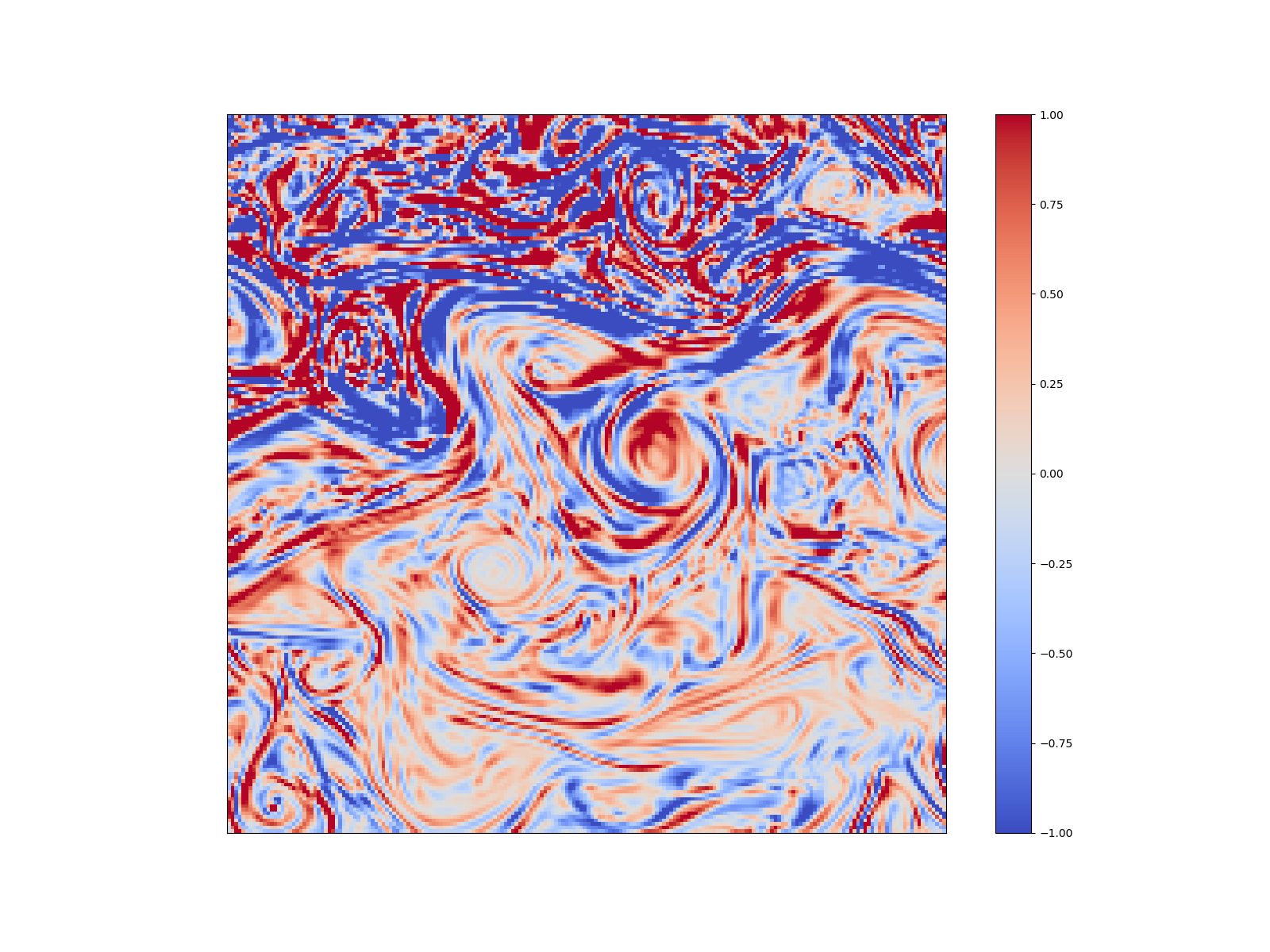}&
    \includegraphics[trim={250 100 300 100},clip,width=2.75cm]{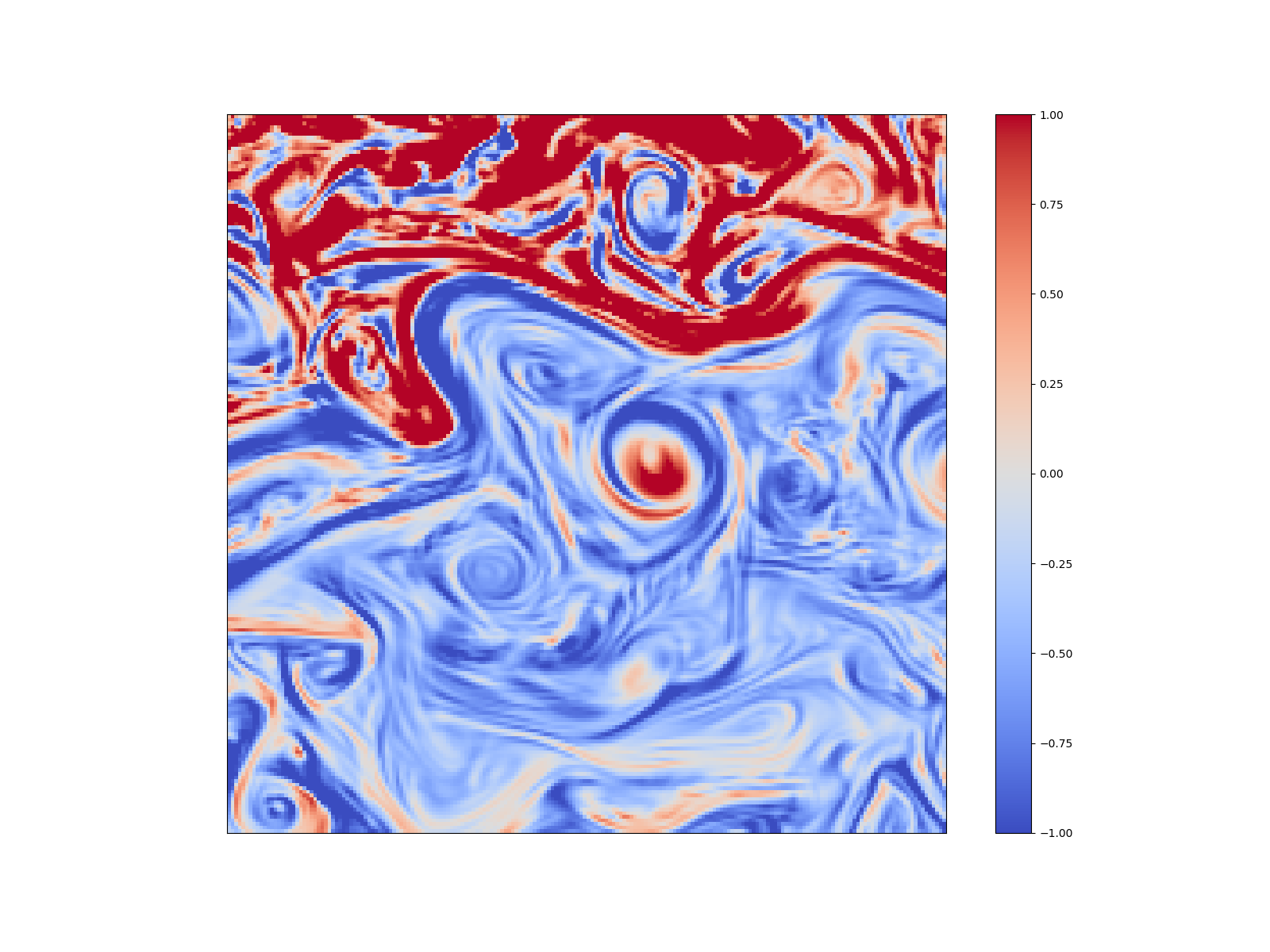}&
    \includegraphics[trim={250 100 300 100},clip,width=2.75cm]{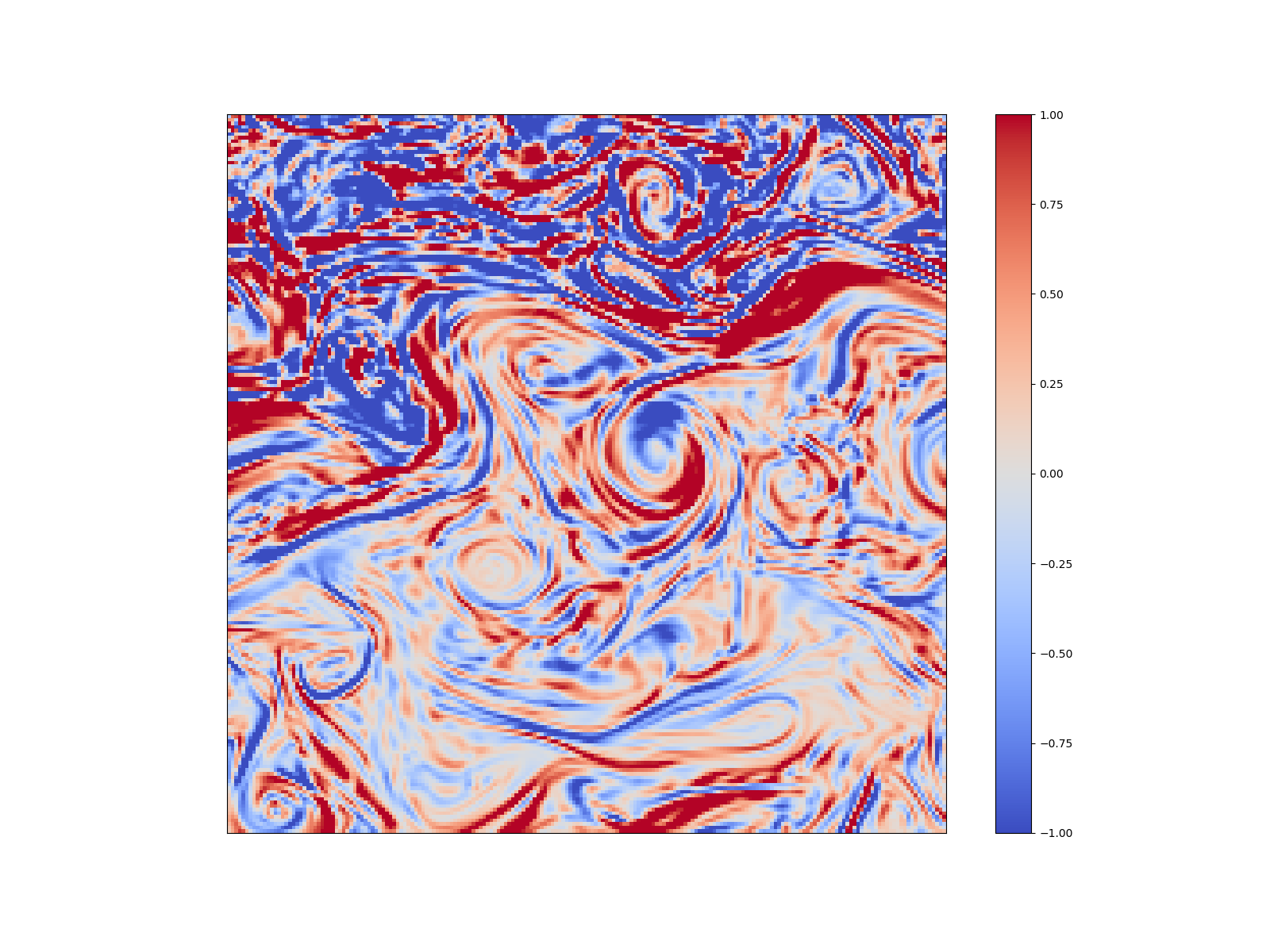}&
    \includegraphics[trim={250 100 300 100},clip,width=2.75cm]{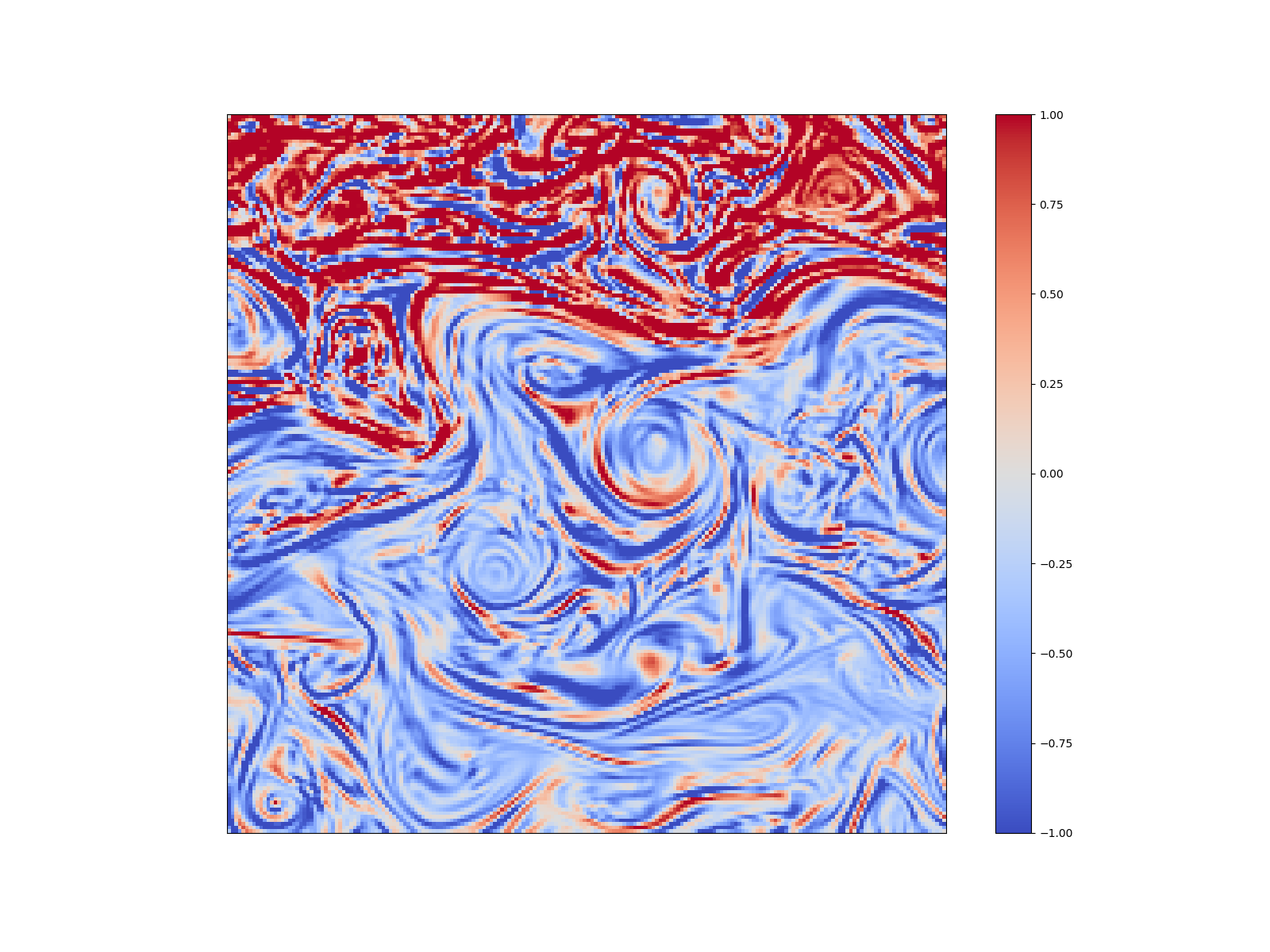}&
    \includegraphics[trim={250 100 300 100},clip,width=2.75cm]{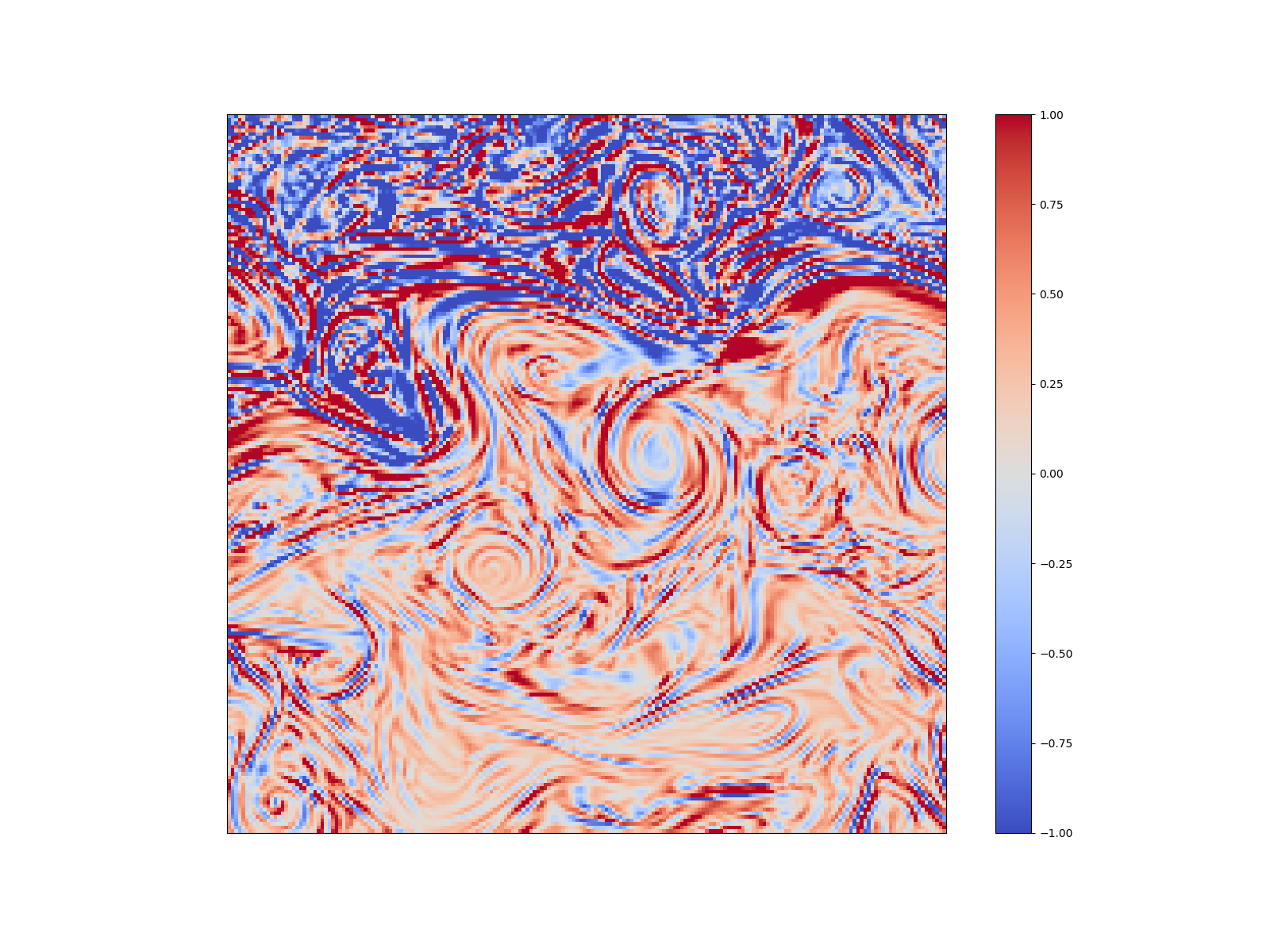}\\
    \end{tabular}
    \caption{{\bf Learnt SST features extracted on October 25$^{th}$ 2012:} we depict the maps of the learnt features extracted from the associated input 7-day SST field sequence for the best 4DVarNet schemes with multi-modal observation terms (\ref{eq: multimodal obs}) using linear (top) and non-linear (bottom) operators. From  \ref{tab:res MM model} and Tab.\ref{tab:res obs model}, we select a 3-dimensional linear configuration and a 20-dimensional non-linear one.
    }
    \label{fig:non-linear features}
\end{figure*}

\subsection{Impact of the resolution of SST observations}

We further evaluate how the resolution of SST observations affects the reconstruction of the SSH. We may remind that satellite-derived microwave SST data typically lead to relatively low-resolution data with a typical $1/4^\circ$ horizontal resolution, but almost gap-free observations on a daily scale \cite{donlon_operational_2012,ocarroll_observational_2019}. By contrast, infrared satellite sensors lead to much higher-resolution observations even below $1/100^\circ$, however at the expense of possibly large missing data rates due to the sensitivity to the cloud coverage \cite{donlon_operational_2012,ocarroll_observational_2019}. Here, using the trained multimodal model with gap-free $1/20^\circ$ SST fields, we assess the reconstruction performance when providing during the evaluation procedure with coarsened versions of the SST data. The coarsening proceeds as follows. We apply an average pooling by a factor ranging from 2 to 10 followed by a linear interpolation onto the original $1/20^\circ$. As such, we simulate SST pseudo-observations with different resolutions from  $1/20^\circ$ to $1/2^\circ$. We synthesize the associated reconstruction performance in Tab.\ref{tab:res SST res}. As expected, the lower the resolution of the SST, the lower the reconstruction performance. This indicates that all spatial scales in SST fields from $1/20^\circ$ to $1/2^\circ$ contribute to inform fine-scale SSH patterns. This is in line with previous studies based on the SQG theory \cite{isern-fontanet_potential_2006}, which considered a scale-invariant hypothesis. Interestingly, up to a  1/4$^\circ$, we report a very significant improvement w.r.t. the altimeter-only baseline (e.g., 4.00 days vs. 5.30 days for the resolved time scale). These results support the potential application of the proposed framework to operational satellite-derived L4 SST products \cite{ocarroll_observational_2019}, which typically resolve horizontal scales between  $1/20^\circ$ and $1/4^\circ$ on a daily resolution. 

Future work could explore how gappy SST observations would affect the reconstruction performance. In this respect, we could consider the SSH-SST state-space formulation with multimodal observation terms to address the  joint interpolation of SST and SSH fields. We could also explore multimodal configuration with different SST data sources for instance from microwave and infrared satellite sensors. As pointed out above, one may consider larger training datasets to overcome overfitting issues.

\begin{table*}[tb]
    \footnotesize
    \centering
     \begin{tabular}{|C{1.5cm}|C{1.5cm}|C{1.5cm}|C{1.5cm}|C{1.5cm}|C{1.5cm}|C{1.5cm}|}
    \toprule
    \toprule
    \bf SST resolution &\bf $\mu$ & $\lambda_x$ ($^\circ$)& $\lambda_t$ (days)&$\tau_{SSH}$&$\tau_{\nabla SSH}$&$\tau_{\Delta SSH}$\\
    \bottomrule
    \toprule
     1/20$^\circ$  & 0.97 & 0.50 & 2.47 &\bf 83.8\%  & \bf 81.4\% & \bf 91.8\% \\
     1/10$^\circ$ & 0.97 & 0.55 & 2.54 &  82.0\% &  78.1\%  & 90.1\%\\ 
     1/5$^\circ$ & 0.96 & 0.67 & 3.87 & 77.7\% &  72.2\% & 87.0\%\\ 
     1/4$^\circ$ & 0.96 & 0.68 & 4.00 & 75.7\% & 69.8\%  & 85.5\%\\ 
     1/2$^\circ$ & 0.95 & 0.88 & 5.65 & 61.5\% &  53.1\%  & 77.3\%\\ 
    \bottomrule
    \bottomrule
    \end{tabular}
    \caption{{\bf Reconstruction performance of the proposed  4DvarNet
    framework depending on the resolution of the SST data}: we report the  performance metrics used in Tab.\ref{tab:res all} for the best 4DvarNet model trained with the full-resolution SST data (1/20$^\circ$) applied with SST data filtered and subsampled at 1/2$^\circ$, 1/4$^\circ$ and 1/8$^\circ$. }
    \label{tab:res SST res} 
\end{table*}

\section{Discussion}
\label{sec: conclusion}

This paper has introduced a novel multimodal learning-based inversion framework for the reconstruction of space-time sea surface dynamics from irregularly-sampled multi-source satellite data. Reported numerical experiments support its relevance compared to the state-of-the-art approaches to exploit SST-SSH synergies and improve the reconstruction of finer-scale sea surface dynamics. We  discuss in this section how the proposed framework relates to and complements previous works according to three different aspects: computational imaging for geoscience, deep learning for data assimilation, deep learning and spaceborne earth observation.

{\bf Computational imaging for geoscience:} End-to-end learning strategies have become the state-of-the-art approaches for a variety of computational imaging problems, including among others  denoising \cite{zhang_plug-and-play_2021,wei_tuning-free_2020}, super-resolution \cite{dong_image_2016} and inpainting issues \cite{xie_image_2012,liu_image_2018}. While numerous applications to the observation and monitoring of geophysical processes exploit  state-of-the-art deep learning schemes, the underlying physical laws naturally advocate for the design of physics-aware approaches. This is particularly true for space-time interpolation issues with very high missing data rates as targeted in our study to the reconstruction of sea surface dynamics from satellite data. We may also point out that video inpainting remain a challenge for deep learning \cite{kim_deep_2019}. Here, we exploit a classic inverse problem formulation to design our neural architecture. This allows us to make explicit the definition of the observation operators and the dynamical prior. As supported by our numerical experiments, the former is key to fully exploit SSH-SST synergies to improve the reconstruction of sea surface dynamics at finer scales. The proposed multimodal 4dVarnet scheme also relates to deep unfolding schemes \cite{mccann_convolutional_2017}. Our neural architecture implements an iterative gradient descent of the trainable variational cost. Rather than considering a reaction-diffusion formulation \cite{chen_learning_2015} or an optimization scheme associated with proximal operators \cite{kobler_total_2020}, our framework exploits the embedded automatic differentiation of the variational cost with a trainable gradient-based solver to speed up the optimization. This strategy also allows us to train jointly linear and non-linear observation operators with the dynamical prior and the solver. This is expected to contribute to reducing inversion biases \cite{fablet_learning_2021,holler_bilevel_2018,liu_bilevel_2019} and improving the overall interpretability of the neural architecture.

{\bf Deep learning and data assimilation:} the proposed framework can be regarded as a neural resolution of a variational data assimilation formulation. While data assimilation schemes
\cite{evensen_data_2009} are widely used in geoscience for the reconstruction of space-time dynamics, they require the explicit knowledge of the underlying dynamics and of the observation operators in (\ref{eq: multimodal obs}). In the considered case-study, the classical choice would be to consider an ocean general circulation model \cite{evensen_data_2009,benkiran_assessing_2021} and identity observation operators with covariance priors. This results in a much more computationally-demanding inversion scheme compared with our approach. Besides, such schemes implemented in
operational systems typically lead to reconstruction performance for sea surface dynamics in the same range as the operational baseline considered in our experiments \cite{taburet_duacs_2019}. Our study illustrates the versatility of deep learning frameworks to define a state-space formulation which only comprises the variable of interest (here, the SSH possibly complemented by the SST). The significant improvement over model-driven approaches such as \cite{le_guillou_mapping_2020,ubelmann_reconstructing_2021} 
stresses the ability and relevance to learn the underlying variational representation from data. This seems particularly appealing to explore multimodal synergies between different geophysical tracers, when the derivation of physical laws to relate the processes of interest reveal complex. In such a context, the proposed framework may also contribute to the identification of such laws from data as the calibration of the observation operators is a by-product of the training process.     

{\bf Deep learning and space oceanography:} Spaceborne earth observation and ocean remote sensing greatly benefit from deep learning advances. Given orbiting characteristics as well as the sensitivity of satellite sensors to the atmospheric conditions, spaceborne earth observation data result in an irregular space-time sampling which may involve very large missing data rates as illustrated here for satellite altimetry data. While optimal interpolation schemes remain the state-of-the-art processing for a wide range of operational gap-free satellite-derived products \cite{taburet_duacs_2019,donlon_operational_2012}, deep learning schemes emerge as relevant approaches \cite{george_deep_2021,fablet_end--end_2021,barth_dincae_2020}. This study further supports their relevance to best exploit multimodal synergies which are not easily accounted for with optimal interpolation schemes. The ever-increasing availability of observation data and numerical simulations also greatly contribute to the development and evaluation of learning-based and data-driven approaches as illustrated by the considered experimental setting based on an open data challenge\footnote{\url{https://github.com/ocean-data-challenges/2020a_SSH_mapping_NATL60}}.  We could apply and extend the proposed framework to other space-time geophysical products such as ocean colour \cite{volpe_mediterranean_2019,sathyendranath_ocean-colour_2019}, sea surface turbidity \cite{vient_data-driven_2021,renosh_construction_2017}, sea and land surface temperature \cite{barth_dincae_2020}, sea surface currents \cite{ciani_ocean_2021,ocarroll_observational_2019}... Future challenges also involve joint calibration and interpolation issues for future satellite missions \cite{febvre_joint_2022} as well as multimodal synergies between satellite data and other remote sensing and in situ data sources such as drifters \cite{sun_impacts_2022,baaklini_blending_2021}, underwater acoustics data \cite{cazau_wind_2019}, moored buoys \cite{fujii_observing_2019}, argo profilers \cite{cossarini_towards_2019,dortenzio_biogeochemical_2020}... Especially, the latter might provide new ways to better monitor the interior of the ocean which cannot be directly observed from space. 

\section*{\bf Acknowledgements}
This work was supported by LEFE program (LEFE MANU project IA-OAC), CNES (grant OSTST DUACS-HR) and ANR Projects Melody and OceaniX. It benefited from HPC and GPU resources from Azure (Microsoft Azure grant) and from GENCI-IDRIS (Grant 2021-101030).  




\bibliographystyle{IEEEtran}

%





\end{document}